\documentclass[%
 reprint,
nofootinbib,
 amsmath,amssymb,
 aps,
]{revtex4-2}

\usepackage{aasmacros}
\usepackage{natbib}
\usepackage{graphicx}% Include figure files
\usepackage{dcolumn}% Align table columns on decimal point
\usepackage{bm}% bold math
\usepackage{newtxtext,newtxmath}
\usepackage{subcaption}
\usepackage[T1]{fontenc}
\usepackage{makecell}
\usepackage{animate}
\usepackage{mathtools}
\usepackage{url}
\usepackage[hidelinks,colorlinks]{hyperref}
\usepackage{xcolor}
\hypersetup{
    colorlinks=true,
    linkcolor={blue},
    citecolor={blue},
    urlcolor={blue}
}
\DeclarePairedDelimiter{\evdel}{\langle}{\rangle}
\newcommand{\ev}{\evdel}
\def\hlinewd#1{%
\noalign{\ifnum0=`}\fi\hrule \@height #1 %
\futurelet\reserved@a\@xhline} 
\usepackage{graphicx}	% Including figure files
\usepackage{amsmath}	% Advanced maths commands

\newcommand{\overbar}[1]{\mkern 1.5mu\overline{\mkern-1.5mu#1\mkern-1.5mu}\mkern 1.5mu}

\begin{document}

\preprint{APS/123-QED}

\title{Till the core collapses: the evolution and ensemble properties of self-interacting dark matter subhalos}% Force line breaks with \\

\author{Zhichao Carton Zeng$^{1,2,3}$}\email{E-mail: zeng.408@osu.edu}
\author{Annika H. G. Peter$^{1,2,4,5}$}
\author{Xiaolong Du$^{6,7}$}
\author{Shengqi Yang$^{6,13}$}
\author{Andrew Benson$^{6}$}
\author{Francis-Yan Cyr-Racine$^{8}$}
\author{Fangzhou Jiang$^{6,9,10}$}
\author{Charlie Mace $^{1,2}$}
\author{R. Benton Metcalf $^{11,12}$}

\affiliation{$^{1}$Department of Physics, The Ohio State University, 191 W. Woodruff Ave., Columbus OH 43210, USA \\
$^{2}$Center for Cosmology and Astroparticle Physics, The Ohio State University, 191 W. Woodruff Ave., Columbus OH 43210, USA\\
$^{3}$Department of Physics and Astronomy, Mitchell Institute for Fundamental Physics and Astronomy, Texas A\&M University, College Station, TX 77843, USA \\
$^{4}$Department of Astronomy, The Ohio State University, 140 W. 18th Ave., Columbus OH 43210, USA\\
$^{5}$School of Natural Sciences, Institute for Advanced Study, 1 Einstein Drive, Princeton, NJ 08540\\
$^{6}$ Carnegie Observatories, 813 Santa Barbara Street, Pasadena CA 91101, USA\\
$^{7}$ Department of Physics and Astronomy, University of California, Los Angeles, CA 90095, USA\\
$^{8}$ Department of Physics and Astronomy, University of New Mexico, 210 Yale Blvd NE, Albuquerque NM 87106, USA \\
$^{9}$ TAPIR, California Institute of Technology, Pasadena, CA 91125, USA \\
$^{10}$ Kavli Institute for Astronomy and Astrophysics, Peking University, Beijing 100871, People’s Republic of China \\
$^{12}$ Dipartimento di Fisica \& Astronomia, Universitá di Bologna, via Gobetti 93/2, 40129 Bologna, Italy \\
$^{12}$ INAF-Osservatorio Astronomico di Bologna, via Ranzani 1, 40127 Bologna, Italy\\
$^{13}$ Los Alamos National Laboratory, Los Alamos, NM 87545, USA
}

\date{\today}% It is always \today, today,

\begin{abstract}
One of the hottest questions in the cosmology of self-interacting dark matter (SIDM) is whether scatterings can induce detectable core-collapse in halos by the present day. Because gravitational tides can accelerate core-collapse, the most promising targets to observe core-collapse are satellite galaxies and subhalo systems.  However, simulating small subhalos is computationally intensive, especially when subhalos start to core-collapse.
In this work, we present a hierarchical framework for simulating a population of SIDM subhalos, which reduces the computation time to linear order in the total number of subhalos. With this method, we simulate substructure lensing systems with multiple velocity-dependent SIDM models, and show how subhalo evolution depends on the SIDM model, subhalo mass and orbits.
We find that an SIDM cross section of $\gtrsim 200$ cm$^2$/g at velocity scales relevant for subhalos' internal heat transfer is needed for a significant fraction of subhalos to core-collapse in a typical lens system at redshift $z=0.5$, and that core-collapse has unique observable features in lensing. 
We show quantitatively that core-collapse in subhalos is typically accelerated compared to field halos, except when the SIDM cross section is non-negligible ($\gtrsim \mathcal{O}(1)$ cm$^2$/g) at subhalos' orbital velocities, in which case evaporation by the host can delay core-collapse. This suggests that substructure lensing can be used to probe velocity-dependent SIDM models, especially if line-of-sight structures (field halos) can be distinguished from lens-plane subhalos. Intriguingly, we find that core-collapse in subhalos can explain the recently reported ultra-steep density profiles of substructures found by lensing with the \emph{Hubble Space Telescope}.
\end{abstract}

\maketitle

%\tableofcontents

\section{Introduction}

Self-interacting dark matter (SIDM) refers to a group of dark matter models that allows for self-interactions among dark matter particles in addition to gravity, typically in the form of localized two-body scatterings \citep{spergel00, dm17, tulin17, buckley17, sidm22}. Recently, it has gained  attention as an interesting candidate for dark matter \citep{kaplinghat16, Kamada17, sameie20, Kaplinghat20, correa21, zzc22, correa22, joy22, nadler23, silverman23, dnyang22}, potentially reconciling the observed larger diversity of rotation curves in dwarf galaxies than found in simulations with cold dark matter (CDM; \cite{kdn14, oman15, errani18, read19, relatores19, santos20, hayashi20, li20}). SIDM has received major interest in its potential to tackle small-scale dark matter problems because it has a unique two-phase dark-matter halo time evolution. First, the momentum exchange among dark matter particles leads to  thermalization and thus to an isothermal, cored (constant, non-diverging) density distribution in the center of the halo, shallower than its CDM counterpart (first invoked in 2000's for the cusp-core and missing satellite problems; \cite{spergel00, kochanek00, dave01}).  Following this phase, and on a much longer timescale, the halo transitions to the core-collapse phase. The core serves as a heat bath, slowly and continuously transferring energy to the outer parts of the halo because of the negative temperature gradient in the halo.   This causes dark matter particles in the core to fall towards the halo center, becoming even hotter than before and thus steepening the negative temperature gradient \citep{Colin02}. 
Through this runaway process, the initial core becomes hotter and denser as it gives up heat to the halo outskirts, because of this negative heat capacity of self-gravitating systems.
This second phase of an SIDM halo is referred to as the core-collapse or the gravothermal collapse process \citep{balberg02}, the end state of which is predicted to be possible progenitors of supermassive black holes \citep{balberg02b, pollack15, shapiro18, wxfeng21, hyxiao21, Meshveliani23}. The inner density profile of an SIDM halo can therefore possibly span a wide range during its evolution, from a shallow core to an ultra-high-density core embedded in a steeply cuspy halo. 

However, the timescale for core-collapse can be very long. For example, galaxy clusters currently set the most stringent constraints on SIDM, based on observations of the central densities of clusters \citep{rocha13, elbert18, andrade22}, cluster ellipticity \citep{Peter13, robertson19, McDaniel21, shen22, despali22, robertson23}, and cluster mergers and galaxy offsets \citep{randall08, kim17, robertson17, cross23, fischer23}. The SIDM (constant) cross section per unit mass $\sigma/m$, which is the key parameter to characterize the strength of this self-interaction, is constrained to $\sigma/m \lesssim \mathcal{O}(1)$ cm$^2$/g. 
The core-collapse timescale of typical dark matter halos with such cross sections would be at least of order $\mathcal{O}(10^2)\ \rm Gyr$ \citep{koda11, essig19, zzc22}, much longer than the Hubble time. Additional degrees of freedom in SIDM models are required in order to satisfy both the low cross sections preferred by observations at cluster scales and the high cross sections that would produce a diversity of density profiles on dwarf galaxy scales \citep{valli18, kaplinghat19, sameie20, correa21, dnyang22}. A popular choice is to make $\sigma/m$ velocity-dependent \citep{mv12, banerjee20, col21, joy22, turner21, correa22, neev23}.  A generic feature from well-motivated  particle physics models is that cross section is higher at the velocity scales relevant for dwarf-sized halos while insignificant ($\leq 1\ \rm cm^2/g$) at cluster scales. We will demonstrate a few of these velocity-dependent SIDM models in this work.

Substructure lensing is another way to probe the density distribution in small halos, even smaller than the ones we use for dwarf rotation curves.  Hence, it potentially has the power of distinguishing among velocity-dependent SIDM models. 
The images of strongly lensed objects, either a quasar which has four discrete images, or a source galaxy which then has a closed shape of a distorted image (Einstein ring), can be perturbed by the existence of dark matter substructures (subhalos) near the center of the main lens. Properties of dark matter subhalos can thus be inferred from the anomalies in these strongly lensed images (quasar source -- quad lens: \cite{Dalal02, metcalf05, metcalf05b, ddxu09, birrer17, dg18, dg20, hsueh20, nierenberg23}; galaxy source -- gravitational imaging: \cite{koopmans05, vegetti10, Meneghetti20, minor21, minor21b, despali22b, ballard23}). Substructure lensing is expected to be especially sensitive to core-collapsed subhalos, whose ultra-compact nature would produce more high-density peaks in the projected density map and thus stronger distortion on lensed images. Recent works have demonstrated the potential of constraining SIDM with substructure lensing \citep{dnyang21, dg21, dg22, nadler23, birendra23}.

In order to utilize  observations to place constraints on SIDM models, theoretical predictions of SIDM subhalos need to fully and consistently capture the key ingredients driving their evolution: the heat transfer and self-gravity within a halo while in isolation, and the interaction between subhalos and their host halos.  The physical interactions between subhalos and the host include tidal \citep{Pennarrubia10, errani23} and evaporation effects (also called the self-interaction ram-pressure stripping; \cite{kummer18, mv19, slone21, nadler20, zzc22}). There is a sizable body of work on the evolution of isolated SIDM halos until core-collapse, using  simulations \cite{koda11, fischer24b, tran24} or less costly analytical methods (e.g., the gravothermal fluid model; \cite{balberg02, koda11, pollack15, essig19, o22, sq22}; isothermal jeans modelling; \cite{robertson21, fz22}), which show good agreement with each other, perhaps up to a calibration parameter \cite{mace25}. 

However, there are few predictions for subhalos \citep{slone21, shirasaki22, neev23, dnyang23, sashimi-sidm}. 
Accurate predictions for small subhalos in massive hosts remain a challenge, especially if subhalos enter the core-collapse regime.  While cosmological simulations that automatically include all the relevant physics are believed to yield the most accurate predictions \citep{mv12, zavala13, robertson19, robles19, banerjee20, dnyang22, correa22}, the simulations get prohibitively expensive to resolve the $10^6-10^7M_\odot$ subhalos expected to be detectable with upcoming lensing surveys \citep{vegetti23}. Zoom-ins of cosmological simulations can focus on simulating a tiny region corresponding to an isolated dwarf galaxy \citep[Cruz et al in Prep.]{xjshen21, xjshen24}, but also with the disadvantage of lacking host halo environments. This challenge of computational affordability becomes even more severe when any of the simulated (sub)halos are in the core-collapse regime, which requires higher resolution in both particle number and time-stepping to ensure accuracy \cite{mace24, fischer24, palubski24}. 

Therefore, in this work we are motivated to design a new simulation framework that allows us to simulate a realistic population of SIDM subhalos from core-formation to core-collapse, with relatively high fidelity and low cost. \cite{zzc22} described and validated a semi-analytical method to track individual subhalos with particles while representing the host as an analytic density distribution and semi-analytically calculating the host-subhalo evaporative interactions.  In this work, we extend the method of \cite{zzc22} to incorporate a realistic subhalo assembly history in a massive host.  We further divide a whole merger tree of subhalos into different groups according to their masses at infall, and simulate with different mass resolution separately, such that the computational cost grows only linearly with the number of subhalos. 

This work is organized as follows: in Sec. \ref{sec:method} we describe our simulation framework in detail. In Sec. \ref{sec:param-space}, we present the SIDM models we study in this work, both a constant cross section one and several models with velocity-dependent cross sections, as well as the specifics of each realization of system we simulate. We present the main results from these simulations in Sec. \ref{sec:results}, with a special focus on probes of subhalo core-collapse.  We describe halo  statistics, lensing statistics, and an analysis of the interplay of different mechanisms in determining these result using  phenomenologically tuned velocity-dependent SIDM models. We summarize and discuss this work in Sec. \ref{sec:summary}. 

\section{Method}\label{sec:method}
In this work, we track and study the evolution of a realistic population of SIDM subhalos in a typical strong lens system. Resolving low-mass subhalos down to $\sim 10^8M_\odot$ at infall (thus $10^6-10^7M_\odot$ at the end of the simulation), which is predicted to be detectable with upcoming substructure lensing surveys, can be very challenging in cosmological simulations. For example, to resolve the inner mass distribution of a subhalo whose infall mass was $10^8 M_\odot$ at present day ($\sim10^7 M_\odot$ if the mass loss were $\sim$90\%) with a thousand-ish particles (generally considered as adequate resolution for subhalos; \cite{vdb18, nadler23, nadler23symphony}), we would require the simulation particle mass to be at most $10^4 M_\odot$.  This  corresponds to $10^9-10^{10}$ particles for the group-sized system with mass $\sim 10^{13}M_\odot$ (typical for strong-lens systems; \cite{birrer17}). Such a computational cost is prohibitive, especially if we want to explore multiple models in the SIDM parameter space or simulate multiple realizations to reduce noise in the predictions. 

This is the motivation for our hierarchical simulation framework.  We use the N-body simulation code \texttt{Arepo} with its SIDM module in this work \citep{Springel10, mv12, mv14, mv19}, as well as a patch to the SIDM module developed in \cite{zzc22} that allows for modelling the host halo analytically while tracking subhalos with N-body particles. We further divide subhalos into different groups according to their masses at infall and simulate them separately with different mass resolution, such that subhalos in different infall mass bins have similar numbers of particles. We are thus able to scale down the total number of particles by two orders of magnitude 
for the same group-sized system, corresponding to a speed boost of about three to four orders of magnitude. 

To simulate a realistic population of SIDM subhalos, we take analytically sampled merger trees of a group-sized system as input, and generate simulations accordingly using our hierarchical framework.
The assembly histories of this population of subhalos are generated by \texttt{Galacticus} \footnote{https://github.com/galacticusorg/galacticus} \citep{Benson:2010kx}, as we show in Sec. \ref{sec:merger_tree}. Details of our hierarchical simulation scheme are elaborated in Sec. \ref{sec:sim_method}. We provide a step-by-step procedure for the whole process in Sec. \ref{sec:step}.

\subsection{Assembly history with Galacticus}\label{sec:merger_tree}

To get the assembly histories of subhalos, we generate $512$ Monte-Carlo merger trees using \texttt{Galacticus} following the algorithm proposed in \cite{Parkinson:2007yh}. The root halo mass is chosen to be $10^{13}M_{\odot}$. The trees are constructed starting from $z=0.5$, a typical redshift of strong gravitational lenses \citep{robertson20}, backward until the leaf nodes reach the mass resolution $2\times 10^7 M_{\odot}$. We then select realizations that have a smooth accretion history to avoid major mergers with a mass ratio larger than $1:10$. This is to ensure that the host properties can be smoothly interpolated along the main branch of the merger tree.  We note that major mergers can be important in the history of lens galaxies, which are mostly ellipticals, but in this work we start with easier-to-model systems, and put off the extension to major mergers to future works.  In total, $37$ trees pass this selection criteria, and we randomly use 4 of them to generate different realizations of simulation in this work. The trees record the time at which each subhalo enters the host's virial radius. We further compute the formation time of each halo, $t_{\rm hf}$, defined as the time when the halo accretes half of its mass at infall.

Each halo in the tree is assumed to have a Navarro–Frenk–White (NFW) profile \cite{Navarro:1995iw} initially
\begin{equation}
\rho(r)=\frac{\rho_0}{r/r_s (1+r/r_s)^2},
\label{eq:NFW}
\end{equation}
with $\rho_0$ the characteristic density and $r_s$ the scale radius. The virial radius of a halo is defined as the radius enclosing a mean density that is $\Delta_{\rm vir}$ times the critical density of the Universe, i.e. $\overline{\rho}(r_{\rm vir})=\Delta_{\rm vir} \rho_{\rm crit}$. In this work, we take $\Delta_{\rm vir}=200$ unless otherwise specified\footnote{Note that the merger trees are built based on the spherical collapse model (including corrections for elliptic collapse). So to be self-consistent, we use a slightly different halo mass definition during the merger tree construction: $\overline{\rho}(r_{\rm vir})=\Delta_{\rm vir,m} \rho_{\rm m}$ with $\rho_m$ the mean matter density in the Universe. The virial density contrast $\Delta_{\rm vir,m}$ is computed from the spherical collapse model \citep{Percival:2005vm}. Then we convert the halo mass to $M_{\rm vir,200}$ assuming an NFW profile.}. The halo concentration $c_{200} \equiv r_{\rm vir,200}/r_s$ is determined following \cite{Ludlow:2016ifl} and \cite{Benson:2019}, which set the concentration based on the mass accretion history of each halo. The host halos we consider in this work have a median concentration of $6.7$ at $z=0.5$.  While the concentration of the host evolves with time as it accretes more mass through mergers, the concentration of each subhalo is set at the infall time. 

To set the initial orbit parameters of subhalos, we randomly place the subhalos on the virial sphere of the host assuming isotropic distribution at the infall time. The radial and tangential infall velocity are drawn from the distribution found by \cite{Jiang:2014zfa}, with the tangential velocity isotropically distributed in the plane tangent to the virial sphere.
Then the orbits of subhalos, characterized by $\overrightarrow{x}(t_{\rm infall})$ and $\overrightarrow{v}(t_{\rm infall})$, are evolved backward to the prior formation time $t_{\rm hf}$ to get the initial condition $\overrightarrow{x}(t_{\rm hf})$ and $\overrightarrow{v}(t_{\rm hf})$. This is because we initialize subhalos with NFW profiles at $t_{\rm hf}$ and need to simulate them with DM self-interaction until $t_{\rm infall}$ to account for their structural evolution prior to infall (see next section for more details). We use Euler's method with a leap-frog integrator to numerically calculate this ``rewinding'' process,  with the host potential determined by the host mass distribution $M_{<r,h}(t)$ that is measured and interpolated from simulation (with dark matter self-interaction when needed; see Sec. \ref{sec:sim_method} and Sec. \ref{sec:step} for more details). 

Throughout this work, we take the cosmological parameters from the Planck 2018 results \citep{Planck:2018jri}: $\Omega_m=0.3153$, $\Omega_b=0.0493$, $\Omega_{\Lambda}=0.6847$, $h=0.6736$, $\sigma_8=0.8111$ and $n_s=0.9649$. The initial linear matter power spectrum used for building the merger trees is computed from the fitting functions given in \cite{Eisenstein:1999jh}.

\subsection{Hierarchical mass grouping of subhalos}\label{sec:sim_method}

For each realization of a host halo and its accreted subhalos, we apply two thresholds on the selection of subhalos: a mass threshold requiring that a subhalo's mass at its infall $M_{\rm infall}\geq 10^8 M_\odot$, and a time threshold requiring that its infall time $t_{\rm infall} \geq 4\ \rm Gyr$ (0 Gyr being the Big Bang). Subhalos within this range of starting masses, which typically experience further mass loss during their orbital evolution, are of interest for current and upcoming substructure lensing observations \citep{vegetti23}. 
The time threshold $t_{\rm infall} \geq 4$ Gyr is chosen  for several reasons.  First, a simulation start time that is the same for all subhalos is advantageous for technical aspects of setting up the simulation suite.  Second, by $t=4$~Gyr, the vast majority of subhalos ($>95\%$) in the merger trees have formed, making this a natural choice (see more discussion in Sec. \ref{sec:summary}).  Third, earlier-accreted subhalos with $t_{\rm infall} < 4$~Gyr experience more mass loss during their orbital evolution, and so are more likely to fall below the detection threshold of substructure lensing. 
For a typical main-lens host halo with mass $\sim 10^{13} M_\odot$ at $z\sim 0.5$, there are around one to two thousand such subhalos that satisfy $M_{\rm infall} \geq 10^8 M_\odot$ and $t_{\rm infall} \geq 4$ Gyr.

We note that these selection criteria, while specific for this work and application to substructure lensing, can be more general with this simulation framework. For example, the mass threshold can be arbitrarily lower since the complexity of our hierarchical framework scales linearly with the number of subhalos.  The infall time threshold can be omitted, at the cost of dividing subhalos into multiple simulations that have different start times  (see Sec. \ref{sec:summary} for more discussion), in the situation where early accreted subhalos are more likely to survive, such as when they are populated with stellar particles and thus have a deep baryonic potential. 

To reduce the computational cost, we introduce a hierarchical simulation scheme to divide these subhalos into different groups, according to their masses at infall: $[10^8, 10^{8.2}) M_\odot, [10^{8.2}, 10^{8.5}) M_\odot, [10^{8.5}, 10^{9}) M_\odot, [10^9, 10^{10}) M_\odot$ and relatively heavy subhalos with masses $\geq 10^{10}M_\odot$. Different mass-groups under this splitting consume similar CPU hours, hence the total computational cost is more controllable. We generate separate simulations for each infall-mass group of subhalos, with different mass resolution, such that the total computational time grows in linear order of the threshold of the subhalo mass function when we try to resolve small subhalos each with at least $\mathcal{O}(10^4)$ particles each at infall. The first four groups all have masses of individual subhalos less than $0.1 \%$ of the host, thus we can neglect the spatial perturbation of the host caused by these subhalos or the dynamical friction on the subhalos (see \cite{zzc22} for a detailed discussion).  For these cases, we simply simulate the host halo with an analytic mass distribution. For the group of subhalos more massive than $10^{10} M_\odot$ at infall, though, we simulate a live host together with the subhalos, since the dynamical friction on the subhalos by the host is no longer negligible. 

We follow the method of simulating SIDM subhalos around an analytic host that is implemented in \cite{zzc22}, where both the tidal effects and the host-subhalo evaporation are self-consistently included. However, unlike in \cite{zzc22} where the analytic host mass distribution is static over time, the host distribution is time-varying, reflecting its mass accretion over time. 
For a host halo within an SIDM Universe, we need to include in its time evolution not only its mass increase history as output from \texttt{Galacticus}, but also the evolution due to dark matter self-interaction, which generally leads to core-creation in its center. Therefore, we set up isolated, host-only simulations at $t=4, 5, 6, 7, 8, 9$ Gyr, with NFW initial conditions and halo parameters from \texttt{Galacticus}, and follow the setup of \cite{zzc22} to simulate each with dark matter self-interaction for 4 Gyrs, until the equilibrium is reached. As the host halo keeps its mass growth, the whole halo does not stay in perfect equilibrium.  However,  the most noticeable difference between an SIDM halo and a CDM halo is in the inner structure, which reaches equilibrium on a relatively short timescale and does not change much with the subsequent mass accretion.  Thus, we consider the treatment of the host in SIDM equilibrium as an appropriate approach. The mass and velocity distribution of these host halos are then measured and interpolated among discrete radial bins and time stamps to build the properties of the analytic host at any time and radius: density $\rho_h(r, t)$, enclosed mass $M_{<r,h}(r, t)$, and radial velocity dispersion $\sigma_{v,h}(r, t)$. 

One source of deviation of our treatment of the host from a cosmological simulation is that we use an NFW profile near the virial radius of the host.  Recent work shows that a halo's density profile steepens relative to NFW at radii within but close to the virial radius \citep{diemer14, more15}.  We argue that it does not have a big impact on the properties of subhalos, since both the tidal effects and evaporation are of low strength at such distance from the host center, and is something we can change in future work. 

We further divide the subhalos based on their potential for mass loss.  Subhalos that have plunging orbits or orbit the inner part of the host for a long time (`heavily-stripping orbits', referred to as h.s.o through the rest of the paper) potentially have more dramatic mass loss than other subhalos with `ordinary' orbits (or o.o. for short). This requires more particles to properly resolve the severely stripped subhalo after its evolution in the host. For the four groups of small subhalos with infall masses $[10^8, 10^{8.2}) M_\odot, [10^{8.2}, 10^{8.5}) M_\odot, [10^{8.5}, 10^{9}) M_\odot, [10^9, 10^{10}) M_\odot$, we select such subhalos and generate high-resolution simulations, separate from their main groups. Our selection is based on two criteria: a) the subhalo's peri-center is closer than 50 kpc from the host center, or b) the subhalo spends more than 2 Gyrs inside a radius of 200 kpc between $t=4$ to 9 Gyr. Subhalos in this `h.s.o.' class with higher particle resolution represent about $20\%$ of the whole population. For the subhalo group with the largest infall masses $\geq 10^{10} M_\odot$, which are simulated together with a live host, we do not have such separate high-resolution runs.

\begin{table*}
	\centering
	\begin{tabular}{lccc} 
		\hline
		Mass group [dex $M_\odot$] & Const. $\sigma/m$ & $v$-dependent \\
		\hline
        $[8.0, 8.2)$  & $ \rm {o.o.} :2\times 10^4$; $\rm{h.s.o.}:1.2\times10^5$ & $\rm {o.o.} :2\times 10^4$; $\rm{h.s.o.}:8\times10^4$ \\
		\hline
        $[8.2, 8.5)$  & $\rm {o.o.}:2\times 10^4$; $\rm{h.s.o.}:1.2\times 10^5$ & $\rm {o.o.} :2\times 10^4$; $\rm{h.s.o.}:8\times 10^4$ \\
		\hline
        $[8.5, 9.0)$  & $\rm {o.o.} :2\times 10^4$; $\rm{h.s.o.}:1.0\times 10^5$ & $\rm {o.o.} :2\times 10^4$; $\rm{h.s.o.}:8\times 10^4$ \\
		\hline
        $[9.0, 10.0)$  & $\rm {o.o.} :2\times 10^4$; $\rm{h.s.o.}:8\times 10^4$ & $\rm {o.o.} :2\times 10^4$; $\rm{h.s.o.}:8\times 10^4$ \\
		\hline
		$\geq 10.0$  & $2\times 10^4$ & $2\times 10^4$ \\
		\hline
	\end{tabular}
	\caption{Minimum number of particles needed to correctly resolve the evolution of subhalos' density profile, for each infall-mass group. For example, for the infall-mass group of $[10^8, 10^{8.2}M_\odot)$ subhalos that are on ordinary orbits (o.o.) with a constant cross section model, we deploy a mass resolution of $m_p=10^8/(2\times10^4)=5\times10^3M_\odot$.  See Appendix \ref{appdx:res} for more details. `o.o.' and `h.s.o.' are abbreviations for ordinary orbits and heavily-stripping orbits.}
    \label{table:res}
\end{table*}

We conducted a detailed convergence test to estimate the resolution needed for each infall-mass group of subhalos (Appendix \ref{appdx:res}). We summarize the resolution results in Table \ref{table:res}, in the form of the minimum number of particles needed per subhalo at the beginning of the simulation $t=4$ Gyr, such that the subhalo's density/mass profile is converged at radii $\gtrsim$ 100 pc, where we conduct analysis on subhalos' properties. For example, for the low-resolution group of subhalos with infall mass between $10^8$ and $10^{8.2} M_\odot$, we find that a minimum of $2\times 10^4$ particles is needed to track a subhalo's evolution accurately.  Thus, the simulation particle mass is $m_p=10^8 M_\odot / (2\times 10^4)$ for this group. 

We find that the minimum resolution depends on the mass of subhalos and SIDM models. Unlike CDM halos that are self-similar and thus scale-free, SIDM halos have an extra length scale -- the core radius or approximately the $r_1$ radius \citep{kaplinghat16, fz22}, within which the particles experience at least one scattering during the simulation. For CDM halos, approximately the system size $r_{\rm vir}$ (and all length scale) scales with $M_{\rm vir}^{1/3}$. For SIDM halos, the approximate core size $r_1$ can be roughly estimated with the solution of $\rho(r_1)v_{\rm rel}(r_1)\frac{\sigma}{m} = 1/\Delta t$, where $\rho$ is the local density, $v_{\rm rel}$ is the mean relative velocity and $\Delta t$ is the halo lifetime. When the halo scales down in mass, its velocity dispersion also decreases, thus $r_1$ is a smaller fraction of $r_{\rm vir}$ than for more massive halos. In other words, the scaling relation $r_1\propto M_{\rm vir}^{\gamma}$ has $\gamma<1/3$ (also see \cite{rocha13}). Thus for a smaller halo, its core mass is also a smaller fraction of the halo mass $M_{\rm vir}$, and we need to invest more simulation particles to make sure we can properly resolve the cored region. This highlights that a mass resolution that is converged for a certain mass range of halos and fixed SIDM model is not automatically guaranteed to be good enough for another system or SIDM model. 

Subhalos have different formation times, which affects how much they evolve before infall onto the host. The time span between their formation and infall is the time they evolve in isolation, which sets the density profile of these subhalos when host effects kick in \citep{nadler20}. 
To include the effect the of subhalos' different formation times, we further split the subhalo mass groups into four subsets, according to their respective formation times.  We assign the simulation starting time $t_{\rm ini}$ as the ensemble binned formation time $t_{\rm ini} = t_{\rm hf} = 0.5$ Gyr to subhalos forming between 0 and 1 Gyr, $t_{\rm hf} = 1.5$ Gyr to subhalos forming between 1 and 2 Gyr, $t_{\rm hf} = 2.5$ Gyr to subhalos forming between 2 and 3 Gyr, and $t_{\rm hf} = 3.5$ Gyr to subhalos forming between 3 and 4 Gyr. For the subhalos that form later than 4 Gyr but still infall by the end of the simulation, we enforce the latest time bin $t_{\rm hf} = 3.5$ Gyr to be their formation time, so that they can co-evolve with other subhalos in the simulation.  The initial position and velocity of each subhalo at the starting time are calculated by rewinding its position and velocity at infall, as generated by \texttt{Galacticus}, back in time as described in Sec. \ref{sec:merger_tree}. 
These subsets of simulations are simulated until $t=4$ Gyr, and then merged back again as the initial conditions for the main simulations of each group, which start from $t=4$ Gyr and ends at $t=9$ Gyr. 

It is important to track the core-collapse in subhalos on-the-fly with the simulation, for both physical and computational reasons. On the physics side, the methodology of representing dark matter with macro N-body particles itself becomes questionable when a (sub)halo's core-collapse process starts and its central region reaches the frequent scattering, fluid-like short-mean-free-path regime. On the computational side, as the (sub)halo's central density gets exponentially higher and higher in the runaway process, the simulation timestep $\delta t$ needed to resolve scattering in the central region becomes smaller and smaller, because the pairwise scattering probability $P\propto \rho v_{\rm rel} \frac{\sigma}{m} \delta t$ needs to be always smaller than 1 to be physical. Thus as the (sub)halo core-collapses and dark matter accumulates at the halo center, it becomes expensive to continue the simulation. Therefore, it is essential to track core-collapse in subhalos on-the-fly.

Similar to \cite{zzc22}, at every timestep, we measure the local SIDM density of each simulation particle, and mark the average of the top 50 of these local densities as $\rho_{\rm max50}$, as the tracer of subhalo core-collapse\footnote{The choice of the `top 50' tracer is a trade-off between the computational (and RAM) cost and the random noise in particles' local densities, following the same treatment in \cite{zzc22}}. When $\rho_{\rm max50}$ reaches 5 times its initial value, at least one subhalo is deep into the core-collapsing phase, with central density at least 5 times its initial condition. This `5 times' criterion is chosen to ensure that a core-collapsing subhalo has a high enough central density such that its core-collapse will not be disrupted by evaporation induced by the host halo at a later time. Here we refer to the results from \cite{zzc22}, where only in one rare case have they observed a subhalo whose central density increased by a factor of ten but was temporarily disrupted, before it eventually core-collapsed at a later time (see the thin orange line in Fig. 9 of \cite{zzc22}). In \cite{zzc22}, since the simulation setup is targeting a single subhalo, they can simply terminate the simulation when the central density threshold is reached.  Because we simulate ensembles of subhalos in this work, we adopt a different procedure to handle core-collapse.  
When $\rho_{\rm max50}$ increases by a factor of five in the simulation, we identify the subhalo which most of the top 50 particles with largest local SIDM densities are associated with at the beginning of the simulation, and turn off the dark matter self-interaction for the whole subhalo, switching all of its particles to CDM. The dense center of this subhalo ceases core-collapsing since the gravothermal process is frozen. We note that with this `freezing' treatment, the central density/mass of the core-collapsed subhalo in our simulation only serves as a conservative, lower limit.  The SIDM scattering remains `on' for all subhalos other than those reaching this criterion. $\rho_{\rm max50}$ is reset each time a subhalo reaches this criterion, such that it can be reused to trace the core-collapse process of other subhalos.

In this work we design this hierarchical simulation scheme to simulate a realistic population of SIDM subhalos, grouping subhalos according to their mass at infall. 
This scheme allows us to resolve small subhalos at a much higher resolution than cosmological simulation in a short amount of computational time, but at the cost of missing part of the interaction among subhalos from different mass groups. As we demonstrate in Appendix \ref{appdx:val}, this tradeoff does not statistically impact the evolution of subhalos. 
We present a step-by-step walkthrough and a flowchart of our simulation scheme in next section.

\subsection{Step-by-step procedures}\label{sec:step}

\begin{figure*}
    \begin{subfigure}[t]{\textwidth}
        \centering
        \includegraphics[width=\textwidth, clip,trim=2.5cm 0cm 0.5cm 0cm]{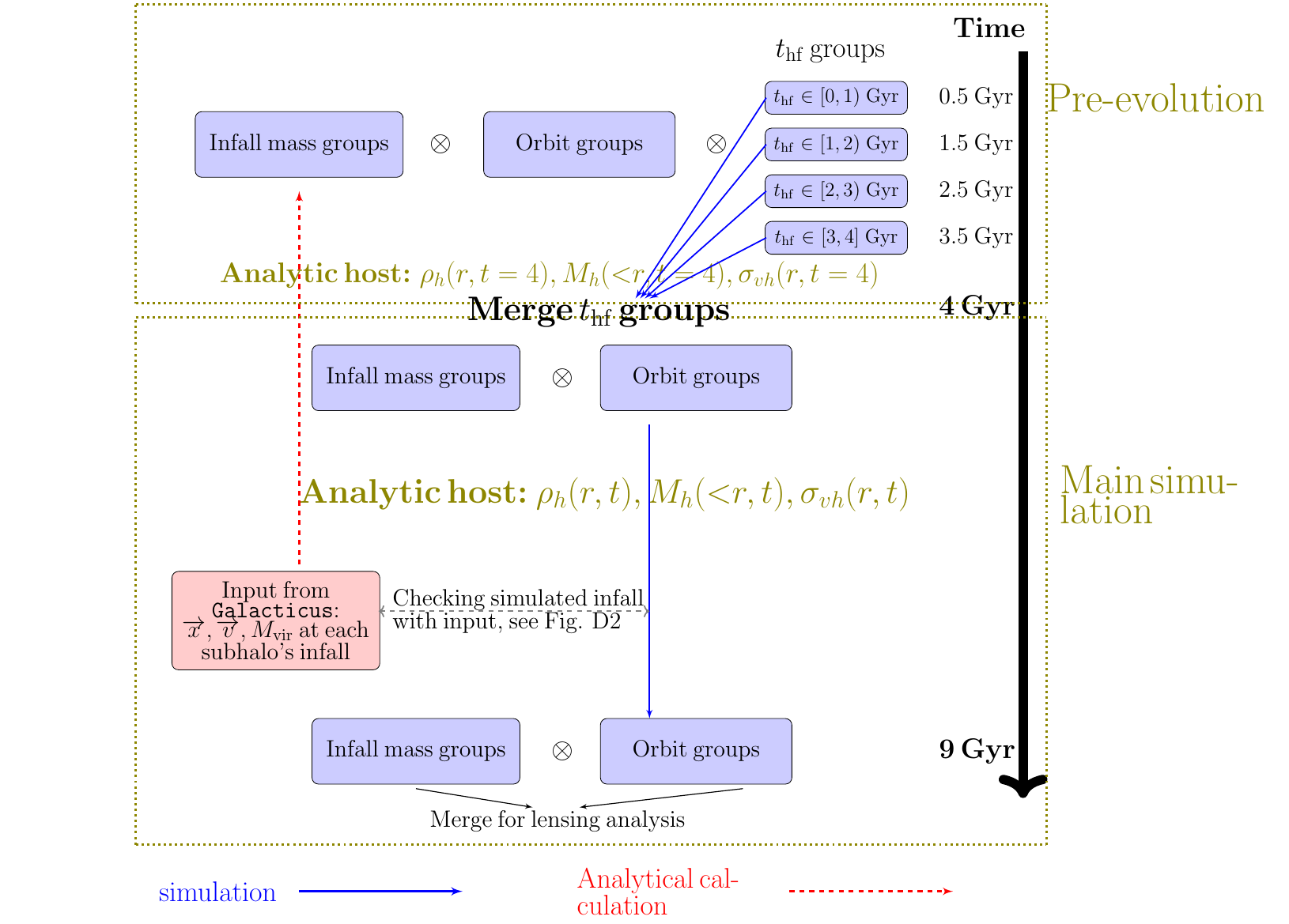}
    \end{subfigure}
    \caption{The workflow chart of our hierarchical simulation framework. See Sec. \ref{sec:sim_method} and \ref{sec:step} for details.}
    \label{fig:flowchart}
\end{figure*}

Fig.~\ref{fig:flowchart} shows a flowchart of our set-up procedure for the simulations. We start from the merger tree catalogues generated by \texttt{Galacticus}, presented as the pink box in the lower left corner of Fig.~\ref{fig:flowchart}. Then we rewind subhalos' information at infall along the red dashed line back to their different formation times. The pre-evolution simulations of each subhalo [mass, orbit, $t_{\rm hf}$] group are then started at $t_{\rm hf}$ and ended at 4 Gyr, as boxed in the upper chunk of Fig.~\ref{fig:flowchart}. The main simulations of subhalo [mass, orbit] groups are then run from 4 Gyr to 9 Gyr in the lower chunk of Fig.~\ref{fig:flowchart}.

\vskip 0.1cm
\textit{Preparation of the analytic host}:

\begin{itemize}
    \item Get the virial mass $M_{200c}$, virial radius $R_{200c}$ and scale radius $r_s$ of the host halo at any time from \texttt{Galacticus}; and the virial mass, virial radius, scale radius, 3-d positions and 3-d velocities of the subhalos at their infall. See Sec. \ref{sec:merger_tree} for setup details of \texttt{Galacticus}. 
    \item To model the gravitational potential and evaporative effects from the host, generate six simulations of isolated host halos with initial conditions that follow NFW profiles, using the characteristic parameters $M_{200c}(t)$, $R_{200c}(t)$ and $r_s(t)$ of the host halo at $t=4, 5, 6, 7, 8, 9$ Gyr respectively. The simulations are run with dark matter self-interaction for each SIDM model until it reaches thermal equilibrium inside the center (about 4 Gyrs).
    \item Measure the density profiles $\rho_h(r)$, enclosed mass profiles $M_{<r, h}(r)$ and velocity dispersion profiles $\sigma_{v, h}(r)$ for the SIDM host halo at six different times. Interpolate among these six discrete times to get $\rho_h(r, t)$, $M_{<r, h}(r, t)$ and $\sigma_{v, h}(r, t)$ at any time $t$ that is between 4 and 9 Gyrs. This concludes all the information we need to build an analytic, smooth host that evolves with time.
\end{itemize}

\textit{Dividing subhalos into multiple groups}:
\begin{itemize}
    \item The whole population of subhalos with infall mass larger than $10^8 M_\odot$ are divided into five groups: $[10^8, 10^{8.2}) M_\odot, [10^{8.2}, 10^{8.5}) M_\odot, [10^{8.5}, 10^{9}) M_\odot$, $[10^9, 10^{10}) M_\odot$ and above $10^{10} M_\odot$, which are denoted as $\{M_{\rm sub}\}_i$. 
    
    \item For each subhalo, the coordinates and velocity at infall, $\vec{x}_{\rm infall}$ and $\vec{v}_{\rm infall}$, together with the profiles of the analytic host at any given time, are used to analytically calculate the subhalo's orbit.
    Subhalos in each infall-mass group, except for the most massive ones with $M_{\rm infall} \geq 10^{10} M_\odot$, are further divided into two sub-groups, according to their orbits (o.o. or h.s.o.). 
    We invest more particles in h.s.o. subhalos and have a separate high resolution run, denoted as $\{M_{\rm sub}\}_{i, \rm h.s.o.}$, while the o.o. subhalos with fewer particles is denoted as $\{M_{\rm sub}\}_{i, \rm o.o.}$. Typically the number of subhalos in $\{M_{\rm sub}\}_{i, \rm h.s.o.}$ counts for $10\%-20\%$ of $\{M_{\rm sub}\}_{i}$. The numbers of simulation particles we use for each $\{M_{\rm sub}\}_{i}$ are listed in Table \ref{table:res}.
    
    \item Subhalos either from $\{M_{\rm sub}\}_{i, \rm o.o.}$ or $\{M_{\rm sub}\}_{i, \rm h.s.o.}$ all have a formation time between 0 and 4 Gyrs, then we divide them into four sub-groups, according to their formation times that are binned at an interval of 1 Gyr. For example, subhalos within $\{M_{\rm sub}\}_{i, \rm o.o.}$ and form between $t=0$ and 1 Gyr are assigned a uniform formation time of $t_{\rm hf} = 0.5$ Gyr, denoted as $\{M_{\rm sub}\}_{i, \rm o.o.}(t_{\rm hf=0.5})$. Each subhalo's initial position and velocity are then determined by its analytically rewinded orbit and the binned formation time.
    
    \item In short, $\{M_{\rm sub}\}_i = \{M_{\rm sub}\}_{i, \rm o.o.} \oplus \{M_{\rm sub}\}_{i, \rm h.s.o.}$; $\{M_{\rm sub}\}_{i, \rm o.o.} = \{M_{\rm sub}\}_{i, \rm o.o.}(t_{\rm hf = 0.5}) \oplus \{M_{\rm sub}\}_{i, \rm o.o.}(t_{\rm hf = 1.5}) \oplus \{M_{\rm sub}\}_{i, \rm o.o.}(t_{\rm hf = 2.5}) \oplus \{M_{\rm sub}\}_{i, \rm o.o.}(t_{\rm hf = 3.5})$
\end{itemize}

\textit{Pre-evolution of subhalos}:
\begin{itemize}
    \item Each $\{M_{\rm sub}\}_{i, \rm o.o./h.s.o.}(t_{\rm hf})$ is run as a separate simulation, which starts at $t_{\rm hf}$ and ends at 4 Gyr. Subhalos in each simulation are initialized with NFW profiles and their own parameters $M_{200c}$, $R_{200c}$ and $r_s$. 

    \item The $\{M_{\rm sub}\}_i$ with most massive subhalos is not split into o.o./h.s.o. runs, but still split into different halo-formation-time groups for the pre-evolution. The host halo, initialized with its $M_{200c}$, $R_{200c}$ and $r_s$ at 4 Gyr, is simulated in an isolated simulation from 0 to 4 Gyr, such that it reaches thermal equilibrium and forms a center core when subhalos begin infalling.
    
    \item We have in total $(4\times2 + 1)\times 4 + 1 = 37$ simulations for the pre-evolution.
\end{itemize}
\textit{Main simulation of subhalos}:

\begin{itemize}
    \item The output snapshots at $t=4$ Gyr from simulations of the same $\{M_{\rm sub}\}_{i, \rm o.o./h.s.o.}$ but four different $\{t_{\rm hf}\}$  are merged into one, serving as the initial condition of our main simulation for each group of $\{M_{\rm sub}\}_{i, \rm o.o./h.s.o.}$. For the group with most massive subhalos, we have five pre-evolution snapshots to merge: four from subhalos with different halo-formation-time, and another one for the isolated host.
    
    \item The main simulation of each $\{M_{\rm sub}\}_{i, \rm o.o./h.s.o.}$ group is then started at $t=4$ Gyr and ended at 9 Gyr. The o.o./h.s.o. simulations of the four low-mass subhalos groups are simulated with an analytical host, while the most massive subhalo group does not, since it automatically includes a live host.
    
    \item The mass growth history of the smooth analytic host already automatically includes the infall of subhalos. To avoid double counting the contribution of subhalos to the host mass, we introduce a time-varying re-scaling factor $\xi$  to uniformly scale down the density and mass profile of the host. $\xi(t)$ equals the ratio between total mass of subhalos in each group that has infall before $t$ and the mass of the host at $t$, and is typically at the level of a few percent or below. Since this re-scaling factor varies with each $\{M_{\rm sub}\}_{i, \rm o.o./h.s.o.}$ simulation, the actual profiles of the analytic host also vary.
    
    \item We have in total $4\times2 + 1 = 9$ main simulations.
\end{itemize}

The validation of this hierarchical simulation scheme is presented in Appendix \ref{appdx:val}, where we aim to quantitatively show that missing the cross-group subhalo-subhalo interaction does not have a significant impact on the overall properties of subhalos. The softening length of any of the aforementioned simulation is set in the same way as of \cite{zzc22} (see their Eqn. 4), which follows the criteria from \cite{vdb18} for CDM subhalos to be reliably traced. 
Results from \cite{mace24} (see also \cite{palubski24, fischer24}) find that these criteria are generally also applicable to SIDM halos.

\subsection{Computational performance}

As we described above and in Table \ref{table:res}, the simulation with highest mass resolution in our hierarchical framework has a particle mass of $\sim 10^3M_\odot$. A full simulation in a cosmological setup that includes all subhalos and the host with total mass $\sim 10^{13}M_\odot$ would have $10^{10}$ particles, while our method reduces the total particle number to $10^8$. This results in the total computational time of $\sim 10000$ CPU hours per realization for SIDM models that do not have many core-collapsed subhalos, and $\sim 25000$ CPU hours for the SIDM model $\{\sigma_0=200, \omega=200\}$ (see next section for definition) that has most core-collapsed subhalos.

\section{SIDM models and parameter space}\label{sec:param-space}

\begin{figure*}
    \begin{subfigure}[t]{0.93\columnwidth}
        \centering
        \includegraphics[width=\textwidth, clip,trim=0.3cm 0cm 0.3cm 0cm]{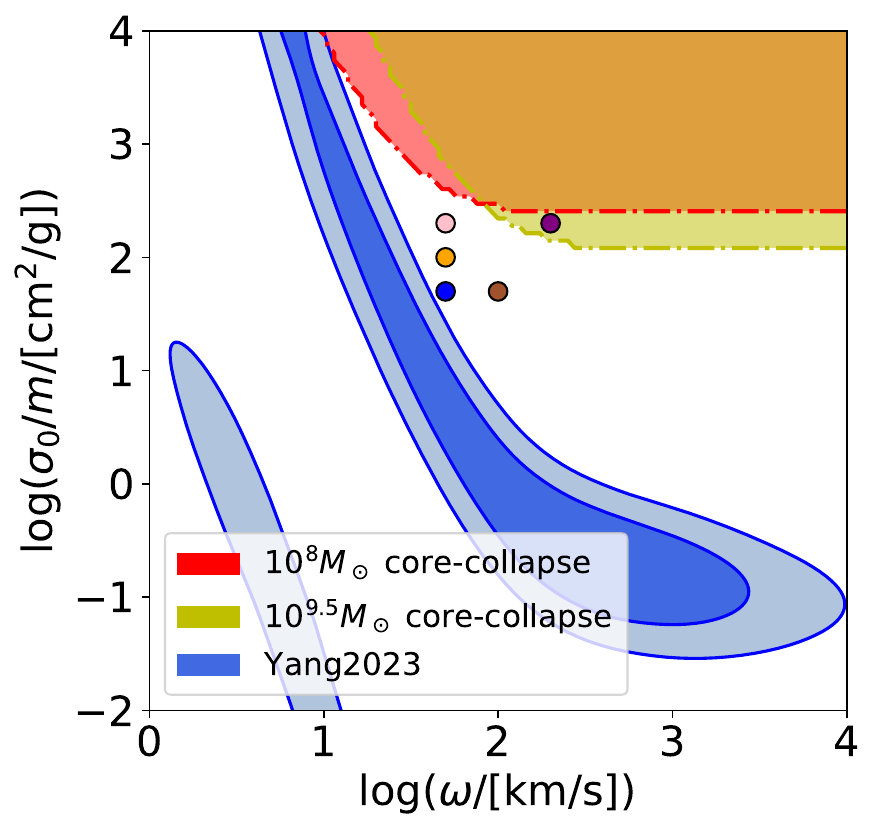}
        \caption{}
        \label{fig:param-space}
    \end{subfigure}
    ~
    \begin{subfigure}[t]{1.05\columnwidth}
        \centering
        \includegraphics[width=\textwidth, clip,trim=0.3cm 0cm 0.3cm 0cm]{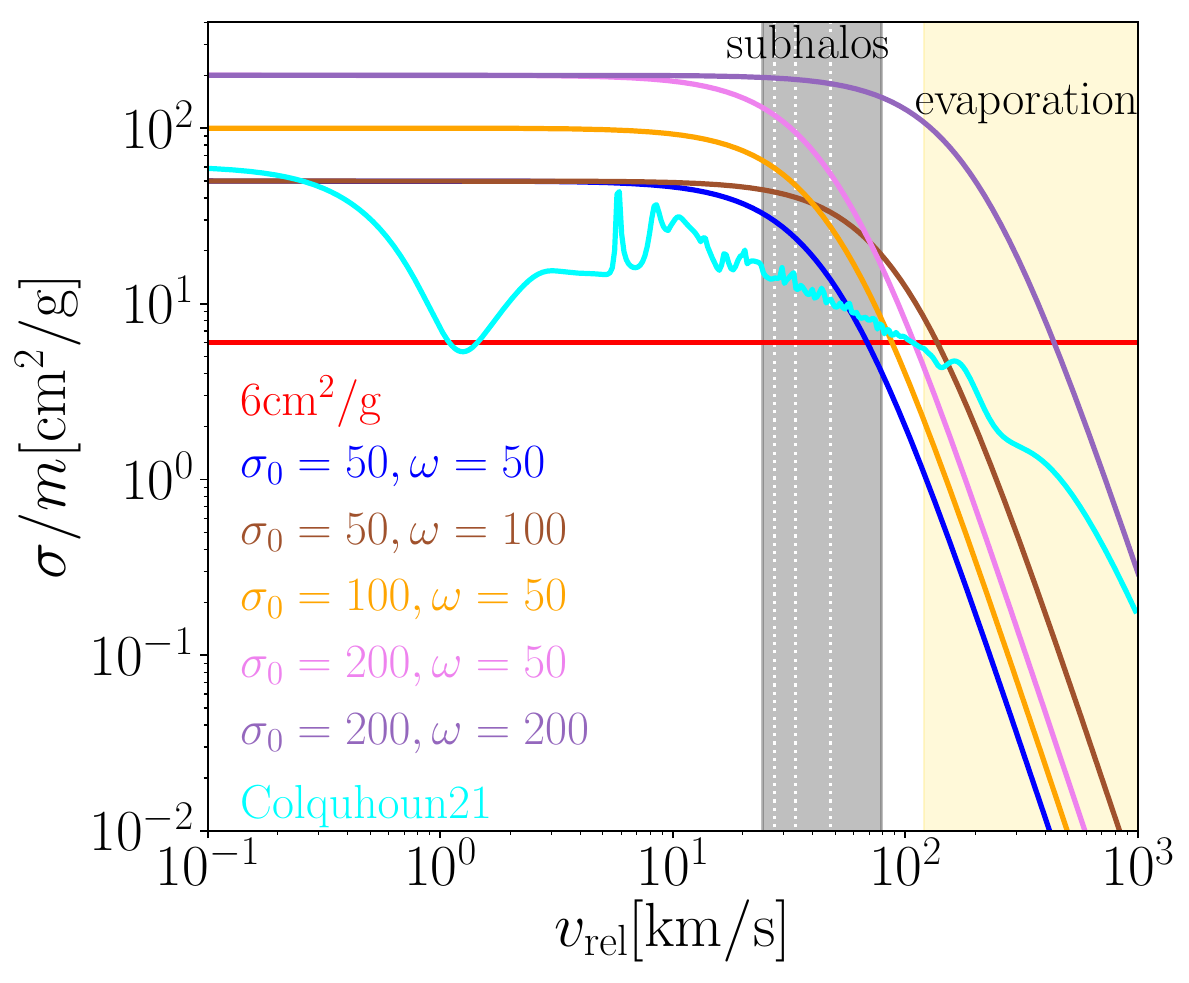}
        \caption{}
        \label{fig:sig-v}
    \end{subfigure}
    \caption{a) Parameter space of the $\sigma_0-\omega$ velocity-dependent SIDM model. The core-collapse region for two typical isolated halos with masses $10^8M_\odot$ and $10^{9.5}M_\odot$ are shown in the red and yellow shaded regions. The parameter space constrained by isolated galaxies and bright cluster galaxies from \protect\cite{sq23} is shown in the blue region. We sample five $\sigma_0-\omega$ models in this work, shown as the scattered dots, for reasons described in the text. b) The $\frac{\sigma}{m}-v_{\rm rel}$ relation of the main SIDM models we use in this work, including the five $\sigma_0-\omega$ models, a constant cross section of 6 $\rm cm^2/g$ and a Yukawa interaction model from \protect\cite{col21}. We also highlight the relative velocity scales most relevant for subhalos' internal heat transfer and host-subhalo evaporation in the gray and yellow shaded regions. Subhalos' $v_{\rm rel}$ is characterized by $\ev{v_{\rm rel}} = C \sigma_v$, where $\sigma_v$ is the typical range of 1-d velocity dispersion of particles for that subhalo mass group and $C\approx 3.8$ (see Sec. \ref{sec:halo-artificial} for details). The subhalos' gray region is further divided by vertical dashed lines for the four low-infall-mass subhalo groups $[10^8, 10^{8.2})M_\odot$, $[10^{8.2}, 10^{8.5})M_\odot$, $[10^{8.5}, 10^{9})M_\odot$ and $[10^{9}, 10^{10})M_\odot$. The evaporation-relevant $v_{\rm rel}$ are subhalos' orbital velocities. An alternate version of b) showing $v_{\rm rel}\times \frac{\sigma}{m}$ is provided in Fig. \ref{fig:sig-vx}.}
    \label{fig:sigma+param}
\end{figure*}

In this work we are interested in simulating SIDM models with large enough cross sections (characterized by $\sigma/m$) such that a non-negligible fraction of the subhalos approach core-collapse  within the simulation time. To quickly scan and locate the relevant cross section parameter space, we adopt the SIDM halo time mapping method introduced in \cite{sq22}. This method takes advantage of the degeneracy between SIDM cross section and halo evolution time in the gravothermal fluid formalism, and can map the gravothermal solution for SIDM halo evolution derived with a constant cross section to arbitrary velocity-dependent cross section cases. Due to the simplicity of the gravothermal fluid formalism, this method is only suitable for modeling isolated and spherically symmetric halos.  We only consider the ``long mean-free-path" regime where the self-interaction mean-free-path is longer than the scale height. 

For our model selection exercise, we consider two subhalos with typical initial NFW parameters \{$\rho_s=2.11\times10^7M_\odot/\mathrm{kpc}^3$, $r_s=0.707$ kpc, $M_{200c}=1.82\times10^8M_\odot$, $c_{200c}=16.97$\} and \{$\rho_s=1.83\times10^7M_\odot/\mathrm{kpc}^3$, $r_s=1.86$ kpc, $M_{200c}=2.80\times10^9M_\odot$, $c_{200c}=16.04$\}, which belong to two infall-mass groups $[10^8,10^{8.2})M_\odot$ and $[10^{9},10^{10})M_\odot$ respectively. 
Assuming that each halo is free from tidal and evaporation effects, we map the parameter space for these two halos to core-collapse within the simulation time of 9 Gyrs, using a generic velocity-dependent SIDM model \citep{dg21, slone21, dnyang22, o22, sq22, sq23, ymzhong23}
\begin{equation}
    \frac{\sigma}{m} = \frac{\sigma_0}{[1+v_{\rm rel}^2/\omega^2]^2},
\end{equation}
where $v_{\rm rel}$ is the relative velocity between two SIDM particles, and $\sigma_0$ and $\omega$ are two free parameters to characterize this model (denoted as the $\sigma_0-\omega$ model hereafter). The core-collapse regions in the $\sigma_0-\omega$ parameter space for these two halos are shown in Fig. \ref{fig:param-space} in red and yellow. As a reference, we also show current constraints on the $\sigma_0-\omega$ model in blue shades ($1\sigma$ and 2$\sigma$) from \cite{sq23}, which utilises low-baryon  rotation curves from isolated galaxies and stellar dynamics of and bright cluster galaxies to constrain SIDM. The mismatch between the red/yellow shaded regions and the blue shades shows that core-collapse is not expected to happen in isolated halos with masses in the range $[10^8, 10^{10}]M_\odot$.

The shaded regions in Fig. \ref{fig:param-space} are calculated for isolated halos.  When halos orbit the group-scale host we consider in this paper, we expect the core-collapse region of subhalos to be different. On one hand, tidal stripping should accelerate core-collapse in subhalos, so the required $\sigma_0$ for core-collapse should be smaller, and the boundary of the core-collapse region should be shifted downwards. On the other hand, the evaporation, whose strength is characterized by the $\sigma/m$ at the subhalo orbital velocity $v_{\rm rel} \sim 10^2 - 10^3$ km/s, can delay or even disrupt the core-collapse process. This implies that the minimum requirement on $\sigma_0$ for core-collapse would be higher for subhalos for $\omega\gtrsim100$ km/s --- subhalos must core-collapse before they experience strong evaporation. As a result of these two effects combined together, we expect the core-collapse region of subhalos to be expanded to cover smaller $\sigma_0$ for $\omega\lesssim100$ km/s, and lifted to larger $\sigma_0$ for $\omega\gtrsim100$ km/s. 

Hence we sample five $\{\sigma_0, \omega\}$ models near the estimated core-collapse boundary of subhalos, as shown in the points in Fig. \ref{fig:param-space}, and generate simulations using the framework described in Sec. \ref{sec:method}. The $\sigma/m - v_{\rm rel}$ relation of these models is shown
in Fig. \ref{fig:sig-v}.  We also include an SIDM model with constant cross section of 6 cm$^2$/g, and a more particle-physics-motivated SIDM model from \cite{col21} (hereafter denoted as Colquhoun21). \cite{col21} assumes a Yukawa potential to represent the interaction among SIDM particles, in analogy to nuclear forces. This Yukawa interaction is characterized by the mediator mass $m_\phi$, the dark matter particle mass $m_\chi$ and the dark fine-structure constant $\alpha_\chi$, which we choose to be 3 MeV, 190 GeV and 0.057 respectively, such that the $\sigma/m -v_{\rm rel}$ relation is consistent with constraints of cluster-sized systems ($\mathcal{O}(1)$ cm$^2$/g) while producing  an interesting diversity for dwarf-sized systems ($\sim 10$ cm$^2$/g).

Recently, \cite{cchan23} reported that the recently observed nano-Hertz stochastic gravitational wave background by Pulsar Timing Arrays (PTA) can be explained by the first-order phase transition of an MeV scale mediator for dark matter self-interaction. The corresponding parameter space of velocity-dependent SIDM favored by PTA results (see Fig. 1 of \cite{cchan23}) covers the main models we choose in this work (five $\sigma_0-\omega$ models and Colquhoun21).

\section{Results}\label{sec:results}
In this section, we explore properties of the subhalos in different SIDM models, with implications for substructure lensing. We use \texttt{Amiga Halo Finder} (AHF; \cite{ahf}) to identify the subhalos in our simulation, and analyze the evolution of subhalos with their averaged inner masses in Sec. \ref{sec:halo1} and with their averaged logarithmic slope of inner density profile in Sec. \ref{sec:halo2}. The averaged inner mass and inner slope serve as two statistical probes for core-collapse in subhalos. In Sec. \ref{sec:halo-artificial}, we repeat the same analysis in Sec. \ref{sec:halo1} and \ref{sec:halo2}, but 
using toy SIDM models designed with special features, which allows us to carefully look into the interplay of relevant physical processes in driving the evolution of subhalos. We use the lensing code \texttt{GLAMER} \citep{glmr1, glmr2} to generate smoothed 2D-projected density maps from our simulation particle data, and conduct analyses relevant for lensing in Sec. \ref{sec:res-lens}. In Sec. \ref{sec:iso-vs-sub}, we compare the core-collapse time and fraction of subhalos and their counterparts in isolation, to highlight the difference between subhalos and field halos, which could potentially be used in the future to break degeneracies in SIDM parameter space. In Sec. \ref{sec:res-real} we show the variance of subhalo properties among realizations of merger systems.

A summary table of the number and fraction of subhalos that reach our core-collapse threshold in each infall-mass group and each SIDM model is provided in Table \ref{table:ccp}.

\begin{table*}
	\centering
	\begin{tabular}{lcccccccccr} 
		\Xhline{4\arrayrulewidth}
		Mass group [dex $M_\odot$] &  \multicolumn{2}{c}{$[8.0, 8.2)$} & \multicolumn{2}{c}{$[8.2, 8.5)$} & \multicolumn{2}{c}{$[8.5, 9.0)$} & \multicolumn{2}{c}{$[9.0, 10.0)$} & $\geq 10.0$\\
		\hline
		SIDM models / orbits & $\rm{o.o}$ & $\rm{h.s.o}$ & $\rm{o.o}$ & $\rm{h.s.o}$ & $\rm{o.o}$ & $\rm{h.s.o}$ & $\rm{o.o}$ & $\rm{h.s.o}$ & \\
		\hline
		$\sigma_0=50, \omega=50$ & 0 & 0 & 0 & 0 & 0 & 1, 0.5\% & 0 & 0 & 0\\
		\Xhline{0.5\arrayrulewidth}
		$\sigma_0=50, \omega=100$ & 0 & 0 & 5, 0.4\% & 3, 1.2\% & 12, 1.3\% & 13, 6.3\% & 4, 0.8\% & 6, 6.2\% & 0\\
		\Xhline{0.5\arrayrulewidth}
		$\sigma_0=100, \omega=50$ & 3, 0.2\% & 13, 3.8\% & 25, 2.0\% & 15, 6.1\% & 10, 1.1\% & 20, 9.8\% & 1, 0.2\% & 5, 5.2\% & 0 \\
		\Xhline{0.5\arrayrulewidth}
		$\sigma_0=200, \omega=50$ & 312, 21.9\% & 91, 26.8\%  & 162, 13.1\% & 59, 23.9\% & 130, 13.7\% & 55, 26.8\% & 15, 2.9\%   & 14, 14.4\% & 0 \\
		\Xhline{0.5\arrayrulewidth}
		$\sigma_0=200, \omega=200$ & 573, 40.2\% & 40, 11.7\% & 304, 24.6\% & 40, 16.1\% & 351, 37.1\% & 54, 26.0\% & 220, 42.6\% & 19, 19.6\% & 34, 28.8\%\\
  
        \Xhline{0.5\arrayrulewidth}
		Colquhoun21 & 0 & 0 & 0 & 0 & 0 & 0 & 0 & 0 & 0\\
		\Xhline{3\arrayrulewidth}
		window1 &  112, 30.8\% & 7, 9.6\% & 75, 25.5\%  & 16, 24.2\% & 116, 47.0\% & 26, 54.2\% & 42, 30.1\% & 5, 27.8\% & 0\\
		\Xhline{0.5\arrayrulewidth}
		window2 & 143, 39.3\%  & 15, 20.5\% & 71, 24.1\% & 20, 30.3\% & 46, 18.6\% & 23, 47.9\% & 0 & 3, 16.7\% & 0 \\
        \Xhline{0.5\arrayrulewidth}
		window3 & 0 & 0 & 0 & 0 & 16, 6.5\% & 3, 6.3\% & 48, 35.0\% & 3, 16.7\% & 0\\
		\Xhline{0.5\arrayrulewidth}
		z3 & 24, 6.6\% & 1, 1.4\% & 22, 7.5\% & 3, 4.5\% & 22, 8.9\% & 5, 10.4\% & 1, 0.7\% & 2, 11.1\% & 0 \\
          \Xhline{0.5\arrayrulewidth}
		z8 & 1, 0.3\% & 0 & 10, 3.4\% & 0 & 10, 4.1\% & 0 & 1, 0.7\% & 0 & 0\\
		\Xhline{4\arrayrulewidth}
        
	\end{tabular}
	\caption{Number and fraction of core-collapsed subhalos in each infall mass group and each SIDM model, in the format of $A, X\%$, where $A$ is the number of subhalos that experience core-collapse before the end of the simulation at $t=9\ \rm Gyr$ and $X$ is the percentage of core-collapsed subhalos at $t=9\ \rm Gyr$. All four realizations are stacked together in this table for the $\{\sigma_0, \omega\}$ and Colquhoun21 model (upper part of this table, see Fig. \ref{fig:sig-v} and Sec. \ref{sec:param-space}), while we only have one realization for the window function and z-type models (lower part, see Fig. \ref{fig:arti-x} and Sec. \ref{sec:halo-artificial} for descriptions). `o.o.' and `h.s.o.' are abbreviations for ordinary orbits and heavily-stripping orbits.}
    \label{table:ccp}
\end{table*}

\subsection{Evolution of subhalos' central mass}\label{sec:halo1}

\begin{figure*}
    \includegraphics[width=1.95\columnwidth]{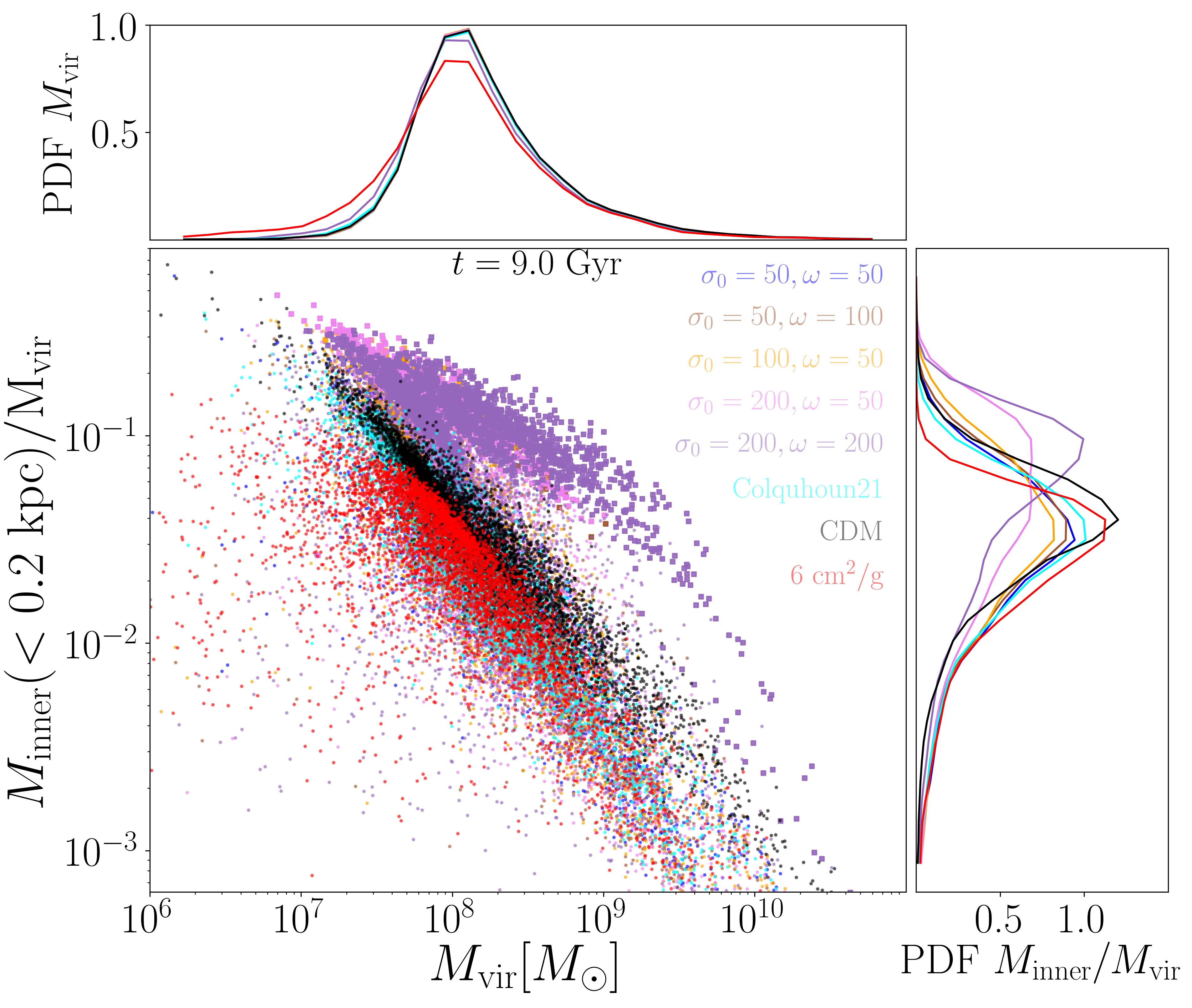}
    \caption{The relation between subhalos' inner mass (within 200 pc) and virial mass, at the end of the simulation. The main scatter plot shows the normalized inner mass $M_{\rm inner}(<0.2\ \rm kpc)/M_{\rm vir}$ and virial mass $M_{\rm vir}$ of each subhalo, of different SIDM models. The subhalos that reach our critetion of core-collapse are labelled with squares, and others with dots. The top and right shoulder panels shows the probability distribution of $M_{\rm vir}$ and $M_{\rm inner}/M_{\rm vir}$ respectively. An animated movie of the full time evolution of this plot is uploaded to \href{https://youtu.be/FA6j8hHZejM}{video1}. Another evolution movie with inner radius defined by 2\% of each subhalo's virial radius instead of a fixed 200 pc is shown in \href{https://youtu.be/4nR-Rxd691c}{video2}. We refer the readers to Table \ref{table:ccp} for detailed core-collapse statistics of each model and mass-group. }
    \label{fig:scatter}
\end{figure*}

In this subsection, we focus on using the central mass of subhalos to track the evolution of subhalos in the host, under the coupled effects of gravity and dark matter self-interaction. The most prominent feature of core-collapse in an SIDM (sub)halo is dark matter increasingly concentrating at the halo center.  Thus, the enclosed mass within a small aperture near the center grows with time in the core-collapse phase. A (sub)halo still in the core-formation stage has a much lower central mass, and a CDM (sub)halo has a central mass between that of core and core-collapsed. 

Here we choose an aperture of 200 pc for the measurement of a subhalo's `inner' mass. This aperture is chosen in order to strike a balance between a small radius to represent the center of subhalos, where we expect most interesting physics of SIDM to happen, and an adequate amount of particles ($\gtrsim$ 50 particles for the massive $[10^9, 10^{10})M_\odot$ and $M_{\rm sub}\geq 10^{10}M_\odot$ groups of cored subhalos, which are most limited by particle resolution in the inner region). 
In Fig. \ref{fig:scatter} we show a scatter plot of this inner mass $M(<200\ \rm pc)$ as a fraction of virial mass for each individual subhalo, at the end of the simulation $t=9\ \rm Gyr$. We label the subhalos that have core-collapsed according to the criterion described in Sec. \ref{sec:sim_method} with squares, and the subhalos that have not core-collapsed with circles.  In the shoulder plots, we show the probability distribution function (PDF) of the subhalos' virial mass $M_{\rm vir}$ (x-axis) and of the inner masses $M(<200\ \rm pc)$ (y-axis). Note that the PDF of $M_{\rm vir}$ looks different from the subhalo mass function in cosmological simulations because we impose a cutoff on subhalos' infall-mass rather than present-day mass. 

From the scatter plot, we  confirm that our labeling criterion  successfully picks out the core-collapsed subhalos, which form a visible outlier cluster in the $M_{\rm inner}-M_{\rm vir}$ space. The $\{\sigma_0=200, \omega=50\}$ (violet) and $\{\sigma_0=200, \omega=200\}$ (purple) models yield core-collapse in a large fraction of subhalos, as also indicated in Table \ref{table:ccp}. The $\{\sigma_0=100, \omega=50\}$ (orange) and $\{\sigma_0=50, \omega=100\}$ (brown) models lead have a small number of core-collapse cases, while the subhalos in the other velocity-dependent SIDM models are not much distinguishable from CDM (black). The constant cross section of 6 $\rm cm^2/g$ (red) has subhalos lower in both inner mass and virial mass compared to CDM, because of the host-subhalo evaporation. 

We find that core-collapse also imprints itself in the shoulder plots. In the top shoulder panel, we find that most of SIDM models we study share a similar subhalo mass function at $t=9$ Gyr.  The outliers are the subhalo mass function for the constant cross section of 6 $\rm cm^2/g$ and $\{\sigma_0=200, \omega=200\}$, where evaporation leads to visible mass loss. Thus, the subhalo mass function is more strongly influenced by evaporation rather than core collapse.  By contrast, the right shoulder panel, showing the distribution of subhalos' inner mass relative to the total mass, is more sensitive to core collapse.   Compared to CDM, the PDF of 6 $\rm cm^2/g$ is shifted as a whole towards lower $M_{\rm inner}$, as all subhalos in this model are still in the core formation phase. By contrast, the models showing significant core collapse in the subhalos, the $\{\sigma_0=200, \omega=50\}$ (violet) and $\{\sigma_0=200, \omega=200\}$ (purple), not only show a boost of subhalos with high inner mass, but also an overall broader distribution of $M_{\rm inner}/M_{\rm vir}$, representing a larger diversity of subhalos.  The tail toward low mass ratios indicates a population of subhalos in the core-formation regime. The other velocity-dependent SIDM models show a slightly larger diversity of subhalo mass ratios than CDM, with a lower peak and broader width in the PDF of $M_{\rm inner}/M_{\rm vir}$. We  show a time-evolution movie of the $M_{\rm inner}-M_{\rm vir}$ plot in Appendix~\ref{appdx:anime}.

\begin{figure*}
    \centering
    \begin{subfigure}[t]{0.32\textwidth}
        \centering
        \includegraphics[width=\textwidth, clip,trim=0.2cm 0cm 0.2cm 0cm]{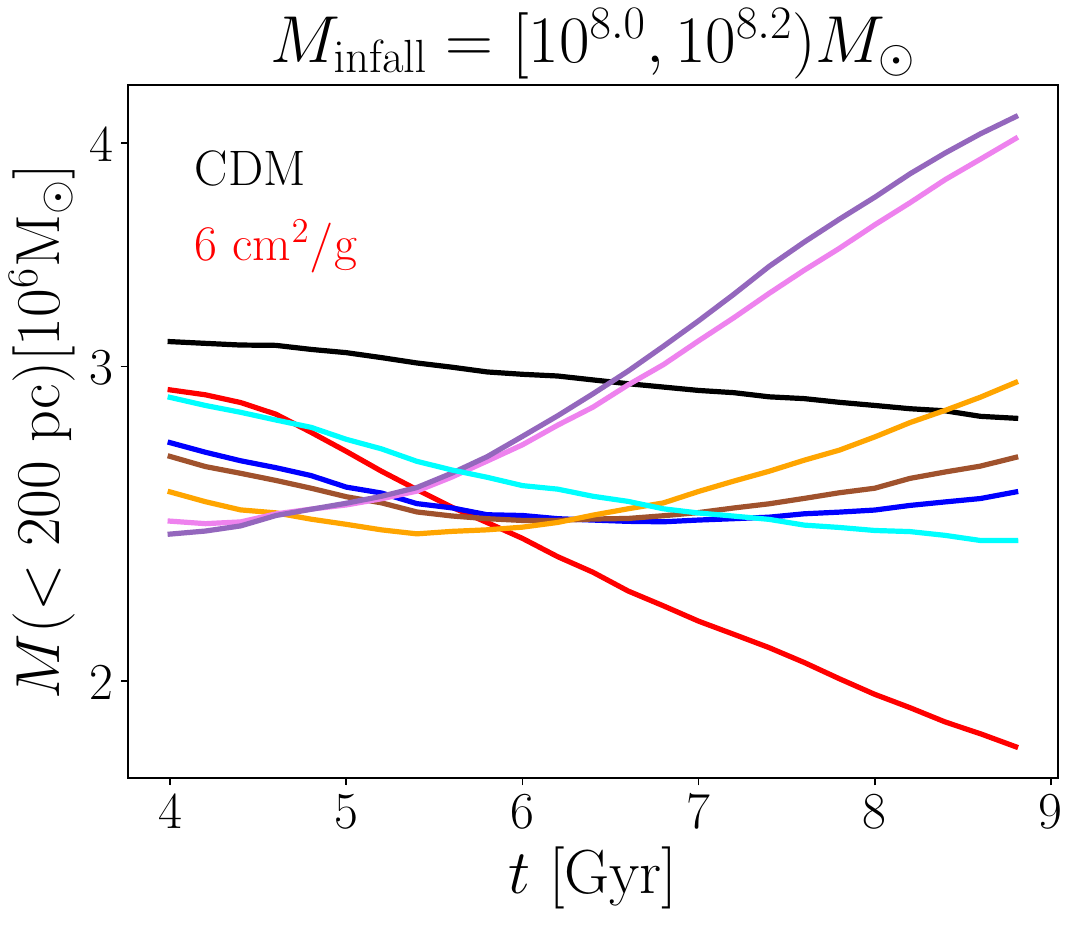}
        \caption{}
        \label{fig:centralm-a}
    \end{subfigure}
    ~
    \begin{subfigure}[t]{0.32\textwidth}
        \centering
        \includegraphics[width=\textwidth, clip,trim=0.2cm 0cm 0.2cm 0cm]{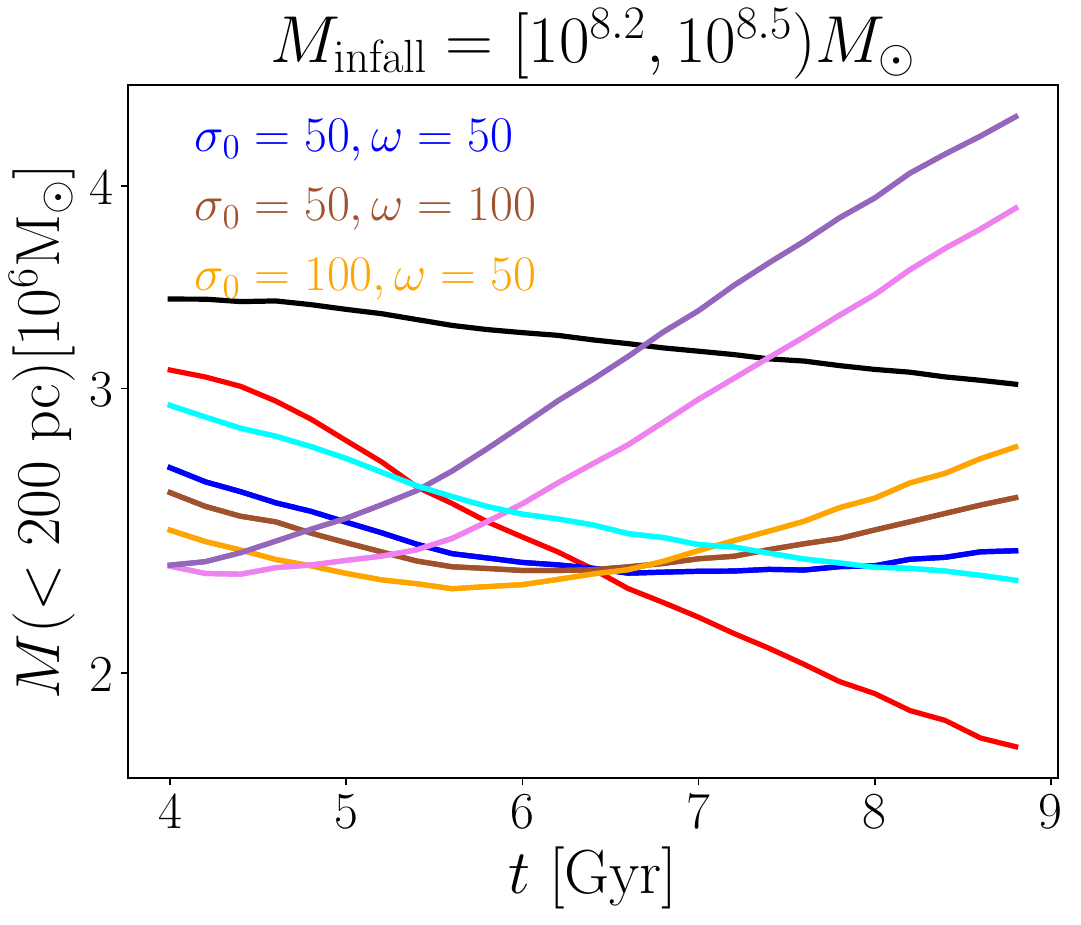}
        \caption{}
        \label{fig:centralm-b}
    \end{subfigure}
    ~
    \begin{subfigure}[t]{0.32\textwidth}
        \centering
        \includegraphics[width=\textwidth, clip,trim=0.2cm 0cm 0.2cm 0cm]{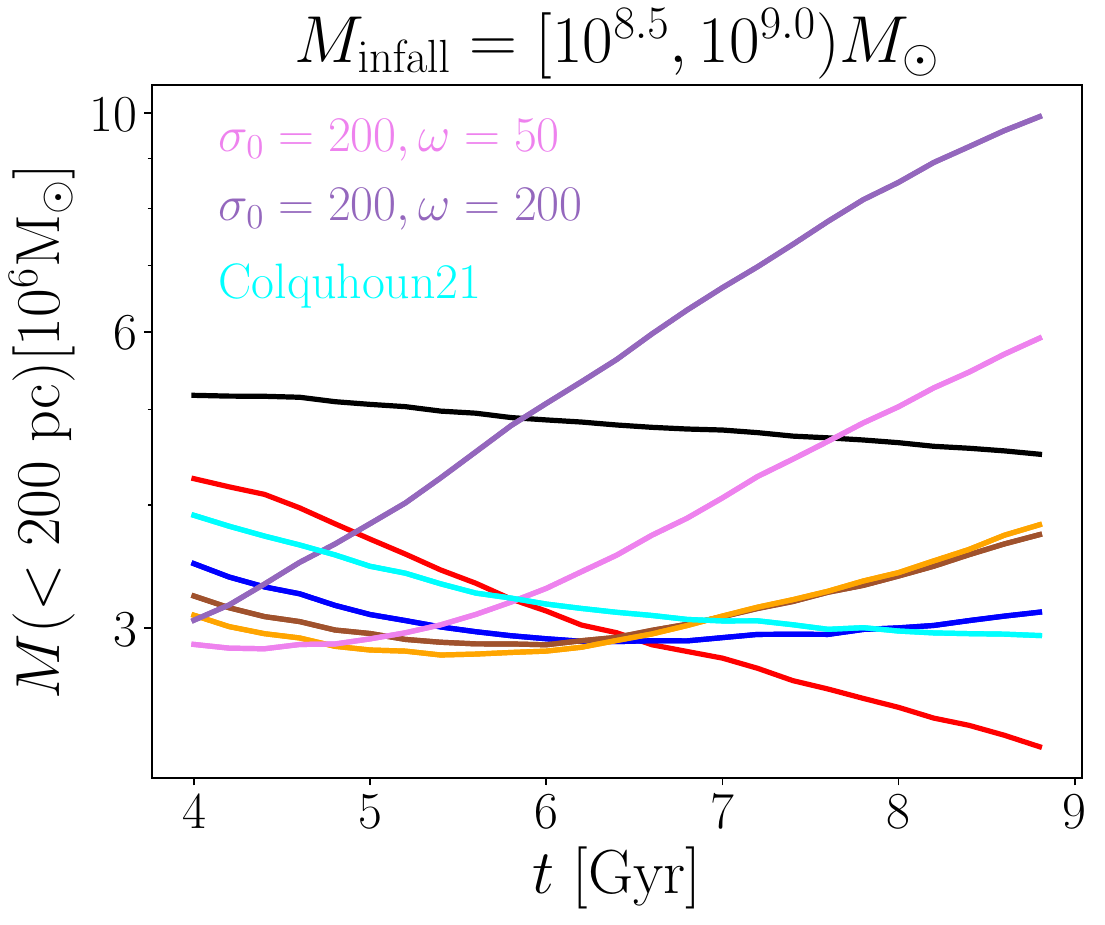}
        \caption{}
        \label{fig:centralm-c}
    \end{subfigure}
     ~
    \begin{subfigure}[t]{0.32\textwidth}
        \centering
        \includegraphics[width=\textwidth, clip,trim=0.2cm 0cm 0.2cm 0cm]{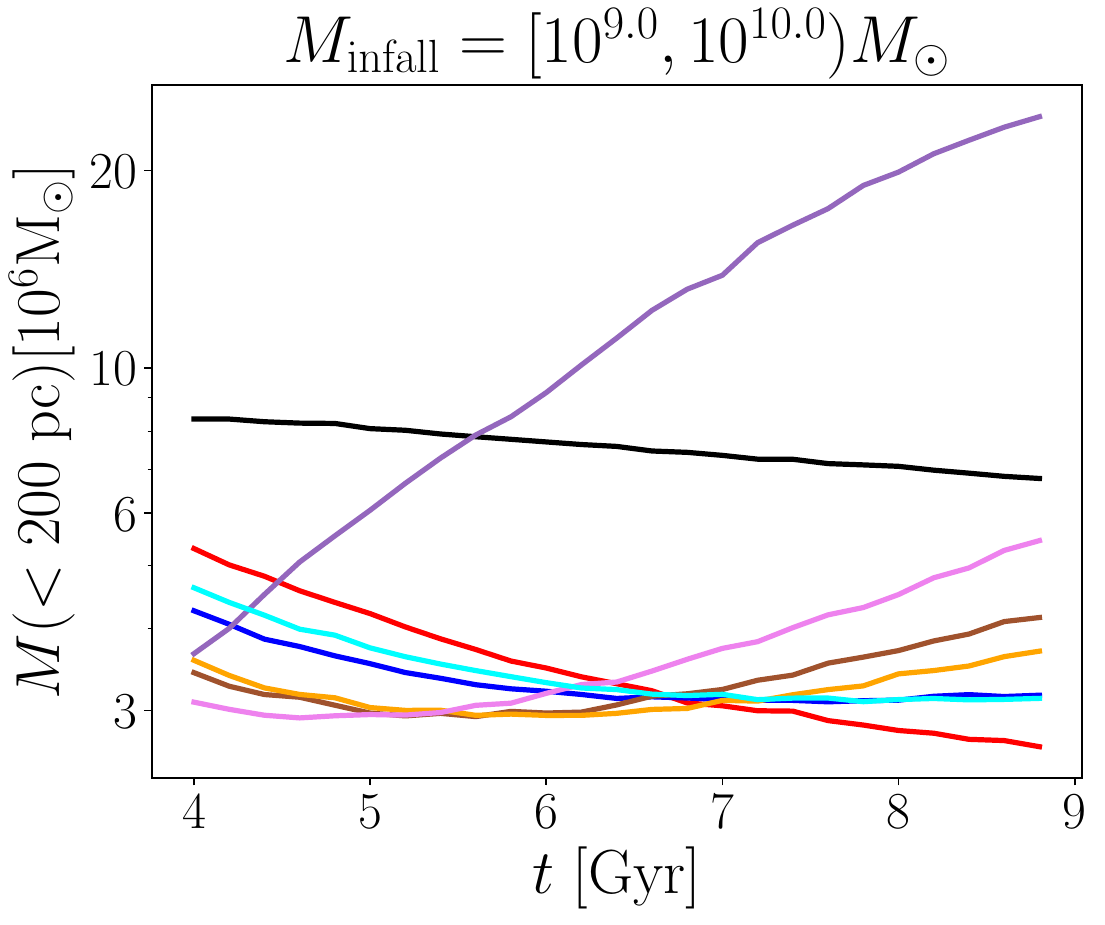}
        \caption{}
        \label{fig:centralm-d}
    \end{subfigure}
    ~
    \begin{subfigure}[t]{0.32\textwidth}
        \centering
        \includegraphics[width=\textwidth, clip,trim=0.2cm 0cm 0.2cm 0cm]{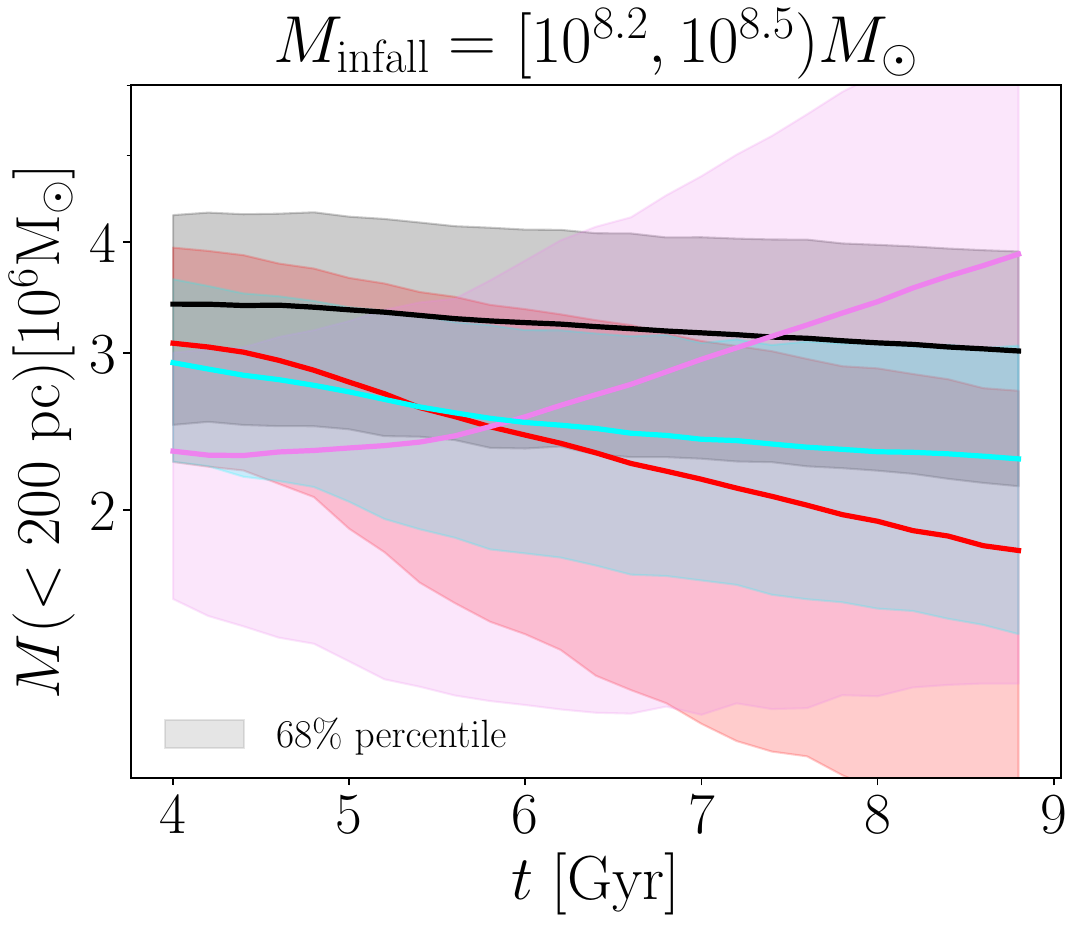}
        \caption{}
        \label{fig:centralm-e}
    \end{subfigure}
    ~
    \begin{subfigure}[t]{0.32\textwidth}
        \centering
        \includegraphics[width=\textwidth, clip,trim=0.2cm 0cm 0.2cm 0cm]{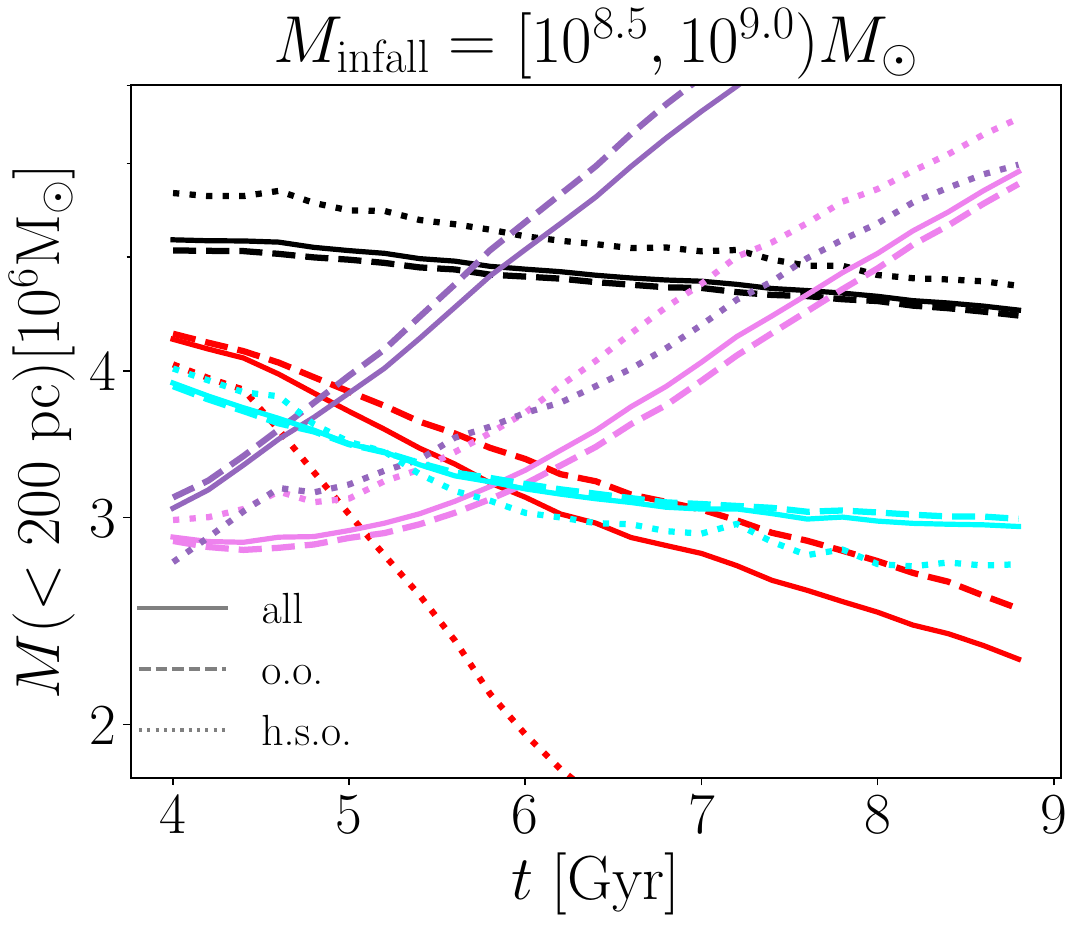}
        \caption{}
        \label{fig:centralm-f}
    \end{subfigure}    
    \caption{Time evolution of the averaged mass of subhalos within an aperture of 200 pc. Panels a) to d) show different infall-mass groups. Panel e) shows the 68\% scatter of the inner mass of selected models. Panel f) further distinguishes suhalos on different orbits (ordinary orbits `o.o.' or heavily-stripping orbits `h.s.o.'). SIDM models have different averaged inner mass at $t=4$ Gyr because of their evolution prior to the 4 Gyr shown on the plots.}
    \label{fig:centralm}
\end{figure*}

We show the time evolution of the averaged subhalo inner mass in Fig. \ref{fig:centralm}. The subplots a) to d) in Fig. \ref{fig:centralm} show the evolution of $M(<200\ \rm pc)$ averaged over all subhalos in each infall-mass group and over four realizations, regardless of the subhalos' orbits. The $\{\sigma_0 = 200, \omega = 50\}$ (violet) and $\{\sigma_0 = 200, \omega = 200\}$ (purple) models have the highest fraction of core-collapsed subhalos, and so their averaged $M(<200\ \rm pc)$ significantly outpace their CDM counterparts (black) at late times in the three lowest infall-mass groups. However, for the large infall-mass group $M_{\rm infall}=[10^9, 10^{10})M_\odot$, only the purple model has a large fraction of core-collapsed subhalos. 
The $\{\sigma_0 = 200, \omega = 50\}$ violet model experiences a drop in cross section for massive subhalos relative to the $\{\sigma_0 = 200, \omega = 200\}$ purple model (see Fig. \ref{fig:sig-v}). On the whole, the subhalos for the other velocity-dependent SIDM models are largely in the core-formation regime, or are only just entering core-collapse near $t=9\ \rm Gyr$, as indicated by the fact that their enclosed masses lie below CDM, gradually ticking up by the end of the simulations.  
The model with a constant cross section of $6\ \rm cm^2/g$ shows that the subhalos are monotonically decreasing in central mass/density, as a combined result of the core-forming stage and the relatively strong evaporation effect from the host halo. The Colquhoun21 model follows a similar trend as this constant SIDM model.  The key differences are that the mass loss in the central region is more gradual than in the constant cross section case, and one starts to see early signs of core-collapse by the flattening trend of inner mass by the end of the simulation.  These results are consistent with the scatter plot in Fig. \ref{fig:scatter}.

The diversity in the scatter plot is also revealed in the scatter in the central mass evolution.  In Fig. \ref{fig:centralm-e}, we show the 68\% percentile scatter of the subhalos' central masses for selected SIDM models. The $\{\sigma_0 = 200, \omega = 50\}$ model (violet) exhibits increasing diversity over time, spanning a wide range of central mass at late times --- from about a factor of 2 smaller (cored) to a few times larger (core-collapsed) than the scatter in the CDM case shown in the black shaded region. The constant cross section with $\sigma/m = 6\ \rm cm^2/g$ also shows a large scatter in subhalo central mass/density (red shaded region), but mostly the subhalos have lower central masses than their CDM counterparts, which is again because of the strong evaporation.  

To investigate the role of orbits in subhalos' evolution, we split the subhalos' central mass evolution by orbital type in Fig. \ref{fig:centralm-f}.
We plot the average evolution of o.o. subhalos with dashed lines and h.s.o. ones with dotted lines. We notice that h.s.o. CDM subhalos lose only slightly more mass relative to the o.o. ones, due to tidal heating and stripping. However, the h.s.o. subhalos simulated with a constant cross section of $6\ \rm cm^2/g$ lose much more mass even at early times, because of the additional strong evaporation. The effect of orbits on the Colquhoun21 model in cyan color is between the two DM models, with evaporation not as strong as the $6\ \rm cm^2/g$ case but still significant. 

The evolution of subhalos in the velocity-dependent models of $\{\sigma_0 = 200, \omega = 200\}$ (purple) and $\{\sigma_0 = 200, \omega = 50\}$ (violet) with significant core-collapsing subhalos is more complicated. Previous work by \cite{nishikawa20} and \cite{sameie20} have argued that the core collapse will be accelerated by tidal stripping, which steepens the negative temperature gradient in the subhalo.  However, \cite{zzc22} find that the host-subhalo evaporation may delay or even halt core-collapse. The overall effect of the host on subhalo evolution can thus be complicated.
In Fig. \ref{fig:centralm-f}, we can see that h.s.o. subhalos have accelerated core-collapse with the $\{\sigma_0 = 200, \omega = 50\}$ (violet) model but experience deceleration with the $\{\sigma_0 = 200, \omega = 200\}$ (purple) model. 
This is because the cross section of the violet model starts to drop for velocities larger than $\sim 20\ \rm km/s$ (see Fig. \ref{fig:sig-v}), and is only $\lesssim {O}(1)\ \rm cm^2/g$ for the orbital-scale relative velocities $100-1000$ km/s relevant for host-sub evaporation.  Thus, for this model, the evaporation is weak and the overall effect of the host is to accelerate core-collapse by tidal stripping. There are two competing effects for the other model.  The cross section for the $\{\sigma_0 = 200, \omega = 200\}$ (purple) model starts to drop for velocities $\gtrsim 200$ km/s, which means a larger cross section for the subhalo mass range $[10^{8.5}, 10^9)M_\odot$ we show in Fig. \ref{fig:centralm-f} than the violet model.  It leads to a higher rate of core collapse for massive subhalos. This model also has a cross section of $\lesssim {O}(10)\ \rm cm^2/g$ for evaporation-related velocities.  Unlike the violet model, evaporation plays a larger role in subhalo evolution than tidal effects, leading to an overall deceleration in subhalo core-collapse.

From Fig. \ref{fig:centralm-f} we can see the coupling between SIDM models, subhalo mass and orbits can be diverse and complicated.  We will further investigate the role of host effects in driving subhalo core-collapse in the next sections.

\subsection{Evolution of subhalos' density slope}\label{sec:halo2}

\begin{figure*}
    \centering
    \begin{subfigure}[t]{0.32\textwidth}
        \centering
        \includegraphics[width=\textwidth, clip,trim=0.3cm 0cm 0.3cm 0cm]{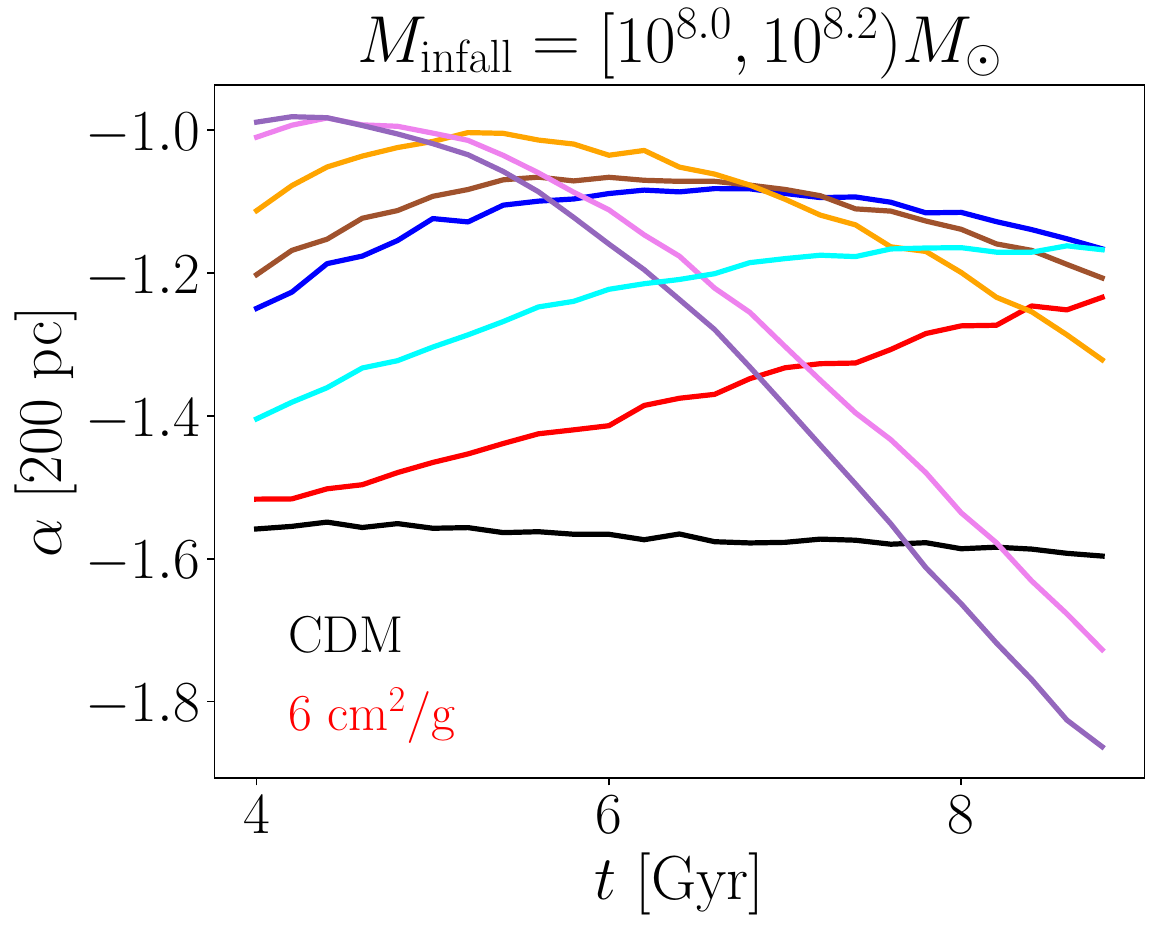}
        \caption{}
        \label{fig:slopea}
    \end{subfigure}
    ~
    \begin{subfigure}[t]{0.32\textwidth}
        \centering
        \includegraphics[width=\textwidth, clip,trim=0.3cm 0cm 0.3cm 0cm]{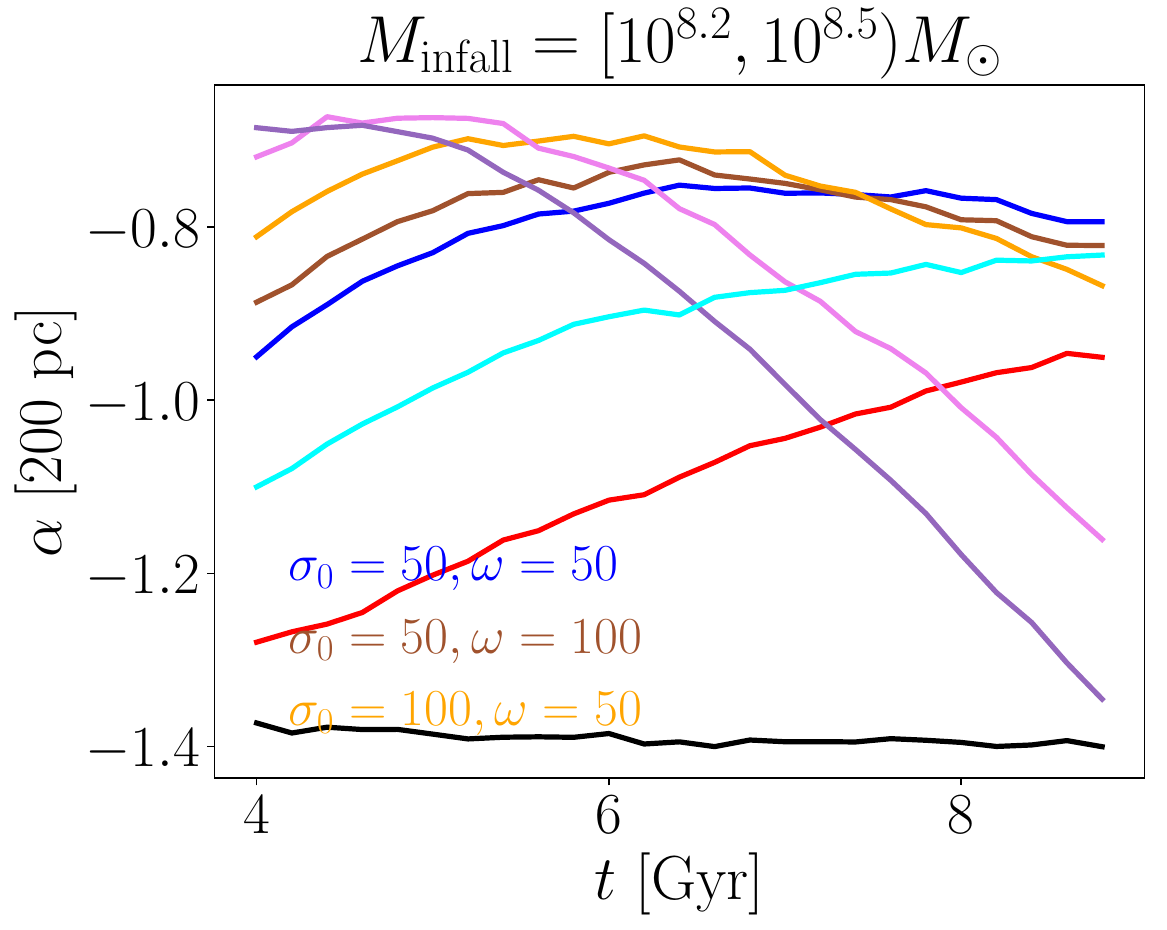}
        \caption{}
        \label{fig:slopeb}
    \end{subfigure}
    ~
    \begin{subfigure}[t]{0.32\textwidth}
        \centering
        \includegraphics[width=\textwidth, clip,trim=0.3cm 0cm 0.3cm 0cm]{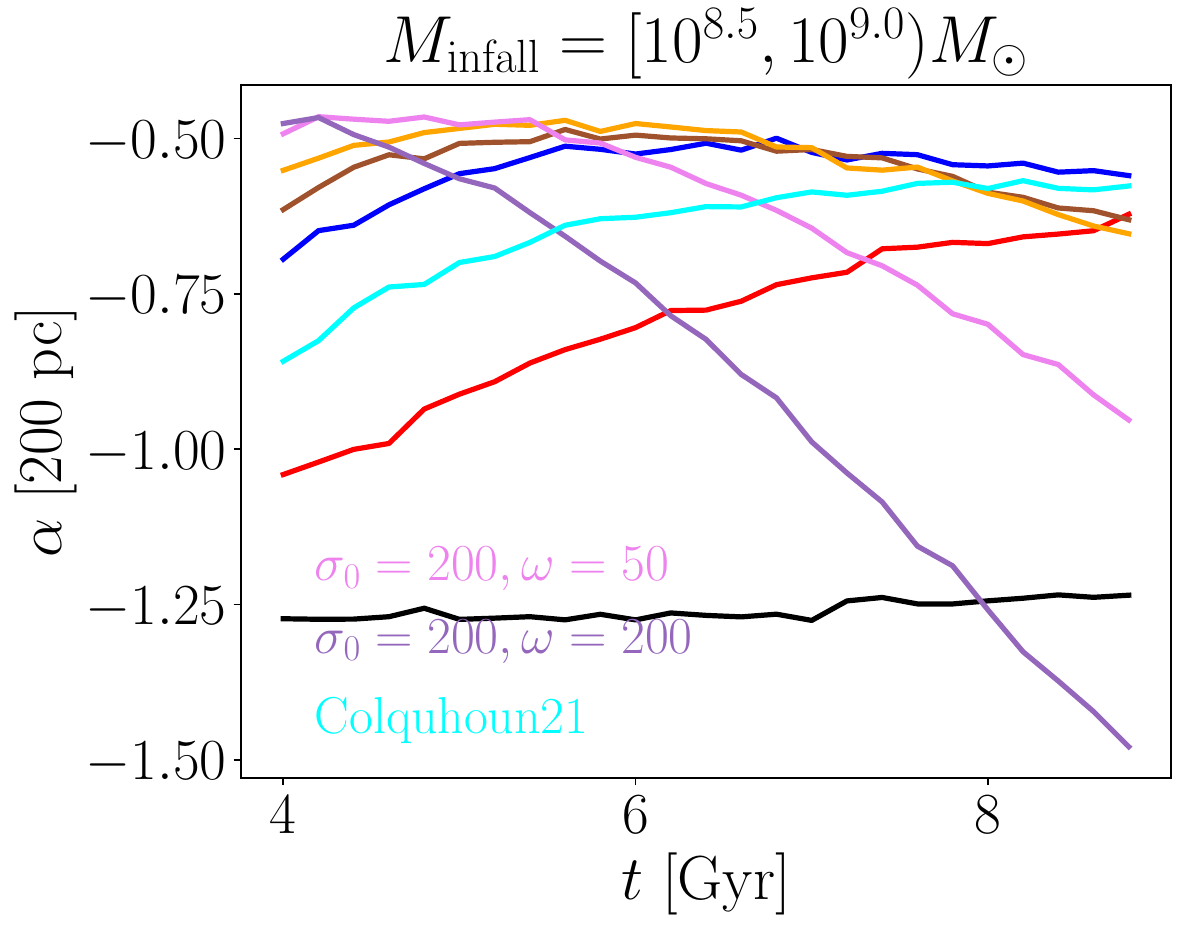}
        \caption{}
        \label{fig:slopec}
    \end{subfigure}
     ~
    \begin{subfigure}[t]{0.32\textwidth}
        \centering
        \includegraphics[width=\textwidth, clip,trim=0.3cm 0cm 0.3cm 0cm]{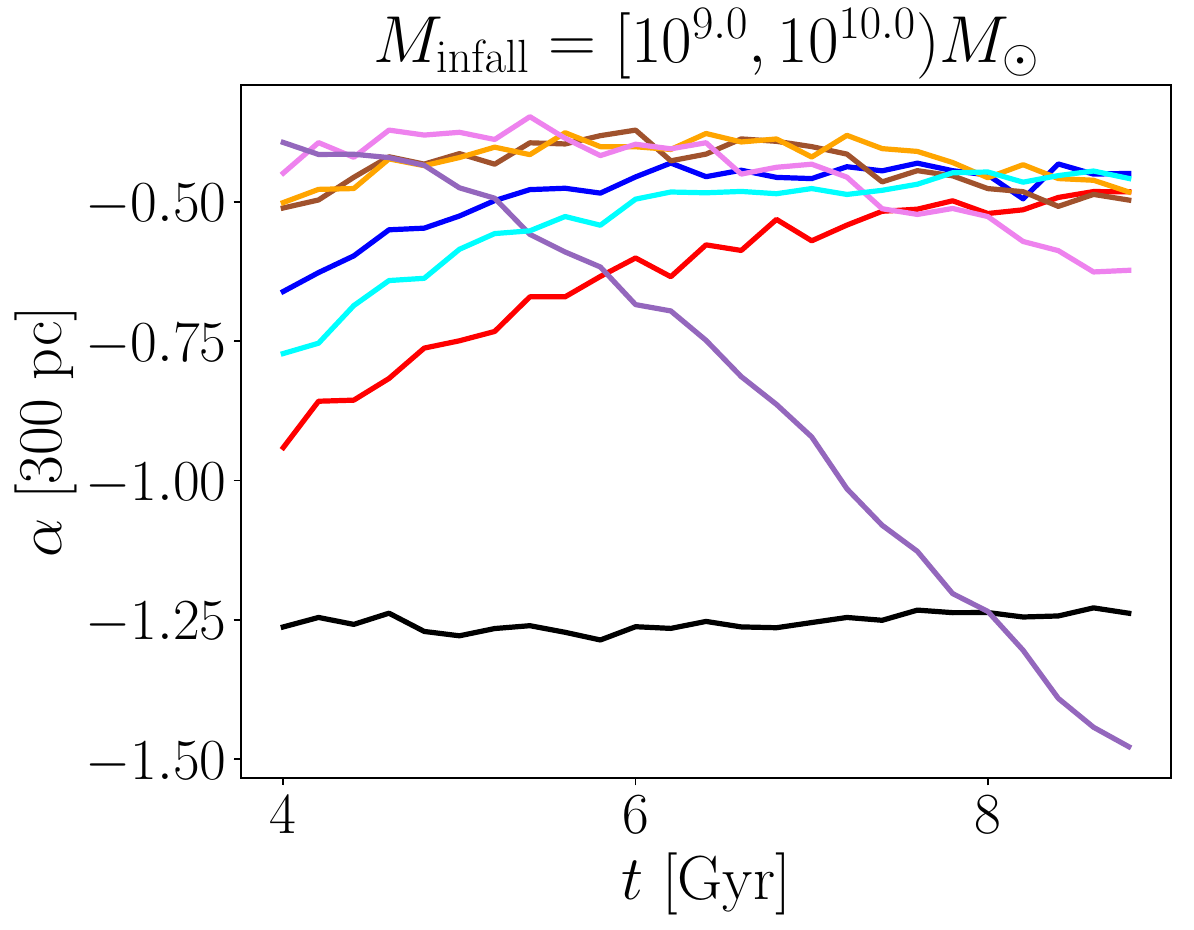}
        \caption{}
        \label{fig:sloped}
    \end{subfigure}
    ~
    \begin{subfigure}[t]{0.32\textwidth}
        \centering
        \includegraphics[width=\textwidth, clip,trim=0.3cm 0cm 0.3cm 0cm]{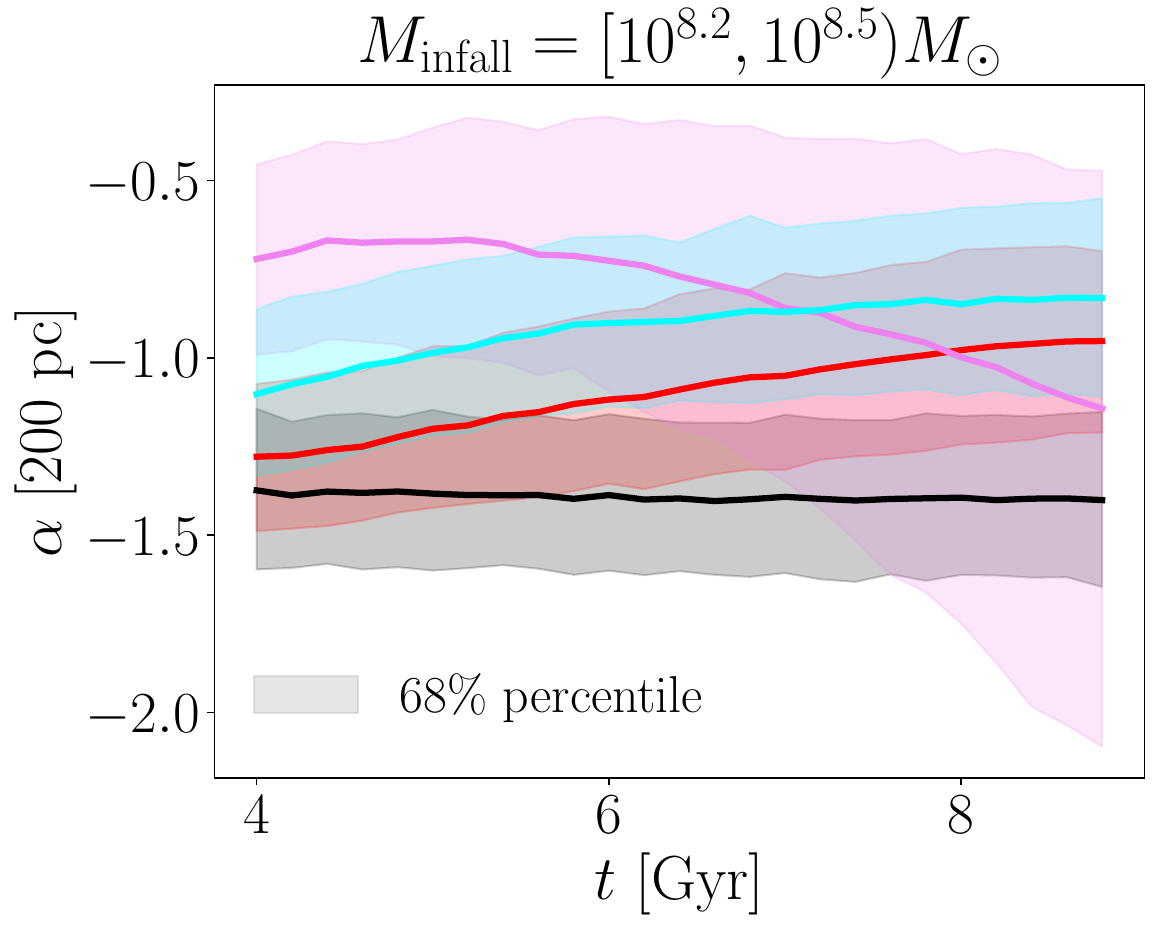}
        \caption{}
        \label{fig:slopee}
    \end{subfigure}
    ~
    \begin{subfigure}[t]{0.32\textwidth}
        \centering
        \includegraphics[width=\textwidth, clip,trim=0.3cm 0cm 0.3cm 0cm]{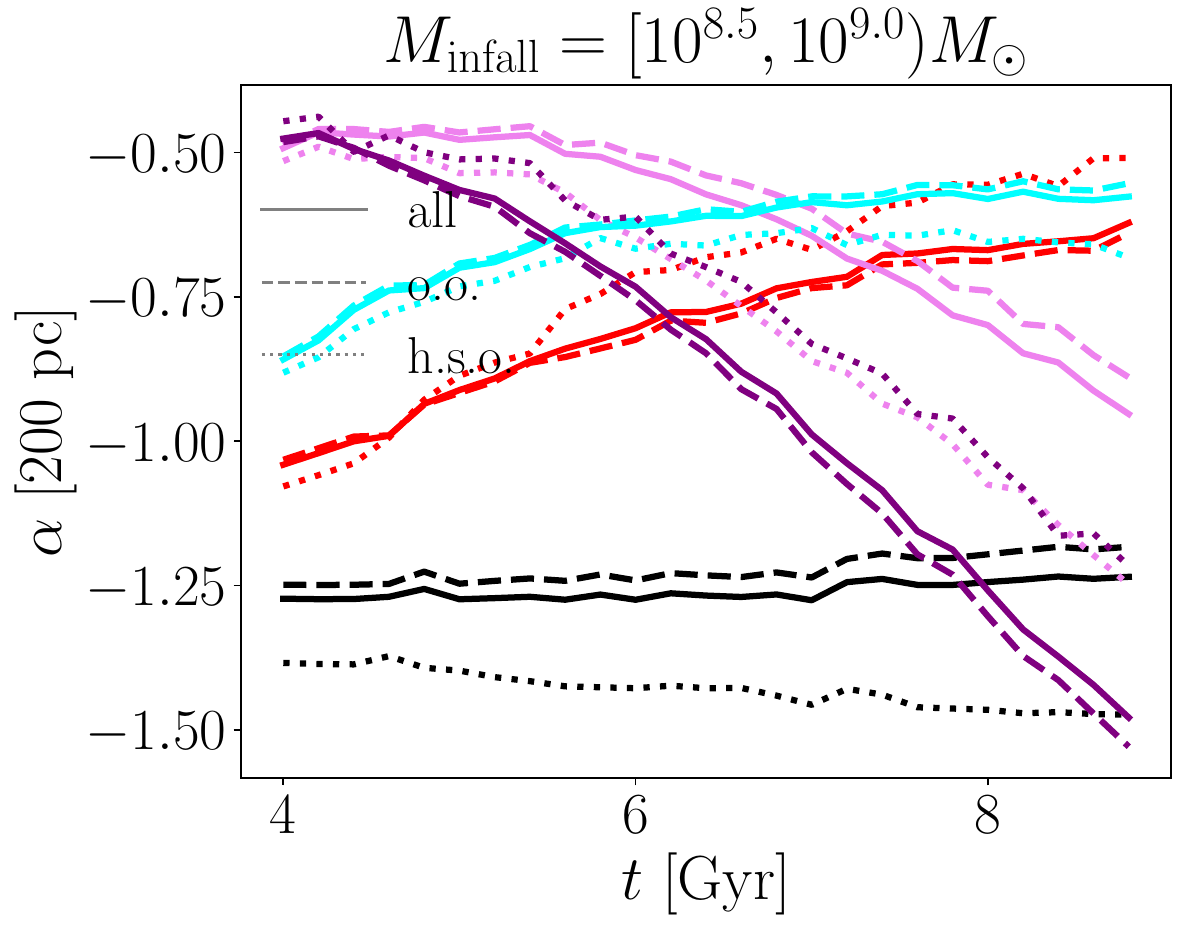}
        \caption{}
        \label{fig:slopef}
    \end{subfigure}    
    \caption{Time evolution of the average slopes of subhalos' density profiles, measured at a certain aperture of $R\pm 50$ pc near the subhalo center, where $R$ is specified in each subplot. a) to d) are the average slopes in each infall-mass group, regardless of the subhalos' orbits. In e) we show the scatter of the slopes in shaded region, for selected SIDM models. f) shows the slope evolution as a function of subhalos' orbits (classified as either ordinary orbits (o.o.) or heavily-stripping orbits (h.s.o.)).}
    \label{fig:slope}
\end{figure*}

Another feature of core-collapse is that the slope of the density profile $\alpha$ \footnote{defined by the scaling between density and radius $\rho(r) \propto r^{\alpha}$, or $\alpha(r)=d \log \rho / d \log r$ for a local measurement near $r$, where the `local' is defined by $\pm$ 50 pc in this work. See Fig. \protect\ref{fig:slope} } steepens significantly outside the core, far steeper than in CDM.  A recent study \cite{sengul22} finds that the slope of a subhalo at roughly $150-500$ pc (see their Fig.2) , called the `region of maximum observability', can be robustly measured with lensed arcs from strong lensing observations (see also \cite{minor17} and \cite{gzhang22} for robust subhalo probes from lensing), thus having the potential of distinguishing different DM models. 
Here, we present the averaged logarithmic slope $\alpha$ of subhalos' density profiles at specific radii to track subhalos' evolution.

We expect $\alpha$ to be a function of position with respect to halo center as well as time, motivating us to consider the slope at specific radii.  CDM halos with NFW profiles have $\alpha=-1$ in the center, smoothly transitioning to $\alpha=-3$ as the radius increases. SIDM halos have cored centers with $\alpha=0$, transitioning to $\alpha=-3$ at outer radii. As core-collapse kicks in, the SIDM core contracts, and the transition region effectively extends to smaller radii. Thus the time evolution of the slope $\alpha$ at a fixed aperture near the center of a subhalo should become more negative (steeper slope) as the subhalo starts to core-collapse. We show the averaged $\alpha$ of subhalos in Fig. \ref{fig:slope}, in the same style as the previous plot of subhalos' central mass panels. Panels a) to d) show the averaged $\alpha$ of each infall-mass group of subhalos, and panel e) demonstrates the scatter in $\alpha$. Panel f) shows the contribution of different orbits in the evolution of $\alpha$. Note that for the four infall-mass groups, we use different radii for measurement of $\alpha$ (200 pc, 200 pc, 200 pc and 300 pc for the four infall-mass groups, as specified in Fig. \ref{fig:slope}), because the simulations of more massive subhalos have more massive particles and lower spatial resolution, which limits our ability to measure the slope at small radii. These radii are within or close to the reported `region of maximal observability' according to \cite{sengul22}. We note that since we fix the aperture of $\alpha$ measurement for each mass group, it is the relative behaviour of $\alpha$ among dark models that matters, rather than the magnitude.

In the a) to d) panels of Fig. \ref{fig:slope}, we find that the averaged $\alpha$ of CDM subhalos remains nearly unchanged. The constant cross section model with $\sigma/m = 6\ \rm cm^2/g$ (red) is still at an early stage of core expansion at the end of the simulations, with the slope increasing (becoming less negative) over time. The velocity-dependent models $\{\sigma_0 = 50, \omega=50\}$ (blue) and  Colquhoun21 (cyan) are also in the core-expansion phase with an increasing (or, at most, starting to flatten) slope. For the three low infall-mass groups in Fig. \ref{fig:slopea} to Fig. \ref{fig:slopec}, the subhalos simulated with $\{\sigma_0 = 100, \omega=50\}$ (orange) and $\{\sigma_0 = 50, \omega=100\}$ (brown) models have decreasing $\alpha$ by the end of the simulation, indicating that a few subhalos are core-collapsed. By contrast, for the $\{\sigma_0 = 200, \omega=50\}$ (violet) and $\{\sigma_0 = 200, \omega=200\}$ (purple) models, we see the signature of core collapse in the fast-dropping $\alpha$. 
For the subhalos with infall mass between $10^9$ and $10^{10}M_\odot$, however, even the $\{\sigma_0 = 200, \omega=50\}$ model only has a mild decreasing trend in $\alpha$.  Only the $\{\sigma_0 = 200, \omega=200\}$ (purple) model still has a large fraction of core-collapsed subhalos, as indicated with the rapid drop in $\alpha$ as a function of time. This is consistent with what we found with the central enclosed masses in the previous section. 

We show the 68 percentile scatter of subhalos' inner slopes of selected SIDM models in Fig. \ref{fig:slopee}. Again, we can see the $\{\sigma_0 = 200, \omega=50\}$ model has the largest and most rapidly increasing scatter in $\alpha$, since there are large populations of both core-collapsing and core-forming subhalos. The scatter in $\alpha$ of the other DM models remains nearly unchanged with time, even for the $\sigma/m = 6$~ cm$^2$/g model that displays an increasing scatter over time in inner mass in Fig. \ref{fig:centralm-e} (red line). The explanation for this is that the drop of inner mass of $\sigma/m = 6$~cm$^2$/g subhalos in Fig. \ref{fig:centralm-e} results from both core-formation and evaporation, the latter tightly correlated with orbits. For the evolution of $\alpha$, though, evaporation does not play as significant a role, since it causes nearly equal mass loss at all subhalo radii and only indirectly contributes to the core-formation by making the central potential shallower. As a result, the $\alpha$ evolution in the $\sigma/m = 6$ cm$^2$/g model is not as strongly affected by the diversity in orbits as the inner mass is, keeping the scatter in $\alpha$ relatively constant.

To investigate the role of different orbits in driving the evolution of $\alpha$, in Fig. \ref{fig:slopef} we show $\alpha$ as a function of orbital type. For the CDM subhalos (black), the o.o. subhalos' slopes remain roughly constant, while the h.s.o. ones are slightly steepened. The latter is a result of the fact that the h.s.o. subhalos lose more mass during close pericenter passages, and the transitioning radius from $\alpha=-1$ to $\alpha=-3$ shifts to smaller radii (see also \cite{Pennarrubia10, green19, abenson22}). The h.s.o. subhalos show slightly faster growth in $\alpha$ than the o.o. ones for the constant cross section $6\rm cm^2/g$ case (red), because the strong evaporation causes extra mass loss and shallower potential in subhalo center, which adds to the core expansion. For the Colquhoun21 model (cyan) that has relatively weak evaporation and core-collapse in subhalos just about to kick in at the end of the simulation, similar to the CDM case, the stronger tidal field for h.s.o. subhalos causes slightly more negative $\alpha$ than the o.o. ones.

The effects of evaporation are apparent in the two models with strong core-collapse shown in the figure.  The $\{\sigma_0 = 200, \omega=50\}$ (violet) case's h.s.o. subhalos grow a steeper averaged $\alpha$ than the o.o. ones, due to the acceleration of core-collapse by stronger tidal stripping.  However, the $\{\sigma_0 = 200, \omega=200\}$ (purple) model, although having the largest core-collapse fraction, shows that h.s.o. subhalos are delayed in core-collapse because the strong evaporation overcomes the acceleration by tidal stripping from the host.

The subhalos' averaged inner slope $\alpha$ tell a story that is consistent with the observations of subhalos' central mass, as described in Sec. \ref{sec:halo1} and the summary Table \ref{table:ccp}.  With upcoming substructure lensing observations, the inner subhalo slopes could be promising as an avenue to setting constraints on SIDM models (also see Sec. \ref{sec:res-lens}).

\begin{figure}
    \begin{subfigure}[t]{\columnwidth}
        \centering
        \includegraphics[width=\textwidth]{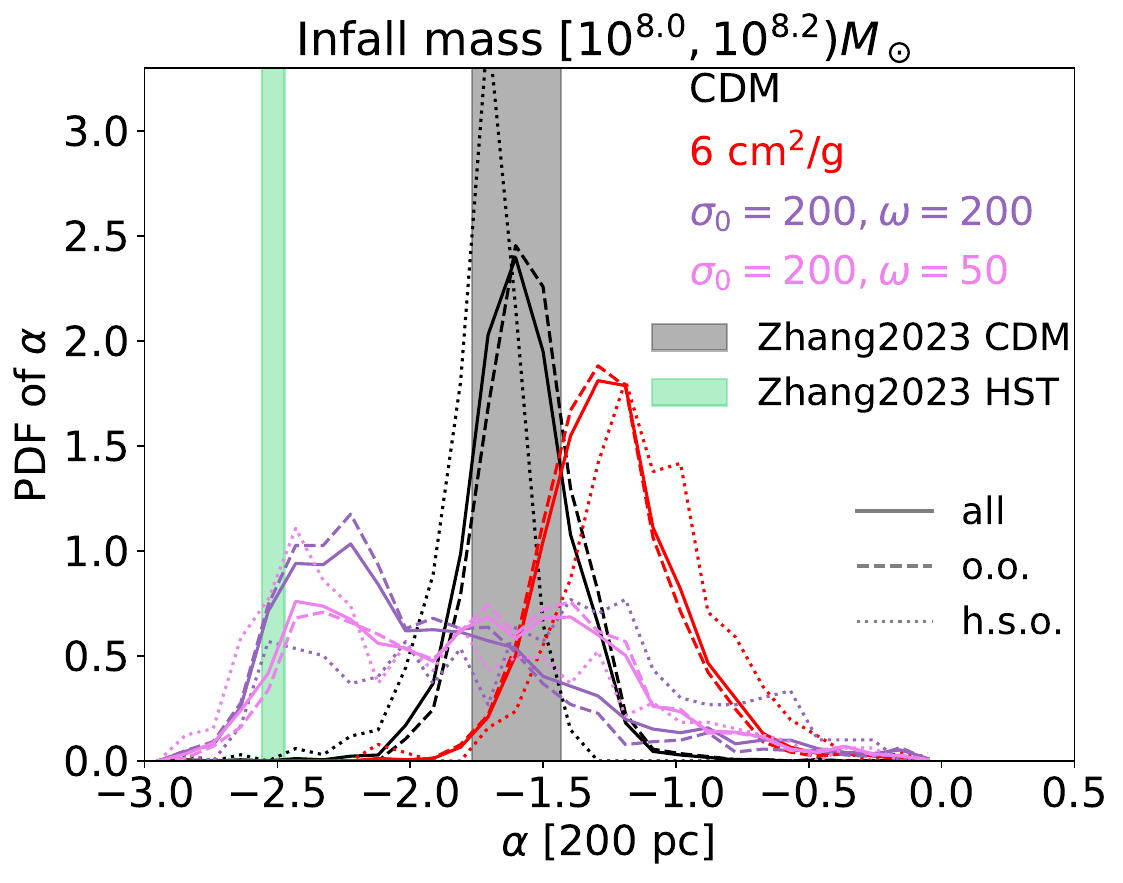}
    \end{subfigure}
    \caption{The probability distribution of the inner slope $\alpha$ for the infall-mass group $[10^{8.0}, 10^{8.2})$. We find that the ultra-steep inner density profiles in subhalos reported in \protect \cite{gzhang23} can be explained by core-collapse in SIDM subhalos. The lines have the same meaning as in Fig.~\ref{fig:slopef}. `o.o.' and `h.s.o.' are abbreviations for ordinary orbits and heavily-stripping orbits.}
    \label{fig:slope-hist}
\end{figure}

Excitingly, \cite{gzhang23} analyzed 13 strongly lensed images by Hubble Space Telescope, measuring subhalos with masses similar to the ones in our simulation.  They found a much steeper overall inner slope than expected for CDM subhalos --- even with extreme tidal truncation (see Fig. 6 of \cite{gzhang23}). As shown in Fig. \ref{fig:slope-hist}, we find that this may be explained by core-collapse in SIDM subhalos.  The probability distribution of $\alpha$ for our two models with highest core-collapse fraction peak at the steep slope reported in \cite{gzhang23}. However, there are two key differences between our results and their measurement. 
First, we use a fixed aperture (200 pc for the infall mass group $10^8-10^{8.2}M_\odot$ shown in Fig.~\ref{fig:slope-hist}), while their radial range for the slope measurement can vary for each subhalo (called Maximal Observability Region; see \cite{sengul22}). On the other hand, the distribution of $\alpha$ of our CDM subhalos does overlap with the CDM lensing mocks in \cite{gzhang23}, suggesting that our results should be comparable. Second, there is not much mass information for the subhalos in the 13 images in \cite{gzhang23} available yet, while in our analysis we have to select subhalos within a certain infall-mass group for comparison. Thus, the comparison shown in Fig. \ref{fig:slope-hist} serves as a first demonstration that SIDM core-collapse is a promising explanation for the observed ultra-steep density profiles in substructure lensing.  More careful investigations are needed to test this intriguing hypothesis.

\subsection{Phenomenological SIDM models}\label{sec:halo-artificial}

\begin{figure*}
    \centering
    \begin{subfigure}[t]{0.32\textwidth}
        \centering
        \includegraphics[width=\textwidth, clip,trim=0.3cm 0cm 0.3cm 0cm]{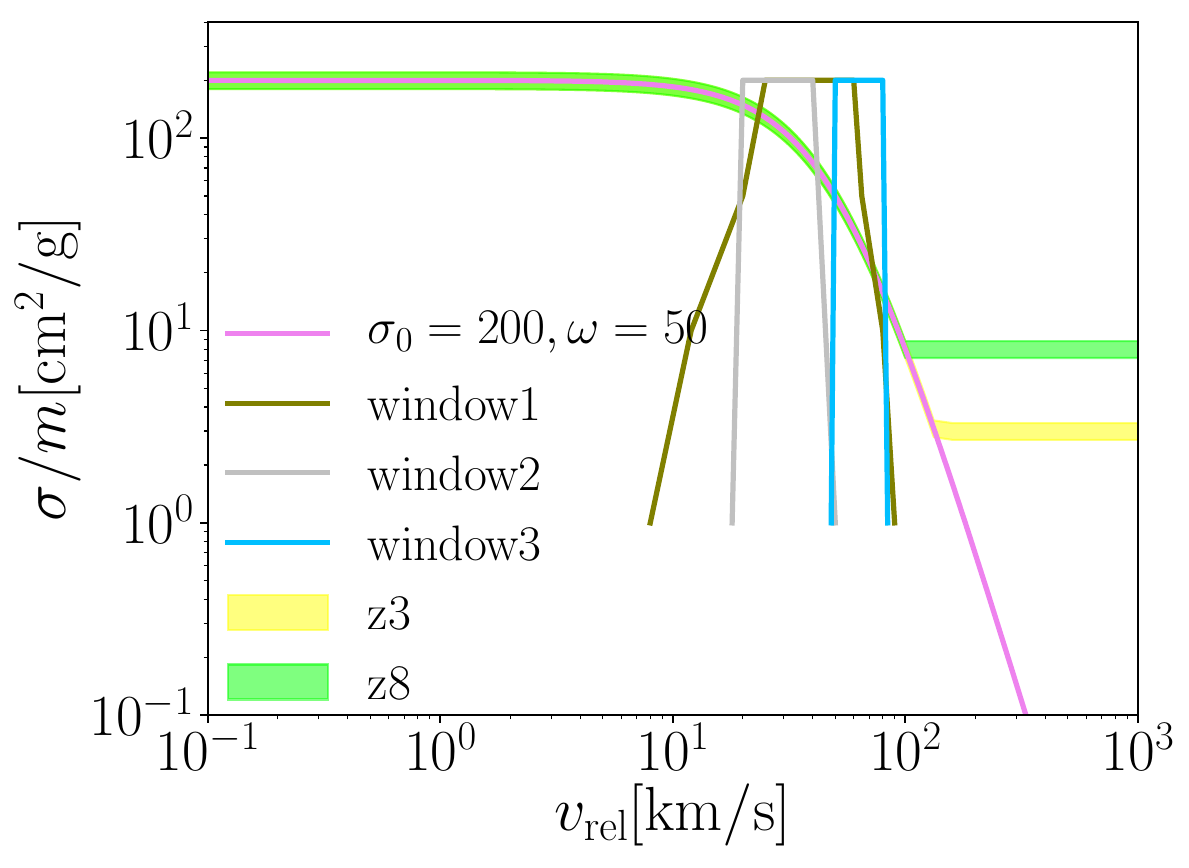}
        \caption{}
        \label{fig:arti-x}
    \end{subfigure}    
    ~
    \begin{subfigure}[t]{0.32\textwidth}
        \centering
        \includegraphics[width=\textwidth, clip,trim=0.3cm 0cm 0.3cm 0cm]{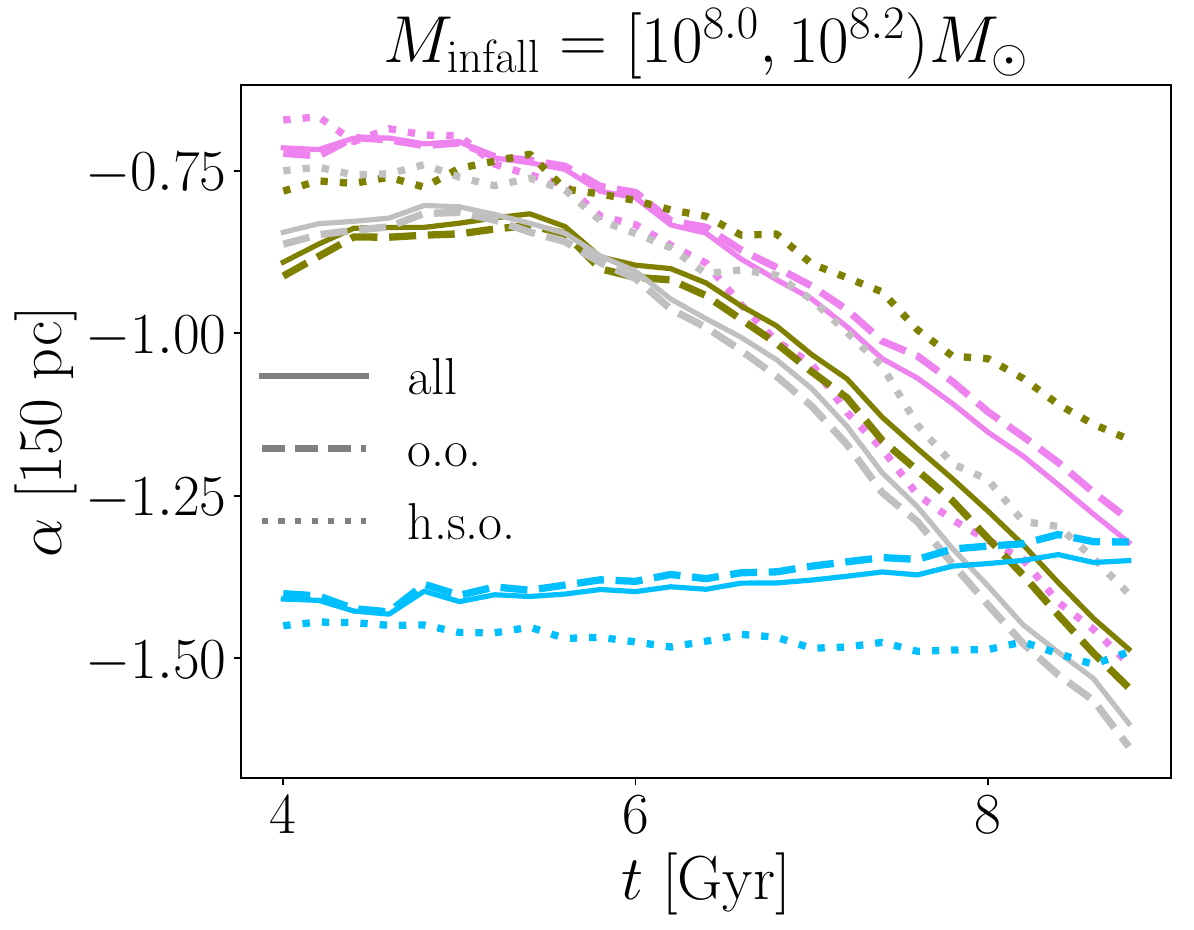}
        \caption{}
        \label{fig:arti-w-8.0}
    \end{subfigure}
    ~
    \begin{subfigure}[t]{0.32\textwidth}
        \centering
        \includegraphics[width=\textwidth, clip,trim=0.3cm 0cm 0.3cm 0cm]{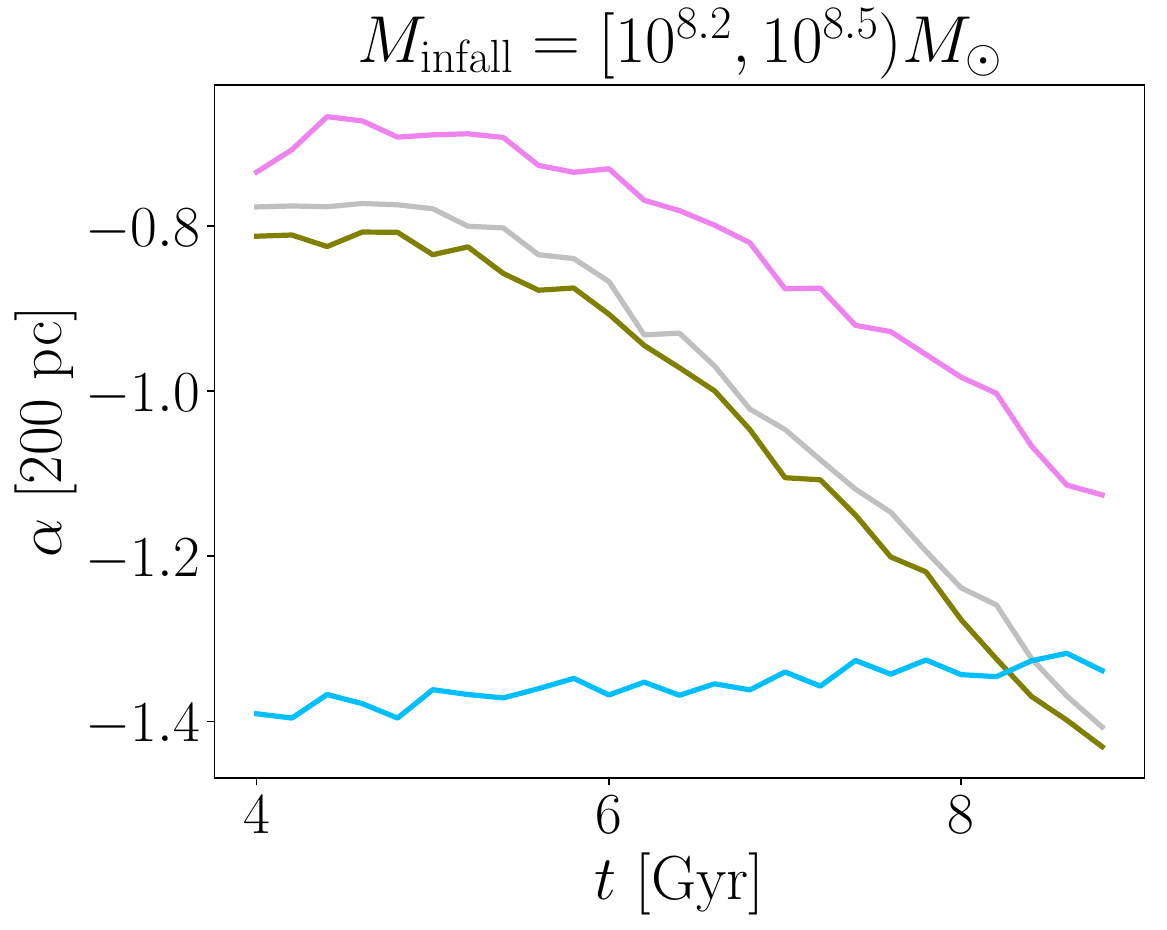}
        \caption{}
        \label{fig:arti-w-8.2}
    \end{subfigure}
    ~
    \begin{subfigure}[t]{0.32\textwidth}
        \centering
        \includegraphics[width=\textwidth, clip,trim=0.3cm 0cm 0.3cm 0cm]{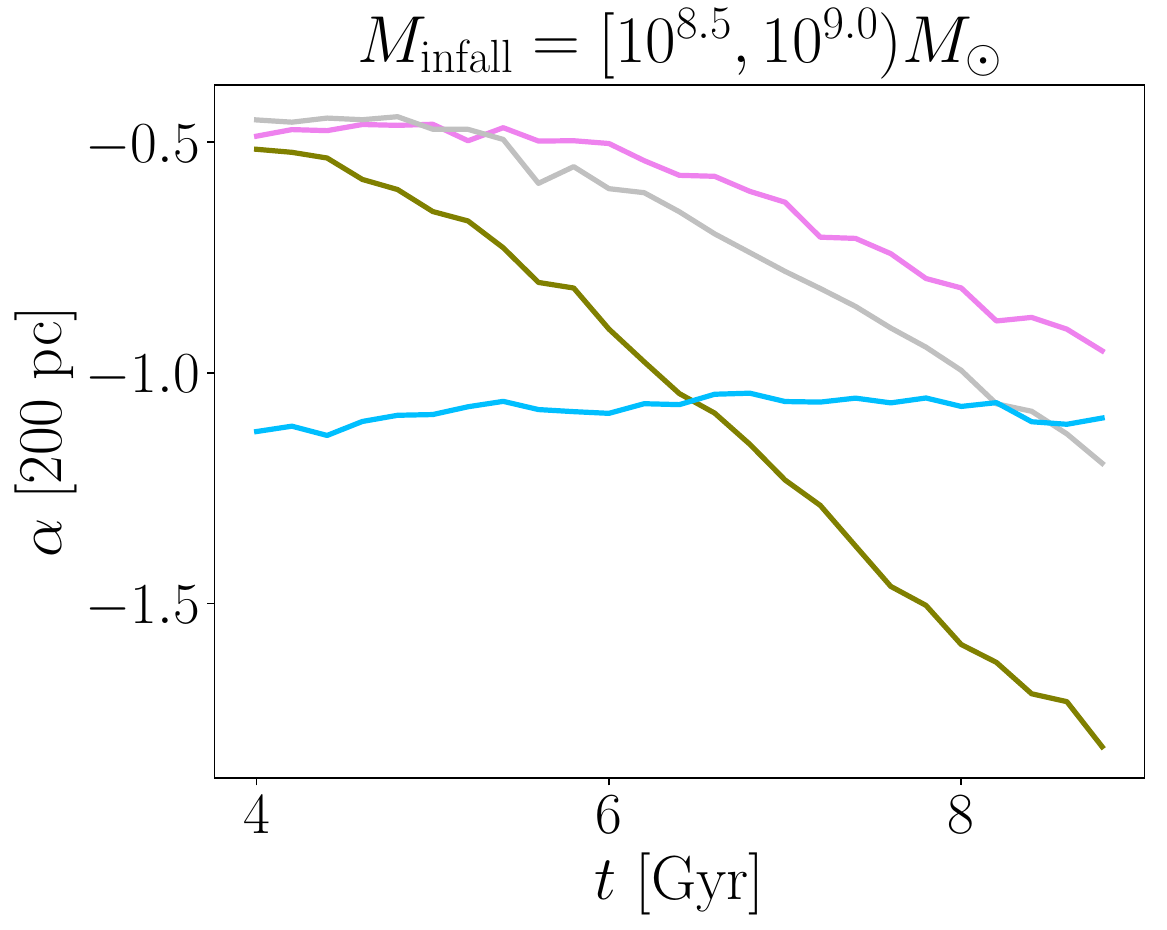}
        \caption{}
        \label{fig:arti-w-8.5}
    \end{subfigure}
     ~
    \begin{subfigure}[t]{0.32\textwidth}
        \centering
        \includegraphics[width=\textwidth, clip,trim=0.3cm 0cm 0.3cm 0cm]{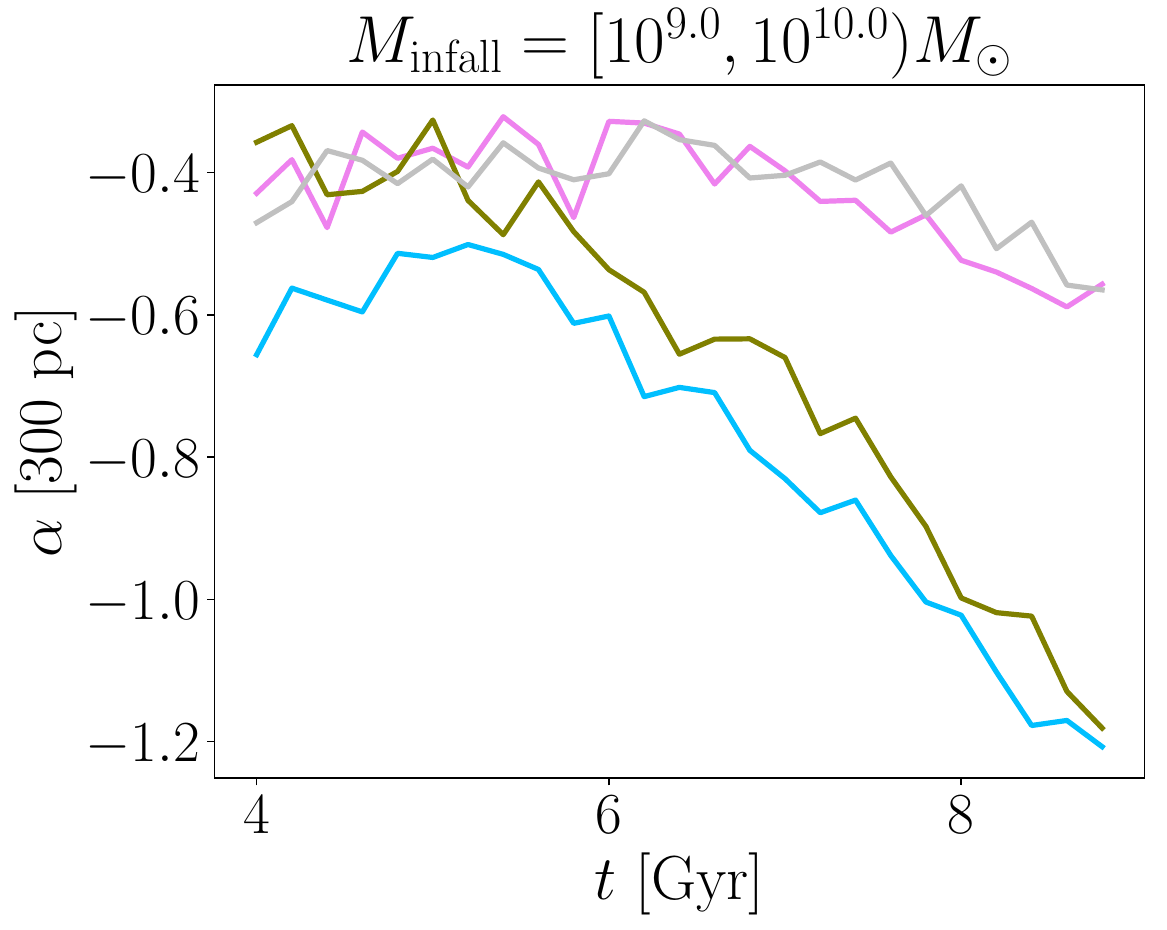}
        \caption{}
        \label{fig:arti-w-9.0}
    \end{subfigure}
    ~
    \begin{subfigure}[t]{0.32\textwidth}
        \centering
        \includegraphics[width=\textwidth, clip,trim=0.3cm 0cm 0.3cm 0cm]{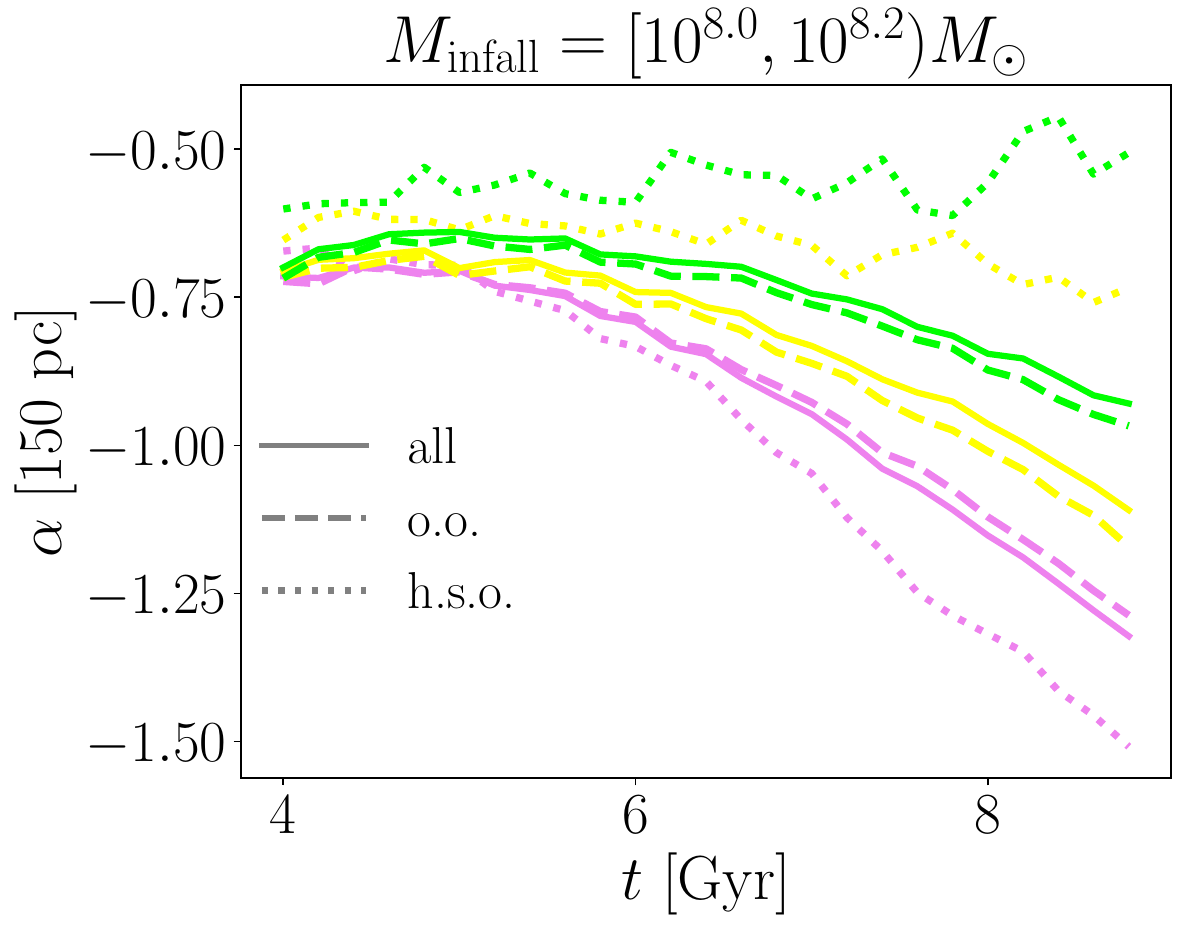}
        \caption{}
        \label{fig:arti-z}
    \end{subfigure}
    ~
    \caption{The time evolution of subhalos' average inner slope, as in Fig. \ref{fig:slope}, but for the phenomenological SIDM models in Sec. \ref{sec:halo-artificial}. Panel a) shows the $\frac{\sigma}{m}-v_{\rm rel}$ relation of the phenomenological models, including window functions and z-type ones. Panels b) to e) show the evolution of averaged inner slope $\alpha$ for subhalos in four different infall-mass groups, with the window function SIDM models. Panel f) shows the evolution of $\alpha$ for z-type models for the lowest infall-mass group $[10^8, 10^{8.2})M_\odot$. In panels b) and f) we also show the contribution by different orbits (ordinary or heavily-stripping).}
    \label{fig:arti}
\end{figure*}

In previous sections, we found that the interplay of competing physical processes have complicated effects on subhalo evolution. In order to explore and disentangle the effects on subhalo evolution, in this section we deploy several toy SIDM models with velocity-dependent features selected to highlight specific physics.  
We choose two categories of phenomenological models here, as shown in Fig. \ref{fig:arti-x}.  We call them the `window-function type' and `z-type' because of their shapes in cross-section--velocity space. Both the window function and z-type are variations of the $\{\sigma_0=200, \omega = 50 \}$ model (violet). 

The `window function' models (olive, grey and light blue colors in Fig. \ref{fig:arti-x}) are meant to mimic resonances.  They have the same constant $\sigma/m= 200\,\rm cm^2/g$, but for a narrow and different range of relative velocities.  The cross section drops to zero when $v_{\rm vel}$ is outside the specified velocity window. These window functions test the halo-mass filtering of core-collapse for resonances. 
Although these window functions serve as toy models in this work, they provide insights to more realistic velocity-dependent SIDM models with complicated resonance peaks, such as the Colquhoun21 model (cyan color in Fig. \ref{fig:sig-v}; \cite{col21}) or in \cite{dg22}. We explain the technical choice for the shape later in this section.

The `z-type' models explore evaporation.  They follow the base model of $\{\sigma_0=200, \omega = 50 \}$ until the cross section drops to $8\,\rm cm^2/g$ (z8, lime color in Fig. \ref{fig:arti-x}) or $3\,\rm cm^2/g$ (z3, yellow color),  remaining constant at increasing velocities. The relevant velocity scale for evaporation is the relative velocity between the subhalo and host, $\mathcal{O}(100)-\mathcal{O}(1000)\ \rm km/s$. \cite{zzc22} showed that for SIDM with constant cross section, the host-subhalo evaporation is almost always strong enough to prevent core-collapse in subhalos (see their Fig. 12). Here we test evaporation in the context where the cross section relevant for heat transfer within the subhalo is much higher than the cross section relevant for evaporation. 

\begin{table*}
	\centering
	\begin{tabular}{lcccc} 
		\hline
		Mass group [dex $M_\odot$] & typical 1-d $\sigma_v$ [km/s] & window1 [25, 60] km/s & window2 [20, 40] km/s & window3 [50, 80] km/s\\
		\hline
        $[8.0, 8.2)$  & 6.4 -- 7.1 & \checkmark & \checkmark  & -  \\
		\hline
        $[8.2, 8.5)$  & 7.2 -- 8.6 & \checkmark  & \checkmark  & -  \\
		\hline
        $[8.5, 9.0)$  & 9.2 -- 11.8 & \checkmark & yes but not many & - \\
		\hline
        $[9.0, 10.0)$  & 13.2 -- 20.8 & \checkmark & - & \checkmark  \\
		\hline
	\end{tabular}
	\caption{This table summarizes the typical velocity scales for each infall-mass group of subhalos, and whether the subhalos typically core-collapse (with fraction $\gtrsim 20\%$) for each window-function SIDM model (marked with a check). 
    Note that the 1-d $\sigma_v$ of each mass group is shown as a range, because the subhalo mass and concentration in each group span a range.  
    Here we show the $\pm 1\sigma$ range of the max 1-d velocity dispersion for each CDM subhalo at $t=4\ \rm Gyr$ (see Fig. \ref{fig:vdisp}). The results of whether subhalos core-collapse in each window-type SIDM model correspond to Fig. \ref{fig:arti-w-8.0} to Fig. \ref{fig:arti-w-9.0} and Table \ref{table:ccp}. } 
    \label{table:arti}
\end{table*}

As shown in Fig. \ref{fig:arti-x}, the `window2' model (grey) has $\sigma/m = 200\ \rm cm^2/g$ for $v_{\rm rel}$ in the range $[20, 40]\ \rm km/s$, `window3' has the same cross section in $[50, 80]\ \rm km/s$, while `window1' covers a broader range of velocities $[25, 60]\ \rm km/s$ and has a less sharp decay outside of the window. As a reference, we mark the typical 1-d velocity dispersion of each infall-mass group of subhalos in Table \ref{table:arti}, as well as whether each mass group sees a large fraction (>10\%, see Table \ref{table:ccp}) of subhalos core-collapsing. The time evolution of average subhalos' inner slope $\alpha$ of each infall-mass simulation and SIDM model is shown in Fig. \ref{fig:arti-w-8.0} to Fig. \ref{fig:arti-w-9.0}, where we can clearly see that these window functions work as a filter to pick out only a certain mass range of subhalos for core-collapse. Window1 covers the broadest range of velocities, so a large fraction of subhalos core-collapse in all the infall-mass groups. Window2 favors relatively low velocities, yielding clear signs of core-collapse in the $[10^8, 10^{8.2})M_\odot$ and $[10^{8.2}, 10^{8.5})M_\odot$ mass groups.  However, the evolution of the inner slope $\alpha$ diverges from window1 (and is more similar to the $\sigma-\omega$ model in violet color) for the $[10^{8.5}, 10^{9})M_\odot$ and $[10^{9}, 10^{10})M_\odot$ mass range.  Window3 favors relatively large velocities.  While the inner slope $\alpha$ hardly evolves for the smallest mass bins we consider (similar to CDM), it evolves rapidly for the massive subhalos in the $[10^9, 10^{10})M_\odot$ infall mass bin. 

A key question for this core-collapse mass filtering is how to quantify the relation between a subhalo's typical velocity (1-d velocity dispersion $\sigma_v$; another alternative is the maximum circular velocity with $ v_{\rm max} \approx \sigma_v/0.64$, see \cite{o22}) and the characteristic velocity relevant for scattering calculations, $\ev{v_{\rm rel}}$. In other words, how to translate between the velocity scale of the halo and the velocity-dependent cross section plot, Fig. \ref{fig:sig-v}? If particle velocities are Maxwell-Boltzmann distributed, one can show that $\ev{v_{\rm rel}} = 2.26 \sigma_v$.  However, this is almost certainly not the right mapping.  As shown in \cite{o22} and \cite{sq22}, what matters is the heat conductivity of the halo, which is a fifth order term of the velocity scale (see Eqn. \ref{eqn:conductivity}). 

Here we use the window-type models to empirically estimate the relation between $\ev{v_{\rm rel}}$ and $\sigma_v$.
We choose the maximal $\sigma_v$ to characterize each subhalo, as shown in Fig. \ref{fig:sigmav1}, and assign its $1\sigma$ scatter among CDM subhalos to be the range of typical $\sigma_v$ for each infall-mass group, as listed in Table \ref{table:arti}. This is because the self-interaction leads to thermalization in halo center, with a nearly uniform velocity dispersion close to the maximal $\sigma_v$ of the CDM initial condition of the halo. 
Assuming a linear relation $\ev{v_{\rm rel}} = C\sigma_v$, we can use the simulation results of which infall-mass bins of subhalos core-collapse in each window-model to estimate a range of $C$. For example, in Table \ref{table:arti} (and also Fig. \ref{fig:arti-w-8.0}), we can see that the subhalos in the $[10^8, 10^{8.2}) M_\odot$ infall mass group, whose CDM typical 1-d velocity dispersion ranges in 6.4-7.1 km/s, have a high core-collapse fraction for the window2 model that operates between 20-40 km/s. Subhalos in this mass group must have $\ev{v_{\rm rel}}$ overlapping with the velocity window 20-40 km/s for core-collapse to happen. This means the highest velocity dispersion with a multiplication factor $C$ must exceed the low-end of the velocity window, i.e. $7.1C>20$, and the lowest velocity dispersion with $C$ must be smaller than the high-end of the velocity window, i.e. $6.4C<40$, where we find a first estimated range $2.8 < C < 6.3$. Similarly, we can estimate a range of $C$ from the combination of each infall-mass bins and window-models. Combining all these estimated ranges of $C$, we estimate $3.0<C<4.2$. We show in Appendix \ref{appdx:window} that an intermediate value $C\approx3.8$  is in remarkable agreement with analytical results from  gravothermal method. We comment that this estimate is specific for the window-type models presented here, and may be different for other models.

In Fig. \ref{fig:arti-w-8.0}, we distinguish the subhalos according to their orbits. Surprisingly, we find an overall \textit{deceleration} of core-collapse for subhalos on heavily-stripping orbits even though the tidal fields are stronger, in the cases of window1 and window2 (statistics can also be found in Table \ref{table:ccp}). 
Since these window functions have only narrow ranges and operate only for $v_{\rm rel}<100\ \rm km/s$, the host-subhalo evaporation is negligible (\cite{zzc22}; see also Fig. \ref{fig:slopef}). Therefore, this overall deceleration of subhalo core-collapse with stronger tidal fields is counter-intuitive.  

To understand why this happens, we must explore tidal effects in tandem with heat transfer in the subhalo. \cite{zzc22} separates the tidal effect into tidal heating and tidal stripping.  The former injects heat globally into the subhalo with the work done by tidal forces \citep{pullen14, sq20}, while the latter strips away dark matter mostly from the outer part of the subhalo when the work done by tidal forces exceeds the binding energy. Tidal heating may heat up the center of the subhalo and delay core-collapse, but only near the pericenter (see Fig. 9 of \cite{zzc22} for an example).  However, tidal stripping indirectly accelerates the core-collapse process by steepening the negative temperature gradient in the subhalo. For the window function cross sections, the delay in subhalo core-collapse due to tidal heating remains since it is purely a gravitational effect.  However, the acceleration from tidal stripping depends on both the temperature gradient and on the cross section relevant for conduction.  Because the particle velocities in the halo outskirts are low relative to the halo centers, the scattering cross section may be effectively zero if the velocity falls out of the operating range of the window-model, even if it is high in the halo center where the typical particle relative speeds are higher. This can quench the core-collapse process since the outward heat transfer is not efficient. 
This scenario is to some extent similar to another example shown in \cite{dnyang22jcap} (see their Sec. 4.3 and Fig. 8), where they find the core-collapse of an isolated halo is delayed by turning off the self-interaction outside the center of the halo. 

We comment that simulating the window function models is currently very time consuming, which is a key driver for why we show only one realization per model in this section.  The high computational cost is because we have to enforce small timesteps to avoid underestimation of scattering probability caused by the non-smooth nature of the window function models. The SIDM-based timestep of each simulation particle in \texttt{Arepo} is determined by $1/[\rho \overbar{\sigma}_v \frac{\sigma}{m}(\overbar{\sigma}_v)]$, where $\rho$ is the local density (`local' defined by the 32 nearest neighbors, as mentioned before), $\overbar{\sigma}_v$ is the locally averaged velocity dispersion, and the velocity-dependent cross section $\frac{\sigma}{m}(v_{\rm rel})$ is calculated using $\overbar{\sigma}_v$. Here the velocity dispersion instead of the actual mean pairwise velocity is used because the latter is a $\mathcal{O}(N^2)$ calculation while the former is $\mathcal{O}(N)$, and they are of the same order of magnitude.  Thus, using $\overbar{\sigma}_v$ to determine the timestep is usually a fair estimate. This scheme generally works well, but for the window function type of SIDM models, feeding the averaged $\overbar{\sigma}_v$ into $\frac{\sigma}{m}(v_{\rm rel})$ sometimes leads to zero when $\overbar{\sigma}_v$ is outside of the window, even if the actual probability of scattering is not zero.  As a result, sometimes the timestep for a particle can be too big and lead to an unphysical probability ($>1$) for pairwise scattering. We apply a temporary patch by enforcing a large enough, constant $\frac{\sigma}{m}$ when determining the timestep, at the cost of much longer simulation time.  Identifying methods to efficiently handle resonances in cross sections will be important for future work.  

Finally, to highlight the effect of evaporation, we investigate the `z-type' cross sections. The evolution of these subhalos' inner slope is shown in Fig. \ref{fig:arti-z}, for the mass group of $[10^8, 10^{8.2})M_\odot$. Only one mass group is shown here because all the mass bins show the same behavior. The strength of the host-subhalo evaporation increases in the order of $\{\sigma_0=200, \omega = 50 \}$ (violet), z3 (yellow) and z8 (lime).  The violet model has a small cross section ($\lesssim 1\ \rm cm^2/g$, sharply dropping) in the velocity range of $\mathcal{O}(100)$--$\mathcal{O}(1000)$ km/s, while z3 and z8 have 3 $\rm cm^2/g$ and 8 $\rm cm^2/g$ respectively. As a result, we can see that the rate and fraction of subhalo core-collapse is delayed by the extra evaporation in the z3 and z8 cases. Moreover, the averaged $\alpha$ of the h.s.o. subhalos of z3 and z8 indicate that they remain in the maximally cored stage.  Thus, with even a constant cross section as small as 3 $\rm cm^2/g$, evaporation causes core-collapse to be substantially delayed. The destruction of core-collapse is slightly weaker for more massive subhalos, because their fractional inner mass loss is less than low-mass subhalos for the same evaporation strength, as can be seen in Table \ref{table:ccp}.  For h.s.o. subhalos with infall-mass bins from [8.0, 8.2) to [8.5, 9.0), the core-collapse fraction of the $\{ \sigma_0 = 200, \omega = 50\}$ model stays constant at $\sim$25\%, but for the z3 model it increases from 1.4\% to 10.4\%. This set of z-type toy models sheds light on velocity-dependent SIDM models that have a mild/slow drop in cross section as $v_{\rm rel}$ increases, showing that it may greatly decrease the likelihood and rate of core-collapse in low-mass subhalos.

In this section we test a few phenomenological SIDM cross sections, namely the window functions and z-types, and show their implications for subhalo core-collapse. The window models are proxies for resonances.  They serve as mass filters where core-collapse only happen in subhalos in a certain range of (infall-)mass.  The z-type models highlight the effects of evaporation on delaying or preventing core-collapse.  Even cross sections of $\mathcal{O}(1)$ cm$^2$/g on the orbital-velocity scale can lead to significant evaporation effects. Simplified as they are, these phenomenological models provide insights to more realistic, physically motivated, and complicated SIDM models.

\subsection{Lensing predictions}\label{sec:res-lens}

\begin{figure*}
    \centering
    \begin{subfigure}[t]{0.48\textwidth}
        \centering
        \includegraphics[width=\textwidth, clip,trim=0.3cm 0cm 0.3cm 0cm]{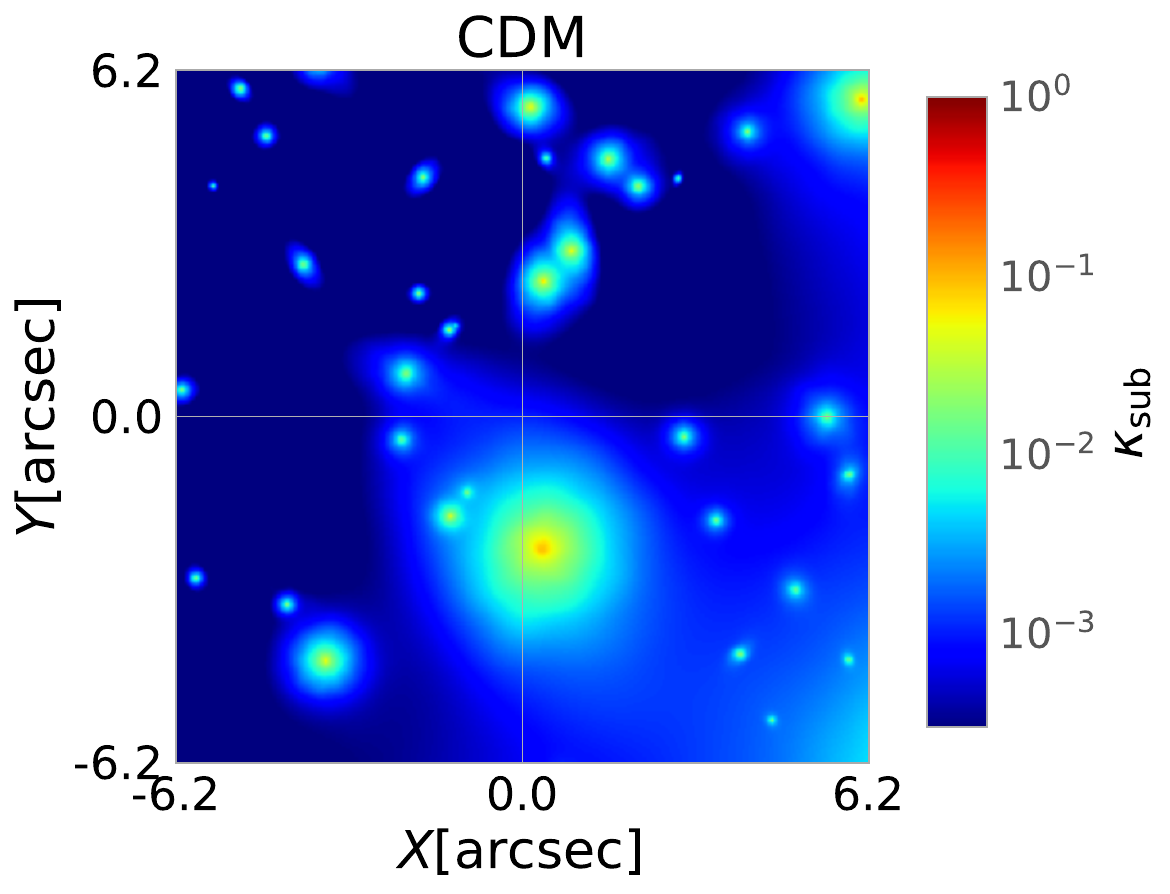}
        \caption{}
        \label{fig:dens-a}
    \end{subfigure}    
    ~
    \begin{subfigure}[t]{0.48\textwidth}
        \centering
        \includegraphics[width=\textwidth, clip,trim=0.3cm 0cm 0.3cm 0cm]{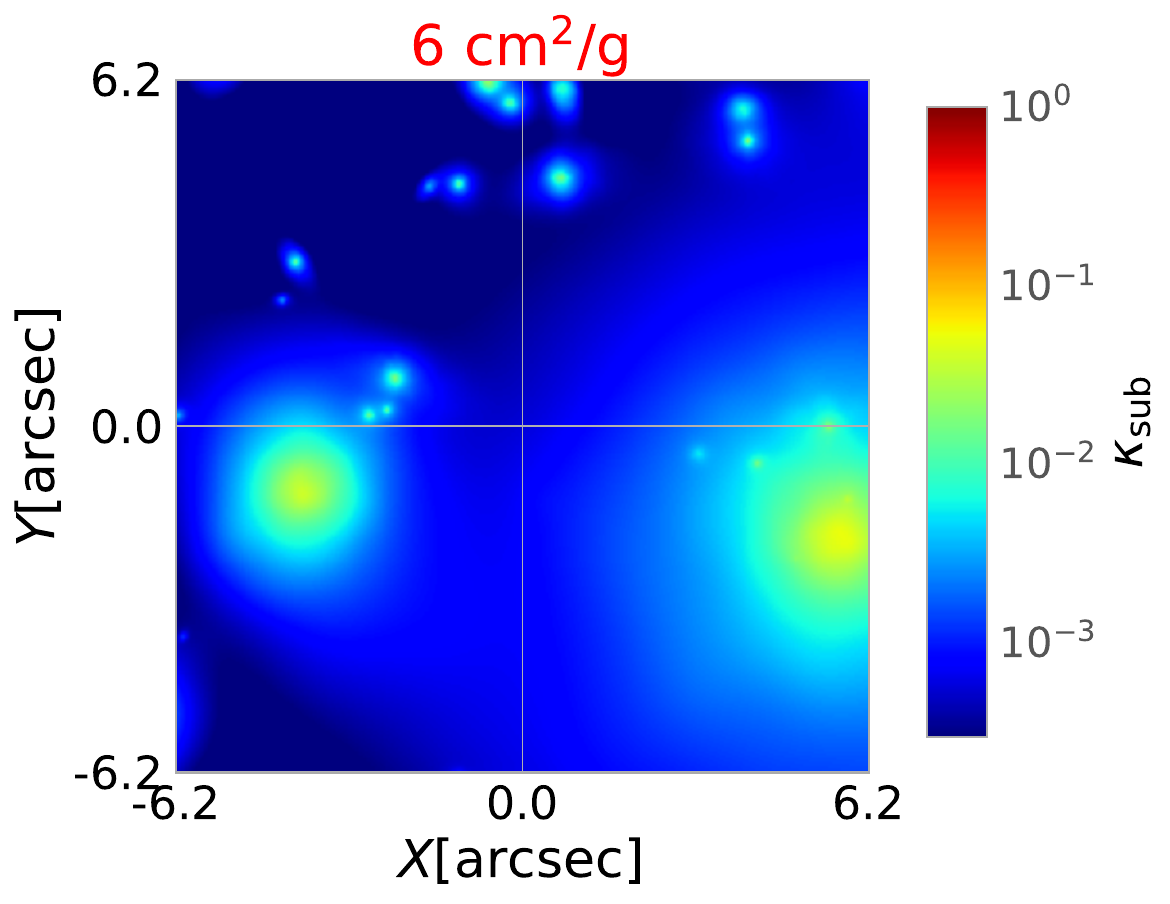}
        \caption{}
        \label{fig:dens-b}
    \end{subfigure}
    ~
    \begin{subfigure}[t]{0.48\textwidth}
        \centering
        \includegraphics[width=\textwidth, clip,trim=0.3cm 0cm 0.3cm 0cm]{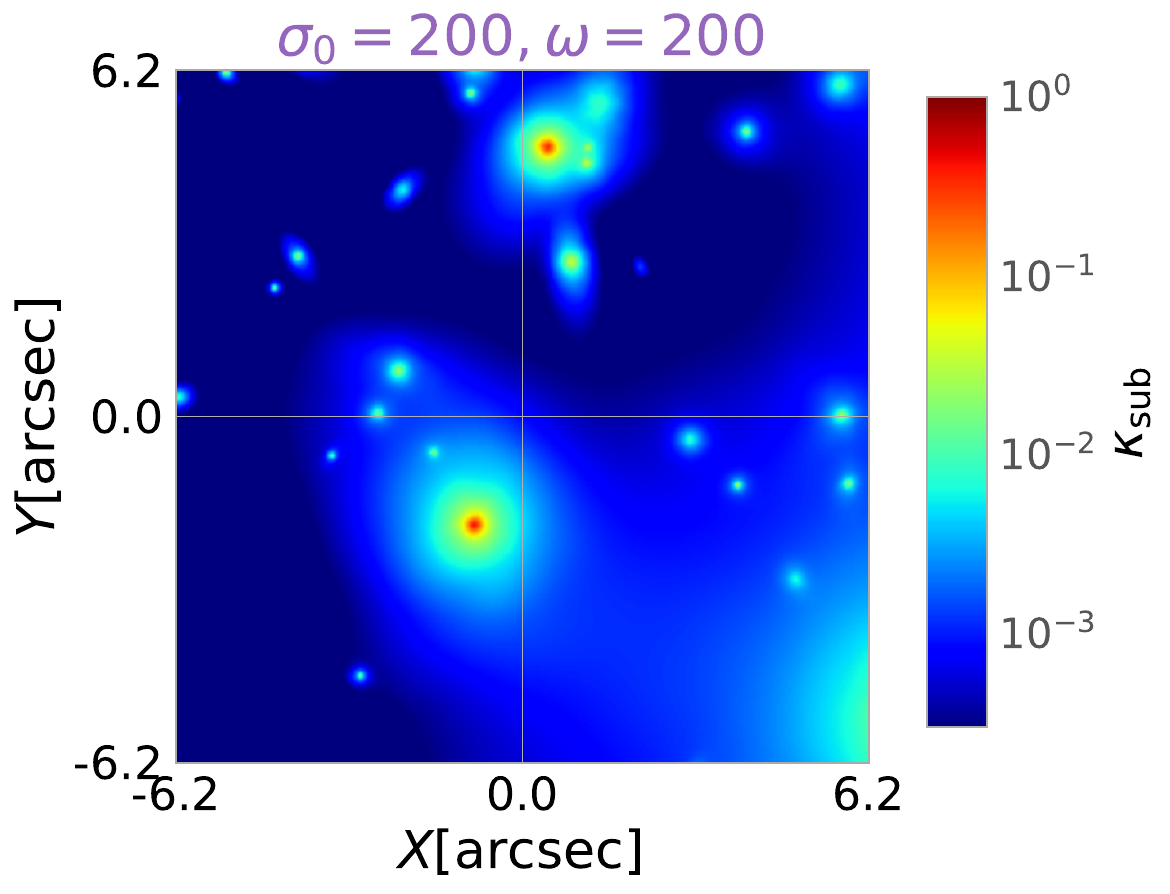}
        \caption{}
        \label{fig:dens-c}
    \end{subfigure}
        ~
    \begin{subfigure}[t]{0.48\textwidth}
        \centering
        \includegraphics[width=\textwidth, clip,trim=0.3cm 0cm 0.3cm 0cm]{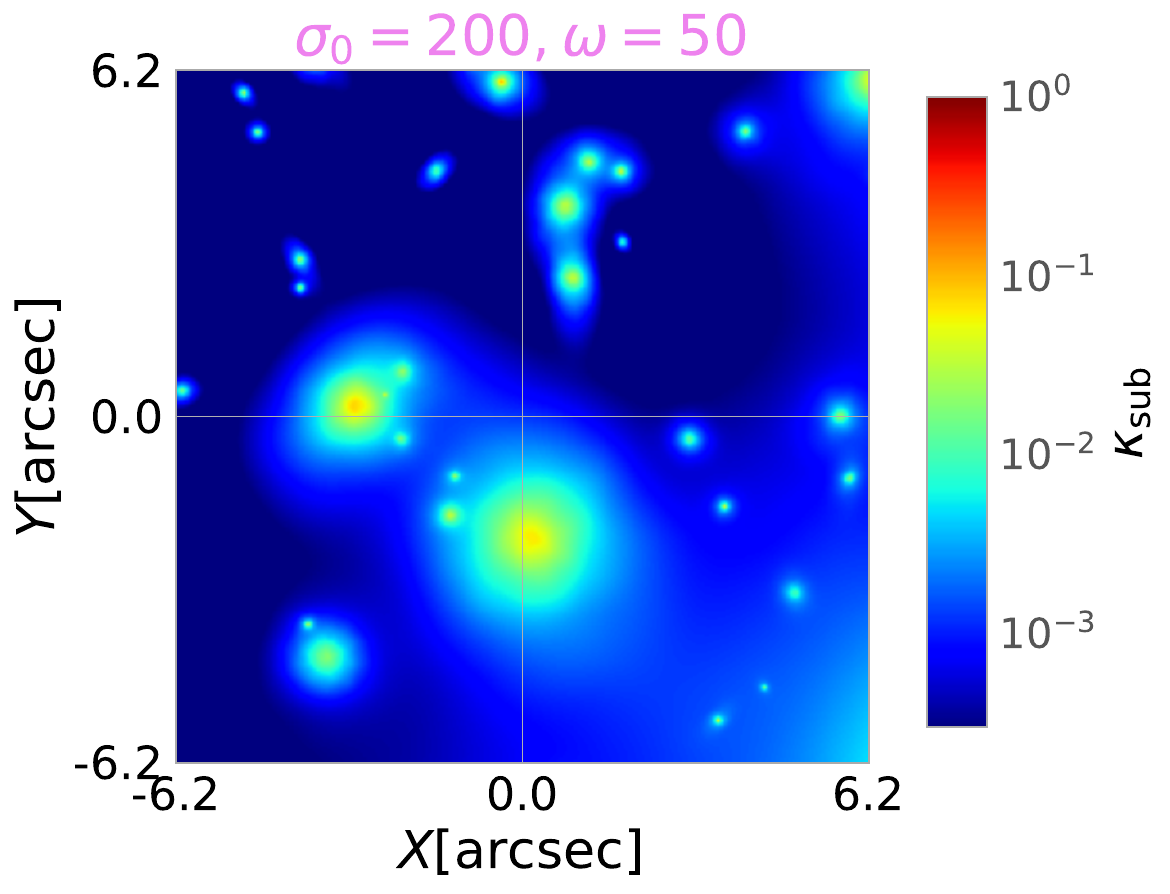}
        \caption{}
        \label{fig:dens-d}
    \end{subfigure}
    \caption{Lensing convergence maps within an aperture of 6.25 arcsec from the host center. Subhalos for both the 6 $\rm cm^2/g$ and $\{\sigma_0=200, \omega=200\}$ models (red and purple in Fig. \ref{fig:sig-v}) experience evaporation from the host halo, thus many of the subhalos are less massive than CDM, having lower $\kappa_{\rm sub}$ (in fact, these two models are the only ones that have visible difference in subhalo mass function than CDM, see the upper shoulder subplot in Fig. \ref{fig:scatter}). On the other hand, the $\{\sigma_0=200, \omega=200\}$ model has a significant fraction of core-collapsed subhalos, leading to higher convergence in some subhalos' center than CDM. The $\{\sigma_0=200, \omega=50\}$ model (violet) also has significant core-collapse, but only in relatively low-mass subhalos, hence its convergence map is similar to that of CDM. }% Another version of this figure, but with a smaller projection radius of 6.25 arcsec that is closer to a typical group-sized lens, is shown in Fig. \ref{fig:kappa-6arcsec}.}
    \label{fig:kappa}
\end{figure*}

\begin{figure*}
    \centering
    \begin{subfigure}[t]{0.32\textwidth}
        \centering
        \includegraphics[width=\textwidth, clip,trim=0.3cm 0cm 0.3cm 0cm]{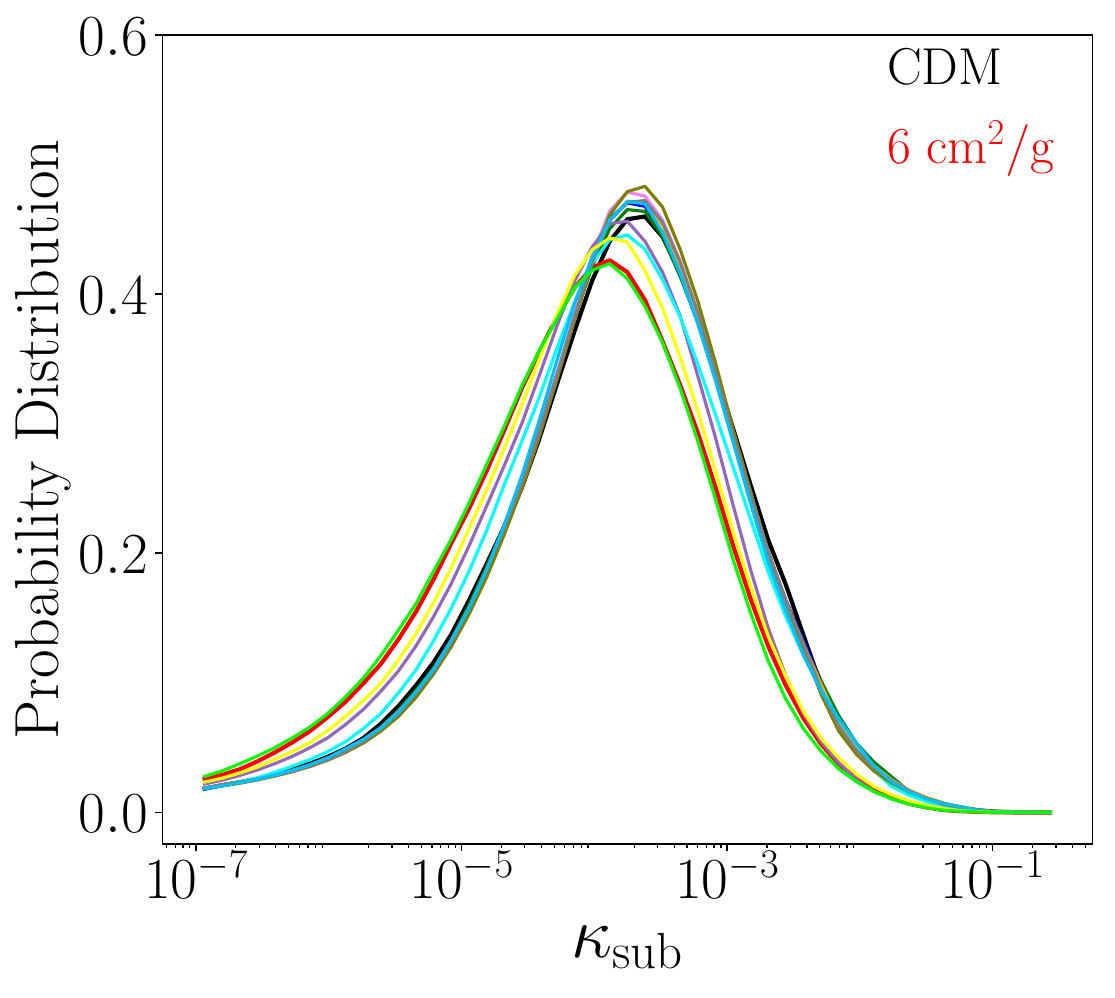}
        \caption{}
        \label{fig:1pta}
    \end{subfigure}    
    ~
    \begin{subfigure}[t]{0.32\textwidth}
        \centering
        \includegraphics[width=\textwidth, clip,trim=0.3cm 0cm 0.3cm 0cm]{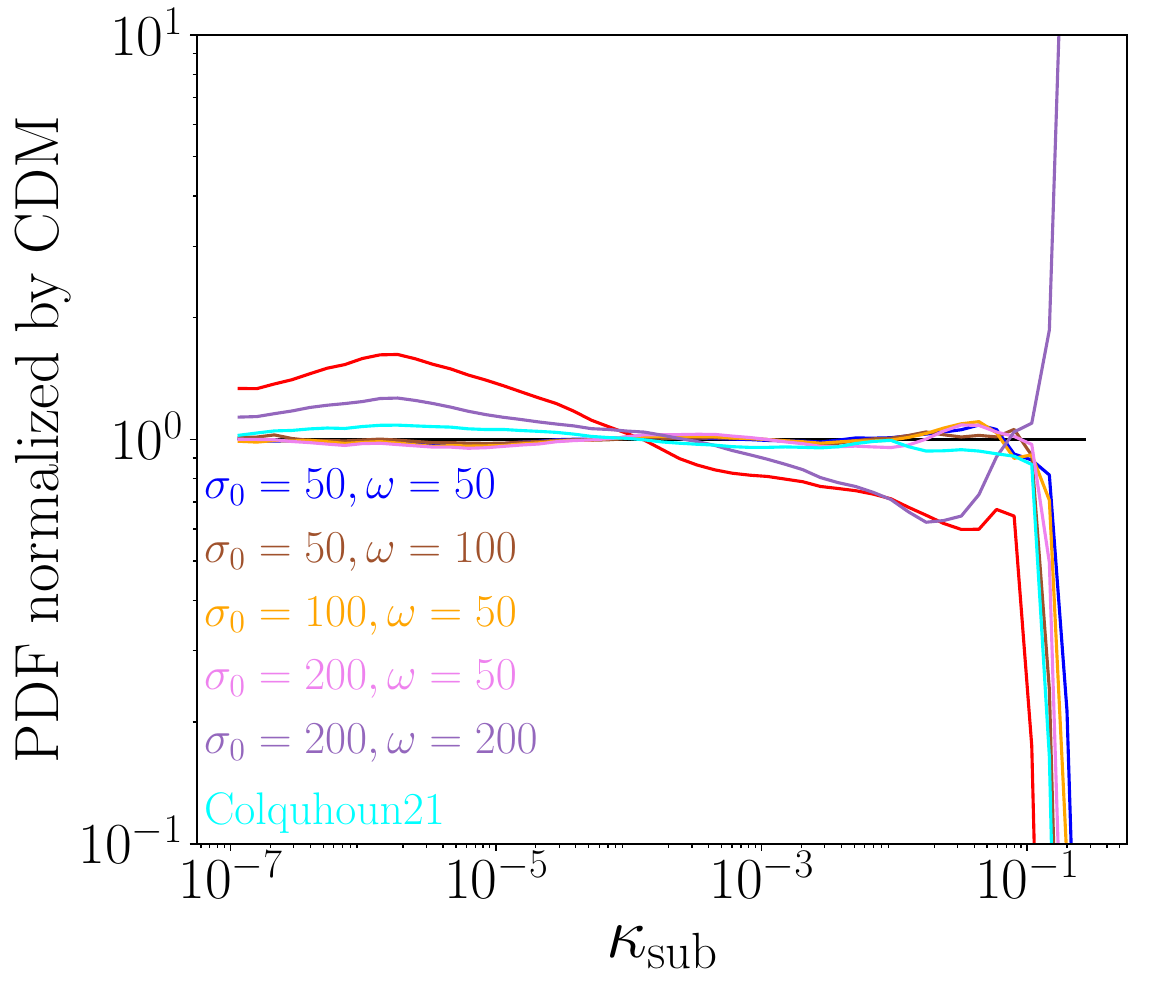}
        \caption{}
        \label{fig:1ptb}
    \end{subfigure}
    ~
    \begin{subfigure}[t]{0.32\textwidth}
        \centering
        \includegraphics[width=\textwidth, clip,trim=0.3cm 0cm 0.3cm 0cm]{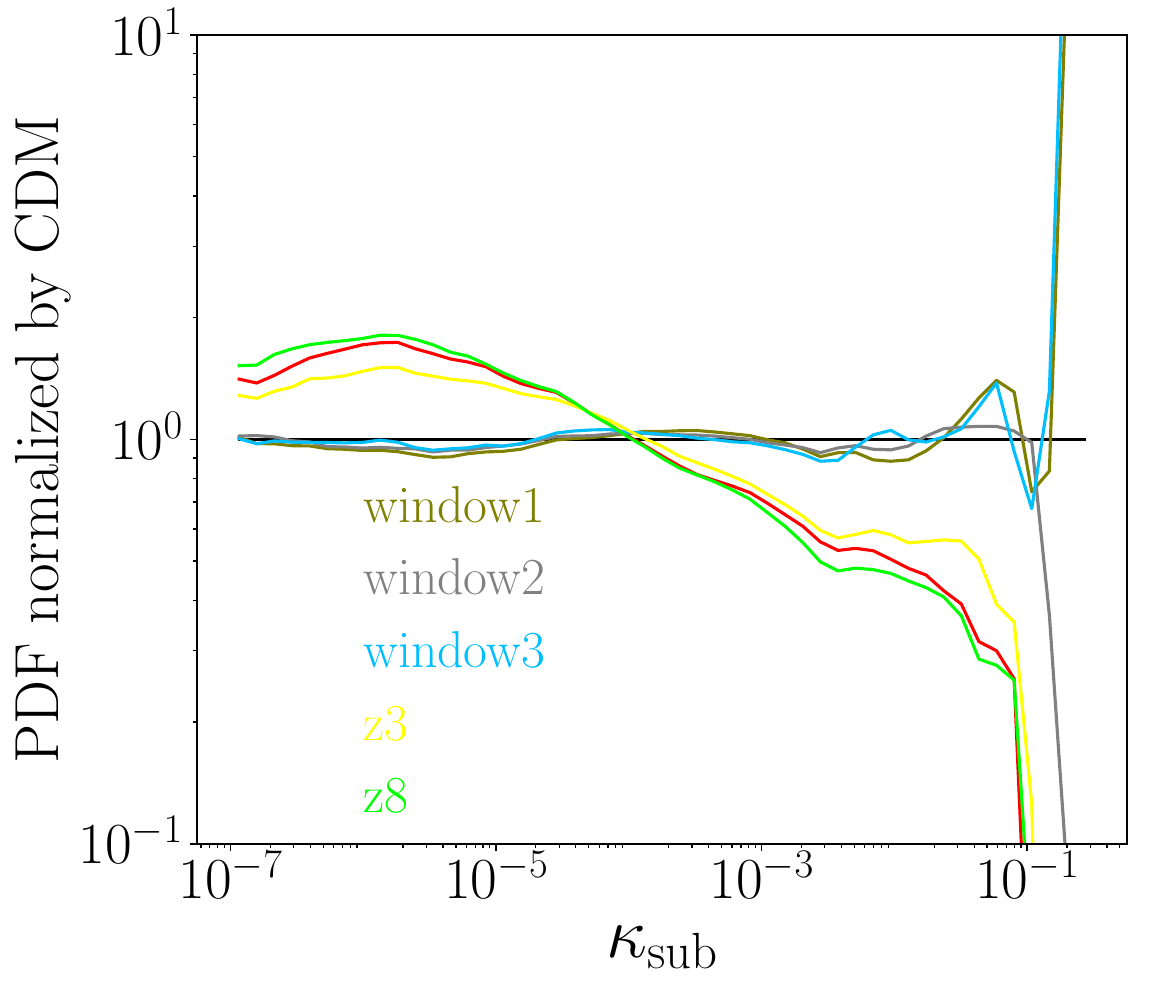}
        \caption{}
        \label{fig:1ptc}
    \end{subfigure}
    ~
    \begin{subfigure}[t]{0.32\textwidth}
        \centering
        \includegraphics[width=\textwidth, clip,trim=0.3cm 0cm 0.3cm 0cm]{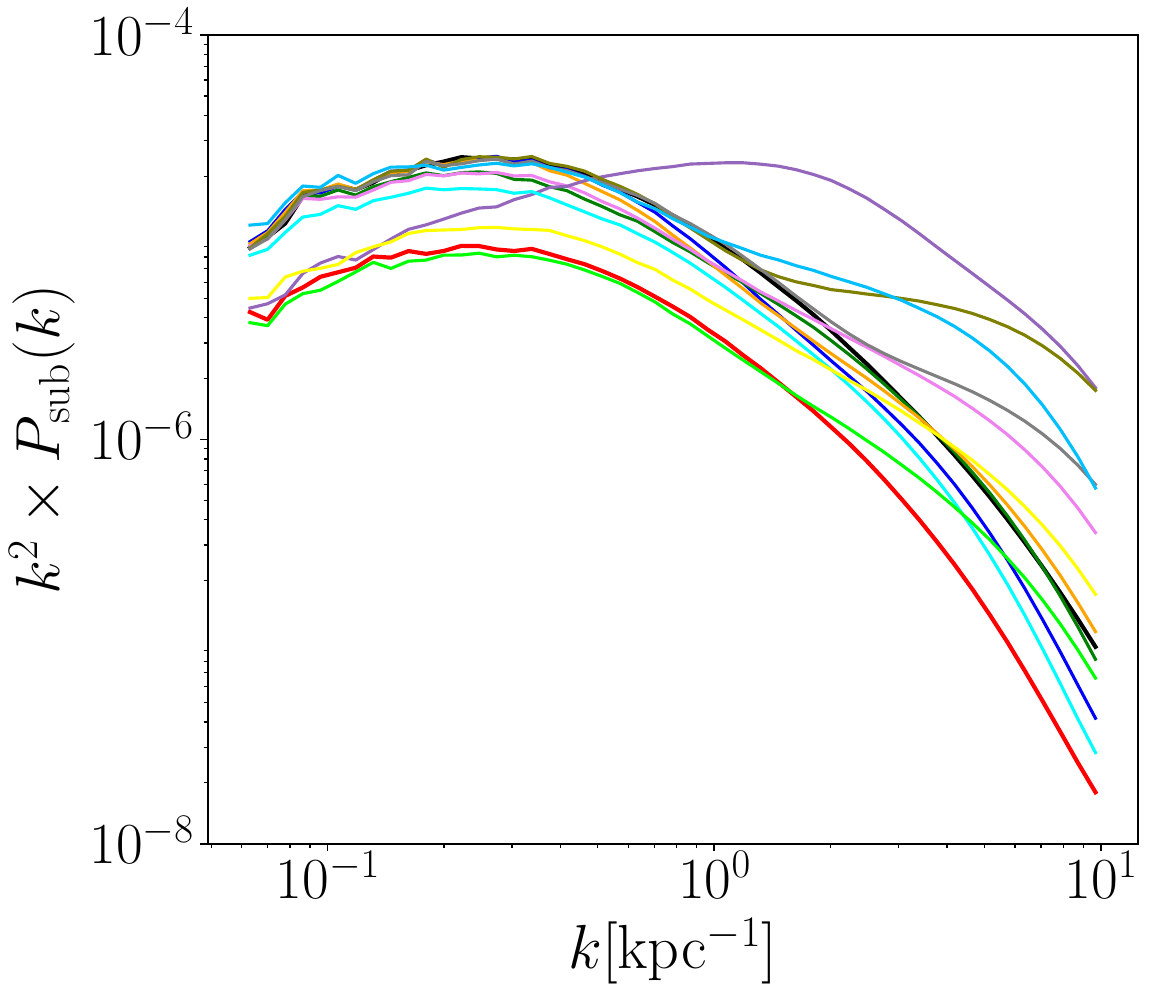}
        \caption{}
        \label{fig:pka}
    \end{subfigure}
     ~
    \begin{subfigure}[t]{0.32\textwidth}
        \centering
        \includegraphics[width=\textwidth]{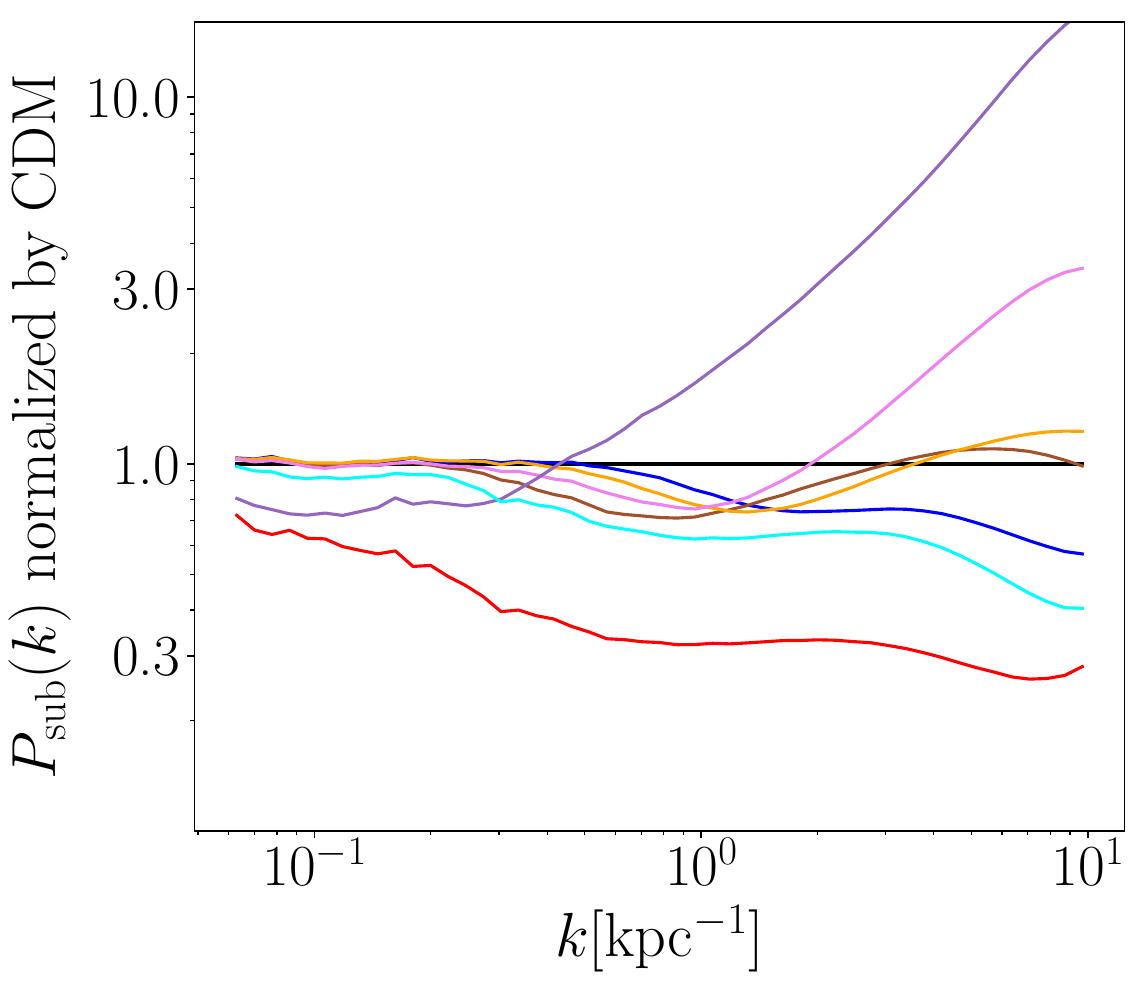}
        \caption{}
        \label{fig:pkb}
    \end{subfigure}
    ~
    \begin{subfigure}[t]{0.32\textwidth}
        \centering
        \includegraphics[width=\textwidth]{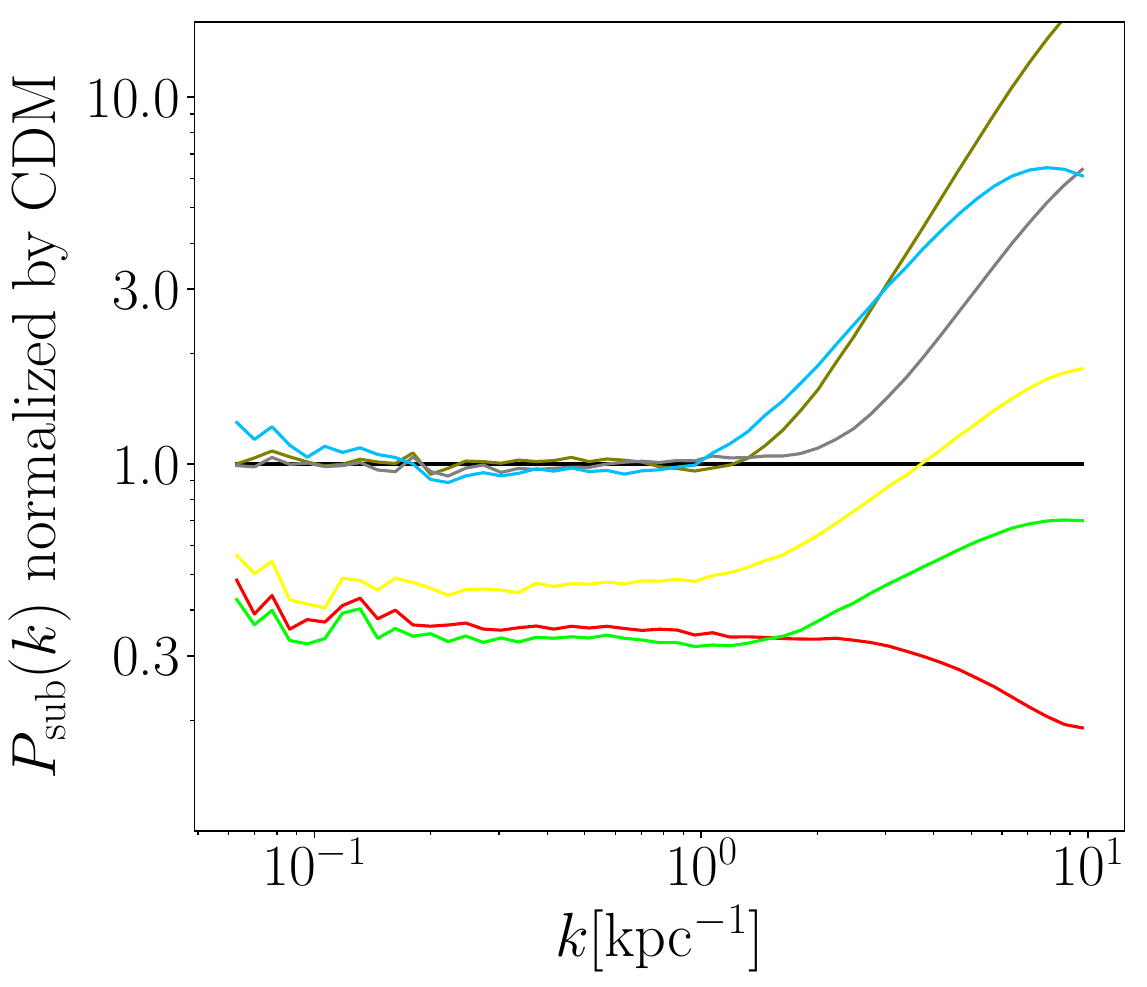}
        \caption{}
        \label{fig:pkc}
    \end{subfigure}
    ~
    \caption{Convergence probability distribution (top row) and the lensing power spectrum (bottom row) of subhalos. Panel a) shows the probability distribution function (PDF) of subhalo convergence $\kappa_{\rm sub}$ (an alternate version showing $\kappa_{\rm sub}^2\times$PDF is shown in Fig. \ref{fig:kappa2-1pt}, which highlights the high $\kappa_{\rm sub}$ bins that are most relevant for substructure lensing). Panels b) and c)  show the PDF of $\kappa_{\rm sub}$ of different SIDM models normalized by the CDM results. Panel d) shows the (unitless) subhalo lensing spectrum $k^2 P_{\rm sub}(k)$. Panels e) and f) show $P_{\rm sub}(k)$ for the SIDM models normalized by CDM results, and grouped in the same way as in panels b) and c). Note that since we run four realizations of simulations for the $\{\sigma_0, \omega\}$, constant SIDM and Colquhoun21 models, and only one realization for the window- and z-type models, b) and c) include the averaged results of four realizations, while the rest are from one realization.} 
    \label{fig:glmr}
\end{figure*}

In this section, we present subhalo statistics that are relevant for substructure lensing, including the projected mass density one-point function and the lensing power spectrum. 

For our calculations, we use the lensing tool \texttt{Glamer} \citep{glmr1, glmr2} to smooth the masses of simulation particles into a pixelized map and calculate the lensing convergence $\kappa$.  Here, $\kappa$ is the 2-d projected mass density $\Sigma$ normalized by $\Sigma_{\rm crit}\equiv c^2D_s/4\pi GD_{ds}D_d$, with $D_s$, $D_d$ and $D_{ds}$ being the (angular) distance  of the source to observer, lens to observer, and lens to source respectively. The smoothing length in \texttt{Glamer} is chosen to enclose 32 of a particle's nearest neighbors, the same as for the SIDM smoothing procedure in \texttt{Arepo}. Without loss of generality, we choose the redshift of the main lens to be $z=0.5$, which corresponds to the $t=9$ Gyr ending time of our simulation, and the redshift of the source to be $z=4$. We use the resolution of 2048$\times$2048 pixels in the map for calculation of the projected mass density, which is well-resolved.  
We generate 64 projections for each simulation to reduce noise (see \cite{despali19} and \cite{brennan19} for discussion on projection variance). For each projection, we only consider the simulation particles that lie within a cylindrical aperture of 25 arcsec ($\sim 150\ \rm kpc$) from the center of the host (main lens), since the substructure lensing is most sensitive to subhalos located near the host center. 
A typical Einstein radius $\theta_E$ of a group-sized system has the range $1 \sim$ 8 arcsec when a central galaxy is included \citep{verdugo14, despali19}, and for cluster-sized systems it increases to $10\sim 30$ arcsec  \citep{gonzalez20, meneghetti23}. We use a relatively big aperture for lensing analysis in this section mainly for more robust statistics, since a smaller aperture yields to larger noise.  We have also tested two smaller apertures of 6.25 arcsec and 2.5 arcsec, shown in Sec. \ref{sec:res-real}.
We only include particles that are bound to subhalos at the end of the simulation (including the most massive subhalos with $M_{\rm infall}>10^{10} M_\odot$ that are simulated with a live host), as determined by the halo finder \texttt{AHF}. This is in correspondence with typical observation pipelines of substructure lensing \citep{vegetti10, dg20}, of which one of the key steps is to reconstruct and remove a smooth mass distribution that represents the main lens. 
Because our simulations do not contain line-of-sight structures, we cannot include any in our convergence maps, even though they are expected to contribute to the lensing signal \citep{ddxu12, despali18, gilman19, birendra23, birendra23a, lazar23}.

An example of the convergence (normalized projected density) map of several DM models is shown in Fig. \ref{fig:kappa}.  Compared to the CDM case, both the 6 cm$^2$/g model and the $\{\sigma_0=200, \omega=200\}$ model have subhalos experiencing strong evaporation effects, hence their subhalos appear to be smaller (less massive). The $\{\sigma_0=200, \omega=200\}$ model also has denser centers in some subhalos because of core-collapse. In contrast, the $\{\sigma_0=200, \omega=50\}$ model that also has a significant fraction of core-collapsed subhalos does not appear to have the same feature of denser centers in this 2D map, because core-collapse only happens in relatively low-mass subhalos for this model (see more discussion later in this section).

We show the probability distribution function (PDF) of the substructure lensing convergence $\kappa_{\rm sub}$ (the $\kappa$ evaluated with dark matter particles that are bound to any subhalos) in the top row of Fig. \ref{fig:glmr}. The PDFs of $\kappa_{\rm sub}$ of all SIDM models (also known as the 1-point function), calculated with \texttt{scipy.stats.gaussian\_kde}, are shown in Fig. \ref{fig:1pta}.  We plot the ratio of these PDFs for $\{\sigma, \omega\}$ and Colquhoun21 models relative to CDM in Fig. \ref{fig:1ptb}, and the ratio of the phenomenological models from Sec. \ref{sec:halo-artificial} relative to CDM in Fig. \ref{fig:1ptc}.  For SIDM models with significant host-subhalo evaporation, such as the constant cross section of 6 $\rm cm^2/g$, and the $\{\sigma_0=200, \omega=200\}$, z3 and z8 models, the abundance of low-$\kappa_{\rm sub}$ pixels is boosted and the abundance of high-$\kappa_{\rm sub}$ pixels is decreased (except when core-collapse comes into play, e.g. the $\{\sigma_0=200, \omega=200\}$ in purple color).  
This is because the strong evaporation removes DM from subhalos globally,  shifting the distribution toward low column-density.

On the other hand, evidence for core-collapse is difficult to discern from the $\kappa_{\rm sub}$ 1-point function.  We find that for the models $\{\sigma_0=200, \omega=50\}$ (violet) and window2 (grey) that have a significant fraction ($>10\%$ of total) of core-collapsed subhalos (see Table \ref{table:ccp}), there is no visible sign of core-collapse at the high-$\kappa_{\rm sub}$ end. 
Their $\kappa_{\rm sub}$ PDFs are overall similar to the CDM case, except for a sharp drop at the highest $\kappa_{\rm sub}$ bins (see Fig. \ref{fig:1ptb} and Fig. \ref{fig:1ptc}).
By contrast, for the other three models that also have a large number of core-collapsed subhalos, $\{\sigma_0=200, \omega=200\}$ (purple), window1 (olive) and window3 (light blue), we find a sharp increase in the PDF at high $\kappa_{\rm sub}$. The difference between the former models and the latter is that the former only have core-collapse in relatively low-mass subhalos ($\leq 10^9 M_\odot$), while the latter see core-collapse in massive ones. Since $\kappa$ is a projected density, in evaluating the line-of-sight integral, not only the 3D density of a subhalo's matters, but also its size (i.e. the length that the column density is integrated along). Thus the highest $\kappa_{\rm sub}$ bins are dominated by the most massive subhalos. 
Therefore, we can conclude that the $\kappa$ distribution is sensitive to core-collapse in massive subhalos, but not in low-mass ones.

The bottom panels of Fig. \ref{fig:glmr} show the substructure lensing power spectrum, which is the 2-point correlation function of $\kappa_{\rm sub}$ in $k$-space. Here we find that the signature of subhalo core-collapse is more visible than in the 1-point function of $\kappa_{\rm sub}$, and even better, has a dependence on the wave number $k$.  In principle, this may allow us to infer at what size or mass scale the subhalos core-collapse. For example, the three models that lead to core-collapse in low-mass subhalos (see Table \ref{table:ccp}), $\{\sigma_0=200, \omega=50\}$ (violet), $\{\sigma_0 = 100, \omega = 50\}$ (orange) and window2 (grey), show $P_{\rm sub}(k)$ beginning to exceed CDM at $k\sim 2-3\ \rm kpc^{-1}$, increasing until $k\sim 10\ \rm kpc^{-1}$. The window3 model (light blue), which only produces core-collapse in massive subhalos with $M_{\rm infall}\geq 10^9 M_\odot$, starts to have stronger clustering than CDM at $k\sim 1\ \rm kpc^{-1}$, reaching a turning point at $k\sim 7\ \rm kpc^{-1}$, before decreasing at smaller scales. The window1 (olive) and $\{\sigma_0=200, \omega=200\}$ (purple) models that have core-collapse in both low-mass and massive subhalos show $P_{\rm sub}(k)$ exceeding CDM monotonically from relatively large scales $k\lesssim 1\ \rm kpc^{-1}$ to smaller scales $k\sim 10\ \rm kpc^{-1}$. The other SIDM models with subhalos mostly still in the core-formation stage, such as $\{\sigma_0=50, \omega=50\}$ (blue) and Colquhoun21 (cyan), correspondingly have lower $P_{\rm sub}(k)$ at small scales of $k\gtrsim1\ \rm kpc^{-1}$. If evaporation is significant, there is suppression of $P_{\rm sub}(k)$ at large scales, which we find in the constant cross section case of 6 $\rm cm^2/g$ (red) and the z8 model (lime). The z3 model (yellow), although suffering from host-subhalo evaporation, manages to produce a few core-collapsing subhalos (see Table \ref{table:ccp}).  Thus, its $P_{\rm sub}(k)$ is lower than CDM at large scales but grows higher at the smallest scales, corresponding to the centers of the low-mass subhalos.  

Our results suggest that the 2-point correlation of $\kappa_{\rm sub}$ is a more sensitive probe of core-collapse in subhalos than the 1-point function, especially for low-mass ones. As shown in \cite{DiazRivero:2017xkd}, this 2-point function directly probes the average density profile of subhalos at different scales, allowing it to directly capture core formation and core-collapse within the ensemble of subhalos populating a lens galaxy. Once line-of-sight halos between the lens source and the observer are included, 2-point function anisotropies may provide additional information on the velocity dependence of the SIDM cross section \citep{birendra23a}. Although more advanced techniques are needed for constraining SIDM models using substructure lensing, such as forward modelling \citep{dg18, dg20} or perturbative analyses \citep{Hezaveh:2014aoa,Cyr-Racine:2018htu}, the results presented here shed some insight on how strong lensing data analysis could be carried out measure the SIDM cross section.

\subsection{Core-collapse fraction of subhalos vs in isolation}\label{sec:iso-vs-sub}

\begin{figure}
    \begin{subfigure}[t]{\columnwidth}
        \centering
        \includegraphics[width=\textwidth, clip,trim=0.3cm 0cm 0.3cm 0cm]{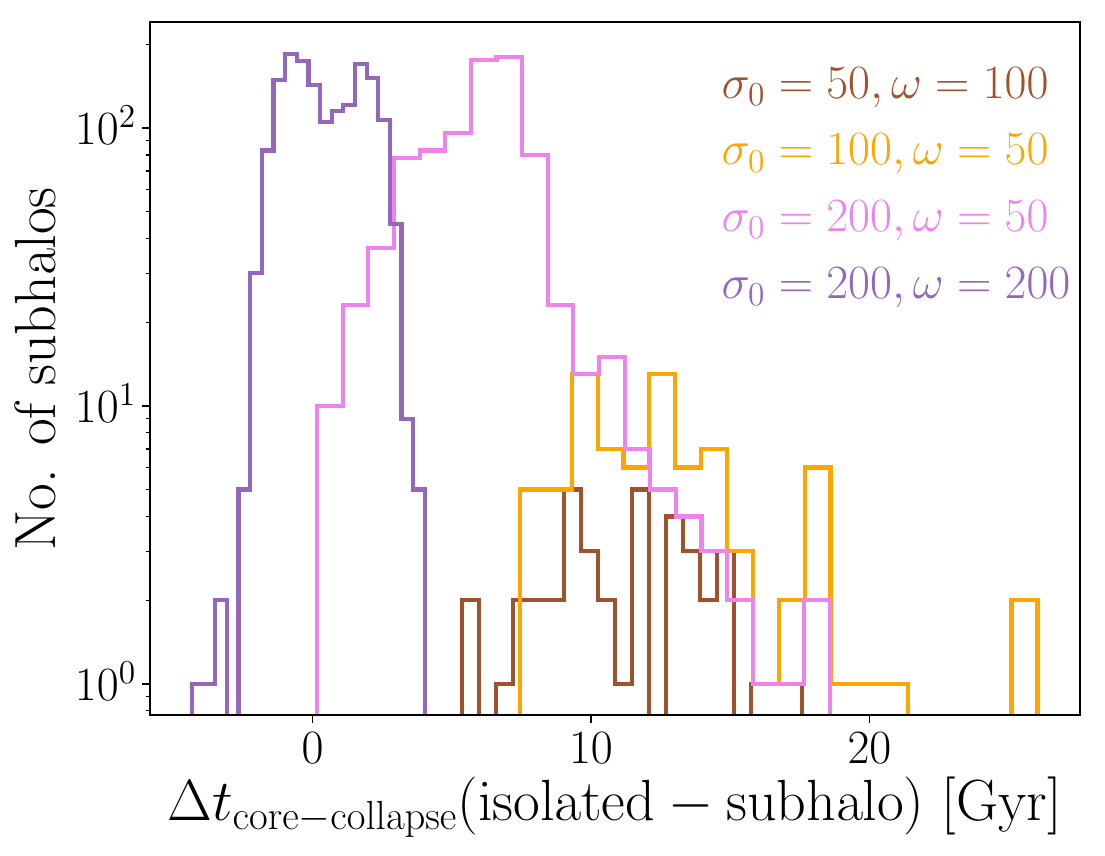}
        \caption{}
        \label{fig:tdiff}
    \end{subfigure}
    ~
    \begin{subfigure}[t]{\columnwidth}
        \centering
        \includegraphics[width=\textwidth, clip,trim=0.3cm 0cm 0.3cm 0cm]{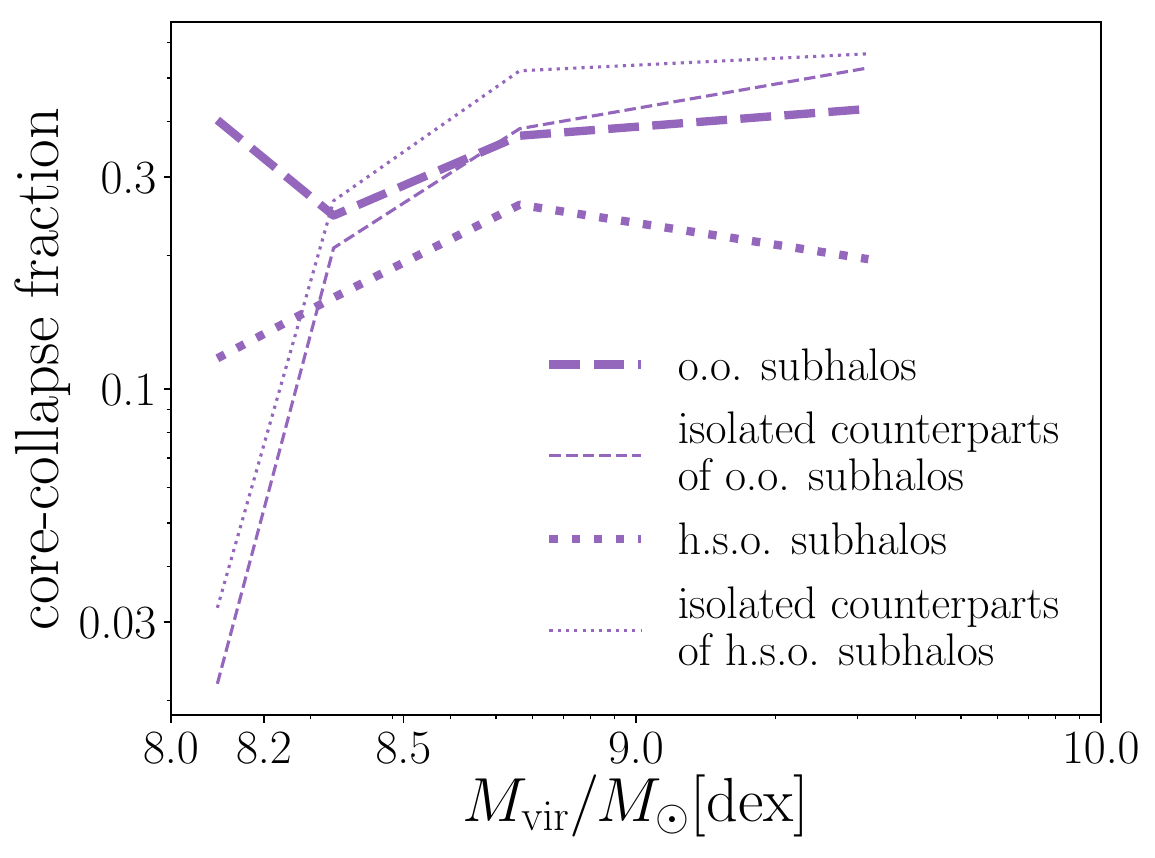}
        \caption{}
        \label{fig:ccp-fraction}
    \end{subfigure}
    \caption{Panel a): The distribution of the core-collapse time of subhalos and their counterparts in isolation $\Delta t_{\rm core-collapse}$(isolated-subhalo), for SIDM models that have core-collapsed subhalos by the end of the simulation $t=9$ Gyr.  Positive values indicate that subhalo core-collapse is accelerated by the host. An alternate version of this plot, with $\Delta t$ rescaled as an acceleration or deceleration factor by the host field, is attached in Appendix \ref{appdx:misc}. Panel b) shows the core-collapse fraction (at $t=9$ Gyr) of subhalos for the $\{\sigma_0=200, \omega=200\}$ model, as a function of subhalos' infall-mass and orbits. `o.o.' and `h.s.o.' are abbreviations for ordinary orbits and heavily-stripping orbits.}
    \label{fig:iso-sub}
\end{figure}

Past work has argued that core-collapse in subhalos can be accelerated by tidal stripping  \citep{nishikawa20, sameie20, correa21} or decelerated \citep{zzc22} by evaporation. In this section, we assess these effects quantitatively by comparing subhalos in simulation to halos with identical initial conditions but in isolation. The evolution of these isolated halos is analytically calculated by the mapping method of the gravothermal model, which utilises the self-similarity of SIDM halos to map between different halos without solving the gravothermal PDEs every time \cite[see also Sec. \ref{sec:param-space}]{sq22}. The different behavior of field halos and subhalos may be used to distinguish between SIDM models.

We show the distribution of the time difference in core-collapse between isolated halos and subhalos $\Delta t_{\rm core-collapse}$ in Fig. \ref{fig:tdiff}. Note that we only include the subhalos that core-collapse before the end of simulation $t=9\ {\rm Gyr}$, because we do not run the simulation beyond that time. 
Thus, Fig. \ref{fig:tdiff} only displays the SIDM models where many subhalos core-collapse in our simulations. We find that the velocity-dependent models that have low evaporation ($\sigma/m \lesssim \mathcal{O}(1) \rm cm^2/g$ for relative velocity $100-1000\ \rm km/s$), $\{\sigma_0=50, \omega=100\}$ (brown), $\{\sigma_0=100, \omega=50\}$ (orange), and $\{\sigma_0=200, \omega=50\}$ (violet), show core collapse accelerated in subhalos relative to the field halos. 

However, for the $\{\sigma_0=200, \omega=200\}$ model (purple) that has large cross sections for both the internal relative velocities of subhalos ($\sigma/m =200\ \rm cm^2/g$ for $< 100\ \rm km/s$) and the subhalo orbital velocities ($\sigma/m \sim \mathcal{O}(10)\ \rm cm^2/g$ for $100-1000\ \rm km/s$), about half the subhalos experience decelerated core collapse. 
Moreover, the relative core collapse time $\Delta t_{\rm core-collapse}$ in the range of -5 to 5 Gyr. The magnitude of $\Delta t_{\rm core-collapse}$ is small compared to other SIDM models, because this purple model has the largest effective cross section for scatterings inside all subhalos.  Of all the SIDM models we explore, it yields the shortest overall core-collapse times (10-20 Gyr for all field halos).

To further understand the role of subhalo mass and orbit in determining the subhalo core-collapse timescale, in Fig. \ref{fig:ccp-fraction} we break down the infall-mass and orbit dependence of the subhalo core-collapse fraction of the purple model $\{\sigma_0=200, \omega=200\}$, at the end of the simulations ($t=9$ Gyr).
Other models are not shown in this figure because their core-collapse fraction for field halos is small. 
We find that field halos (thin dashed and thin dotted) generally have an increasing core-collapse fraction with increasing infall mass.  The isolated counterparts of h.s.o. subhalos (thin dotted) have a slightly higher collapse fraction than the o.o. counterparts (thin dashed) because the h.s.o. subhalos tend to have earlier formation times. 
By contrast, o.o subhalos (thick dashed) have a higher collapse fraction than h.s.o. subhalos (thick dotted).  Despite the overall earlier formation time and stronger tidal stripping experienced by the h.s.o. subhalos relative to the o.o. ones, the h.s.o. subhalos have a noticeably lower core-collapse fraction because of the strong evaporation and tidal heating near the orbital pericenter. 

When compared to their field counterparts, we find a complicated dependence of the core-collapse time on both subhalo (infall) mass and orbits. For the least massive subhalos with infall-mass in range $[10^8, 10^{8.2})M_\odot$, o.o. and h.s.o. subhalos experience accelerated core-collapse. For the $M_{\rm infall}=[10^{8.2}, 10^{8.5})M_\odot$ subhalos, o.o. subhalos experience only mildly accelerated core-collapse times, and h.s.o. subhalos experience delayed core-collapse. For $M_{\rm infall}=[10^{8.5}, 10^{9})M_\odot$ subhalos, o.o. core-collapse fractions are similar to those in the field, but the h.s.o. collapse fraction is lower than in the field. For the most massive group $M_{\rm infall}=[10^{9}, 10^{10})M_\odot$, both o.o. and h.s.o. subhalos have lower collapse fractions than their isolated counterparts. 
Fig. \ref{fig:ccp-fraction} shows that the subhalo core-collapse fraction does not scale monotonically with the strength of the host effects (e.g., the strength of the tidal field or evaporation).
This complicated interaction comes from the competition of at least four physical processes: the heat transfer within the subhalo, tidal stripping, tidal heating, and evaporation \citep{zzc22}, each reacting differently to the change in subhalo mass or orbits. 

In summary, we show that for SIDM models with no or low evaporation (low effective cross section for subhalo orbiting velocities), the presence of a host results in a net acceleration of core-collapse.  However, for SIDM models where evaporation becomes non-negligible, the evolution of subhalos is complicated, and the ensemble properties of the subhalos can change dramatically.
Intriguingly, this complexity implies a high level of diversity in subhalo density profiles and evolution.  In the future, it will be important to determine whether or with what SIDM models the diversity of subhalo density profiles we find in simulations can match the observed diversity of galaxy rotation curves.

\subsection{Realization variance}\label{sec:res-real}

\begin{figure*}
    \centering
    \begin{subfigure}[t]{0.48\textwidth}
        \centering
        \includegraphics[width=\textwidth]{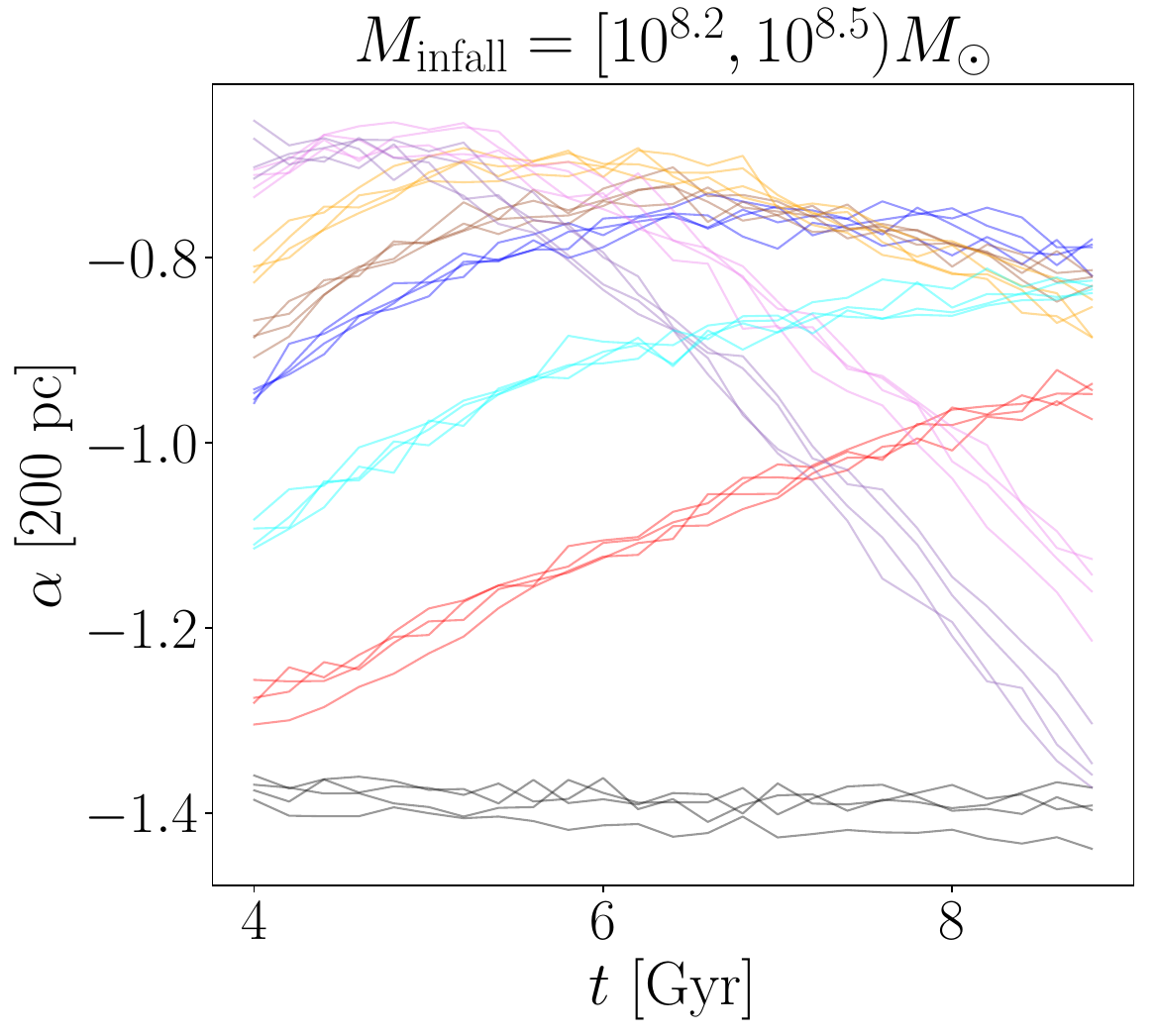}
        \caption{}
        \label{fig:var1}
    \end{subfigure}
    ~
    \begin{subfigure}[t]{0.48\textwidth}
        \centering
        \includegraphics[width=\textwidth]{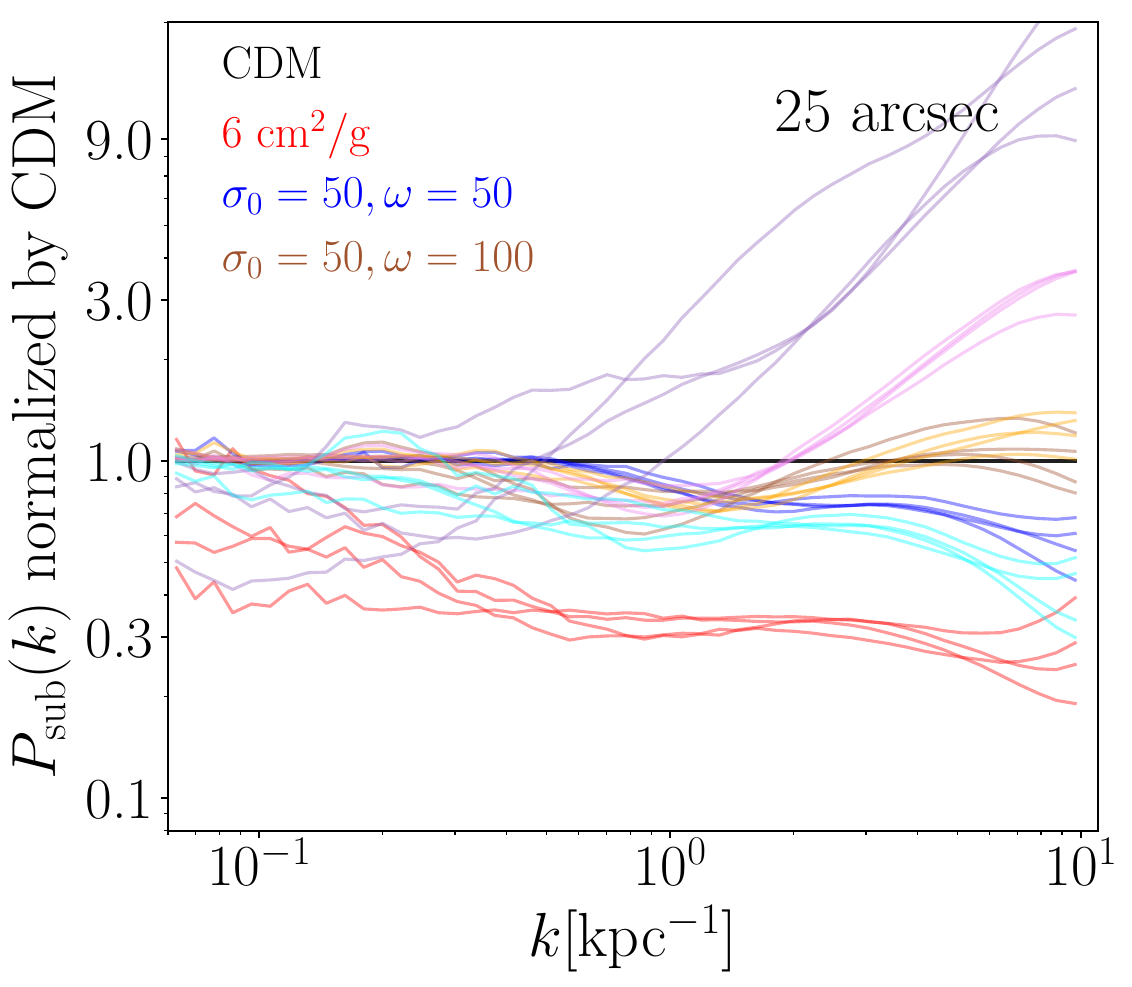}
        \caption{}
        \label{fig:var2}
    \end{subfigure}
    ~
    \begin{subfigure}[t]{0.48\textwidth}
        \centering
        \includegraphics[width=\textwidth]{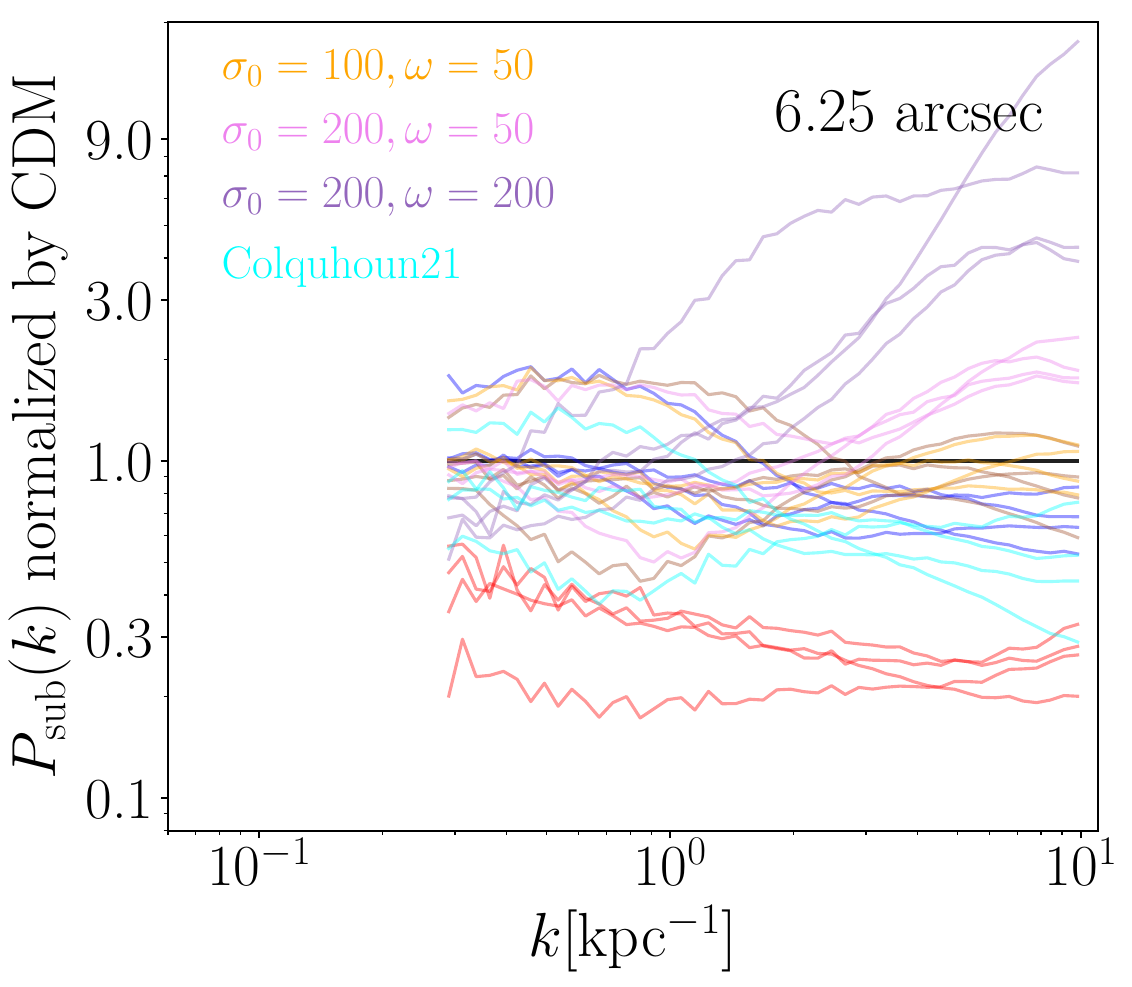}
        \caption{}
        \label{fig:var3}
    \end{subfigure}
    ~
    \begin{subfigure}[t]{0.48\textwidth}
    \centering
    \includegraphics[width=\textwidth]{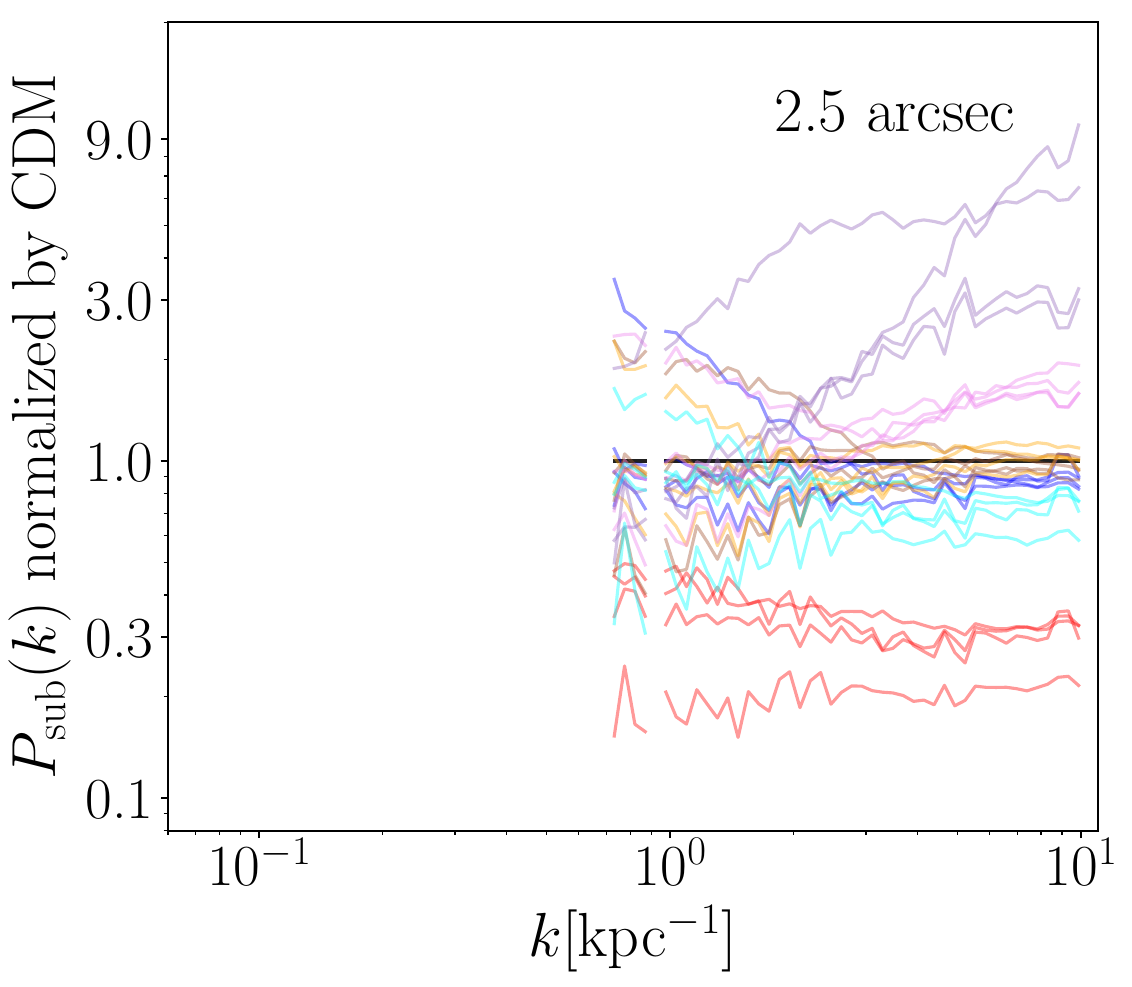}
    \caption{}
    \label{fig:var4}
    \end{subfigure}
    \caption{Realization variance. Panel a) shows the time evolution of subhalos' inner slope $\alpha$, with different realizations of merger trees distinguished. Panel b) shows the substructure lensing power spectrum $P_{\rm sub}(k)$ (normalized by CDM) with a projection aperture of 25 arcsec ($\sim$150 kpc) at the end of the simulation. Panel c) and d) are similar to panel b), but with a smaller projection aperture of 6.25 arcsec ($\sim$37.5 kpc) and 2.5 arcsec ($\sim$15 kpc) respectively.}
    \label{fig:real-var}
\end{figure*}

When connecting simulations to observations, a key ingredient is to estimate the realization variance from theoretical model predictions. Previous work such as \cite{brennan19} find that in the specific case of the substructure lensing power spectrum, simply using multiple projections of a single population of subhalos leads to an underestimate of the variance.  % compared to the actual variance by multiple populations.
In this section, we present the variance among our different realizations of simulated subhalo systems, to shed light on the amount of data/systems we may need in simulations and observations in the future for robust constraints on SIDM. 

We show the averaged inner slope $\alpha$ of the infall-mass group $[10^{8.2}, 10^{8.5})M_\odot$ for each realization and SIDM model in Fig. \ref{fig:var1}. We find the overall variation of $\alpha$ between realizations is minimal. However, the variance among realizations in the substructure lensing power spectrum $P_{\rm sub}(k)$, as shown in Fig. \ref{fig:var2}, is more significant, even if we have averaged the power spectrum over 64 projections to reduce noise. This contrast is because the projection aperture within which we measure the $\kappa_{\rm sub}$ map is only 25 arcsec ($\sim$ 150 kpc), which is about 1/6 of the host virial radius, while the slope $\alpha$ is averaged over all subhalos (of a certain infall mass group) within the host. When we use an even smaller projection aperture of 6.25 arcsec ($\sim$ 37.5 kpc) in Fig. \ref{fig:var3} or 2.5 arcsec ($\sim$ 15 kpc) in Fig. \ref{fig:var4}, the enclosed subhalos become fewer in number, and we find a greater variance among realizations, as expected (see also \citealt{despali19}). The power spectrum for the smaller aperture also yields to more random noise, making it more challenging to estimate a smooth power spectrum at large and intermediate scales. However, the signature of frequent subhalo core-collapse, as indicated by a boost in the power spectrum at smallest scales, is robust.

\section{Summary and discussion}\label{sec:summary}

In this work, we presented a new hierarchical simulation framework for the efficient study of low-mass subhalos in SIDM cosmologies.  Building on the hybrid method described in \cite{zzc22} that tracks a single subhalo with N-body particles and semi-analytically calculates interactions with the host, we simulate realistic populations of subhalos.  To speed up simulations while achieving high resolution, we divide subhalos into different infall-mass groups/bins of simulations with different mass-resolution (hence, the `hierarchical' designation).  With this new framework, we are able to scan through multiple SIDM models and realizations, and conduct post-simulation analysis in the space of observables, in a reasonable amount of computational time.   

Compared to full cosmological simulations, the major advantage of this hierarchical simulation framework is that its computational time grows only linearly with the subhalo mass function. 
This allows us to study the evolution of even very low-mass SIDM subhalos in large hosts
with the high resolution that is prohibitively costly for cosmological simulations.  
The lowest simulation particle mass (thus highest mass resolution) in our hierarchical simulation is about $10^3M_\odot$ in the work we presented here (although this can go to much smaller masses if tracing much smaller subhalos is desired).  However, the total number of particles in our simulations is two orders of magnitude smaller than it would be for a cosmological simulation for the same system, corresponding to a computational speed boost of three to four orders of magnitude. 

We use this framework to explore subhalo evolution for several phenomenological SIDM models, chosen to span a range of cosmologically interesting scenarios.  First, we sample points from the parameter space of a phenomenological $\sigma_0-\omega$ SIDM model, near the predicted boundary within which core-collapse happens in isolated halos of the same range of masses and evolution time (Fig. \ref{fig:param-space}). This class of model, featuring a $\sigma \propto v^{-4}$ scaling at large relative particle velocities, is the analytical solution to particle interactions in Yukawa-type potential in the Born approximation regime \citep{feng09, tulin13, slone21, sq22, ymzhong23}.    For each of these $\sigma_0-\omega$ model and another SIDM model with resonant features (Colquhoun21, which explores beyond the Born regime of Yukawa-type interactions; \cite{col21}), we generate simulations for four realizations of host-subhalo merger trees that are typical for substructure lens systems, with a host of $10^{13}M_\odot$ and subhalos down to $10^8M_\odot$ at infall.  We run until $t = 9\ \rm Gyr$ ($z\sim0.5$), a typical lens redshift.  We summarize the evolution of subhalos in these SIDM simulations in the form of core-collapsed fraction in Table \ref{table:ccp}, and analyze the evolution of observables: the averaged subhalos' inner mass (Sec. \ref{sec:halo1} and Fig. \ref{fig:centralm}) and inner slope  (Sec. \ref{sec:halo2} and Fig. \ref{fig:slope}). We find that when a significant fraction of subhalos are core-collapsed in an SIDM model, the averaged inner mass $M_{\rm inner}$ grows higher than the CDM counterpart, while the averaged inner slope $\alpha$ becomes more negative. For the velocity-dependent SIDM models we select, the subhalo mass functions hardly show any visible difference from CDM, but the probability distribution of subhalos' inner mass $M_{\rm inner}$ depends strongly on the DM model (Fig. \ref{fig:scatter}).  

Excitingly, we find a significantly growing diversity of subhalos' inner mass distribution as core-collapse starts to trigger in subhalos, potentially providing solutions to the observed diversity in dwarf galaxy internal velocity profiles \citep{kdn14, oman15, read19, relatores19, santos20, hayashi20, li20}. This diversity is linked to subhalos' orbital histories and the strength of SIDM-driven evaporation by the host.  By categorizing subhalos by orbits and simulating separately, we find that subhalos that are heavily stripped overall have a higher core-collapse fraction than subhalos with more typical mass-loss histories (and less plunging orbits), when the velocity-dependent SIDM model has negligible cross section at velocity scales $\mathcal{O}(100-1000)$ km/s relevant for evaporation.  This result is consistent with the tidal-field acceleration of subhalo core-collapse times found by previous works \citep{sameie20, nishikawa20, correa21, zzc22}. However, when the evaporation is non-negligible, such as with the $\{\sigma_0=200, \omega=200\}$ (purple) model (Fig.~\ref{fig:sig-v}), we find that core-collapse can be delayed.  
Qualitatively, our work implies that the region corresponding to core-collapse in subhalos in the $\sigma_0-\omega$ parameter space of Fig. \ref{fig:param-space} shifts to smaller $\sigma_0$ for $\omega\lesssim \mathcal{O}(100)$ km/s but bigger $\sigma_0$ for $\omega \gtrsim \mathcal{O}(100)$ km/s relative to isolated halos ( so that halos core collapse before being accreted).

Intriguingly, for the two SIDM models with significant core-collapse in subhalos, $\{\sigma_0=200, \omega=200\}$ (purple) and $\{\sigma_0=200, \omega=50\}$ (violet), we find that the distribution in subhalos' inner slopes can explain the recently reported ultra-steep density profiles of subhalos found in substructure lensing with 13 Hubble Space Telescope images \citep{gzhang23}, as shown in Fig. \ref{fig:slope-hist}.

In addition to these well-motivated velocity-dependent SIDM models, we also investigate toy models that are motivated by specific velocity-dependent features (see Fig. \ref{fig:arti-x}), to explore and highlight the interplay of physical processes in driving subhalo evolution. The window-function type of models, a simplified version of velocity-dependent SIDM models with resonant peaks, work as mass filters that only allow subhalos within a specific mass range to core-collapse. If observed, this filtering effect can greatly help us constrain physical models for SIDM.  Interestingly, even though evaporation is negligible in these models, we find an exception to the rule that tidal fields accelerate  subhalo core-collapse in these window-function models (see Table \ref{table:ccp} and Fig. \ref{fig:arti-w-8.0}).  This is because tides cause mass loss on the outskirts of subhalos, lowering the relative velocities between particles below the velocity range relevant for resonant interactions. It effectively shuts down (or impedes) heat outflow within the subhalo, without which core-collapse cannot proceed. Additionally, tidal heating injects energy to the subhalo center, further delaying its core-collapse process. These two effects together lead to the reverse of the expected tidal-field acceleration on core-collapse process, and may have a similar but less extreme outcome when SIDM models have resonant peaks in velocity-dependence function. 

To explore the effects of evaporation on core-collapse, we design ``z-type'' models that follow the $\sigma_0-\omega$ models below a characteristic velocity, and are constant at $\sigma/m \sim \mathcal{O}(1)$ cm$^2$/g at higher velocities (much higher than the $\lesssim \mathcal{O}(0.1)$ cm$^2$/g of the $\sigma_0-\omega$ models for characteristic subhalo orbital velocities), as shown in Fig. \ref{fig:arti-x}.  These models are designed to test the destructive effect of evaporation on subhalos core-collapse, which has been demonstrated using SIDM with constant cross sections in \cite{zzc22}.  We find that even a relatively weak ($\mathcal{O}(1)\ \rm cm^2/g$) cross section at orbital velocity scales is enough to significantly reduce the subhalo core collapse fraction. 

In addition to the subhalo slopes relevant for the lensing observations of \citeauthor{gzhang23} \cite{gzhang23}, we explore other model-dependent signatures of subhalo core-collapse in substructure lensing statistics. We find that the PDF of the lensing convergence $\kappa_{\rm sub}$ (the 1-point function) is only sensitive to evaporation or core-collapse in massive subhalos ($\gtrsim 10^9 M_\odot$ at infall; Fig. \ref{fig:1ptb} and \ref{fig:1ptc}). 
By contrast, the 2-point correlation function of the convergence map conveys more information about core-collapse. The substructure lensing power spectrum $P_{\rm sub} (k)$ increases when core-collapse kicks off at a scale $k$ corresponding to a specific subhalo mass, and grows higher than the CDM case when the core-collapse fraction is significant (Fig. \ref{fig:pkb} and \ref{fig:pkc}).  Importantly, evaporation leads to an all-scale suppression in the 2-point function. 
While our preliminary analysis of substructure lensing statistics here hints that the 2-point correlation of the subhalo convergence map maybe a sensitive probe of the physics of subhalo core-collapse, more advanced techniques, such as forward-modelling \citep{dg18}, will be needed for a more quantitative constraint on SIDM parameter space with upcoming lensing surveys. 

SIDM models may further be distinguished by their relative effects on isolated halos vs. subhalos. For most of the velocity-dependent SIDM models we test, subhalos have faster core-collapse than isolated halos, consistent with previous work \citep{sameie20, nishikawa20, correa21}. However, when evaporation becomes significant, such as in the $\{\sigma_0=200, \omega=200\}$ model, we find that subhalo core-collapse can either be accelerated or decelerated relative to the field (Fig. \ref{fig:tdiff}). The evolution of (sub)halos in different environments does not scale monotonically, but shows a complicated dependence on subhalo mass and orbits. This highlights that the evolution and properties of subhalos involves a complicated coupling of heat transfer within the subhalo, tidal effects, and host-subhalo evaporation.  

It is important to explore the current limitations of our hierarchical scheme, and a path toward improvement. Here we list the known components that are missing in our method, which do not affect our key results, but which should be addressed in future work.  

\begin{itemize}
    \item \textit{Quiescent merger history} In this work, we deliberately select merger trees with a quiescent accretion history, with no major mergers after $t=4$ Gyr. This is for two reasons: first, although we have validated our hierarchical framework (see Appendix \ref{appdx:val}), this could change when a massive major merger is present (e.g. like the LMC in the Milky Way; see \cite{nadler20b}). Second, to construct an analytic host at any time, we interpolate among the measured density and velocity profiles of a simulated host halo at a finite number of time steps. An underlying assumption in this process is that accreted subhalos contribute to the mass growth of the host in a smooth, spherically symmetric way. This is no longer a valid assumption when a major merger happens. Although it is beyond the scope of this work to explore in more detail, a possible extension to include major mergers in our framework is to model the massive LMC-like subhalo in a similar way as our treatment of the host halo --- a time-varying analytic density distribution, but also with a time-varying center of the potential (see also \cite{nicogc21} for another possible approach of representing LMC with basis function expansions). Since lens galaxies are mostly elliptical which likely have major mergers in their past, this should be an important next-step.

    \item \textit{Pre-infall mass loss of subhalos} As has been described in Sec. \ref{sec:method}, we rewind the position, velocity and virial mass of a subhalo at its infall time back to its formation time, to start the ensemble simulations at a single time.  
    The rewinding is done analytically, typically guaranteeing that the forward-simulated velocity and position at infall time agree with the input ones --- but not necessarily the infall mass.
    This is because the mass loss of subhalos prior to infall is not taken into account during the rewinding process. For most subhalos, such pre-infall mass loss is not significant \citep{Behroozi:2013fqa} because the tidal force from the host is still very weak before subhalos enter the host's virial radius. Furthermore, tidal stripping proceeds outside-in. Typically at the first pericenter passage, where the tidal radius reaches its minimum, a large fraction of mass in the outskirt of subhalos is tidally stripped, which erases any minor effects from the pre-infall mass loss. However, we find that $\sim$10\% of subhalos in our simulation sets have a spurious earlier encounter with the host before their assigned infall events (an example is the red line in Fig. \ref{fig:orbit-check}), which causes unwanted extra mass loss prior to infall and a mismatch between their infall mass in the simulation and the input information from \texttt{Galacticus}. For these cases, the actual infall time is much earlier than that indicated by the merger trees, thus it may have a minor effect on the detailed statistics of subhalos. The inconsistent infall times of these subhalos are mainly due to the way we sample the subhalo orbits in \texttt{Galacticus}. The distributions of subhalo's radial and tangential velocity at infall measured from cosmological simulations \citep{Jiang:2014zfa} do not contain the information about the correlation between subhalo orbits and the merger history of the host or the host's virial radius in the past. Thus for a specific realization of merger trees, when we evolve a subhalo from the infall time backward, it may already fall within the host's virial radius at an earlier time. In this work, since we mainly focus on the relative effects of different SIDM models on subhalos, such effect does not affect our key results. In the future, this problem may be solved with either an updated version of \texttt{Galacticus} that takes into account the specific merger history when sampling the subhalo orbits or a different rewinding calculation which enforces that each subhalo approaches the host from a large distance.

    \item \textit{Accommodating subhalos' formation and infall times} A weakness of the simulation framework presented in this work (or in general any idealized simulations of isolated host-subhalos systems), compared with a fully cosmological simulation, is that dark matter (sub)halos cannot form as a natural result of the growth of the dark matter overdensity field, and must be in place from the beginning of the simulation. In a real Universe, subhalos form and infall at various times --- some subhalos are already infalling to the host while some others are not yet formed. Thus starting the main simulations at the same time can never simultaneously accommodate all the subhalos. As a result, an idealized simulation setup with a population of subhalos has to choose one disadvantage from the following three: a) having a cutoff at subhalos' infall time, such that all sampled subhalos form before this cutoff time and infall after it; b) not accounting for subhalos' evolution prior to infall; c) further splitting the population of subhalos into more simulations, each having different starting times characterised by subhalos' formation time. In this work we choose a), focusing only on subhalos that infall after $t=4$ Gyr. For a simulation scheme with b), all subhalos regardless of their different formation times would be initialized at the same starting time, earlier than the earliest infall event of all subhalos. Such a scheme with b) would work best for CDM, of which the (sub)halos' structure is not expected to evolve prior to infall, but not for SIDM, for which the (sub)halo's central densities constantly evolves with time. For a simulation scheme with c), subhalos need to be further divided into simulations that start according to their different halo formation times. Effectively it would be to extend our pre-evolution runs (see Fig. \ref{fig:flowchart}) to the end of the simulation, without merging the formation time bins at $t=4$ Gyr. 
    More subhalo-subhalo interactions will be missing in simulations with this approach, which is shown to be a minimal effect in the validation Appendix \ref{appdx:val}, regardless of how many simulations the whole population is divided into.
    Hence we suggest that c) might be a better choice for future works.

    \item \textit{Core-collapse in this work as a lower-limit on subhalos' final central density} An issue that all SIDM simulators face is how to treat SIDM when the core enters the short mean free path limit during core-collapse.  In this work, when a subhalo is deep in the core-collapse phase (its central density grows 5 times higher than the highest density in the initial conditions), we freeze the self-interaction of dark matter particles associated with that subhalo.  Otherwise, the timesteps required to simulate interactions become prohibitively small. 
    Most of the relevant works on core-collapse of SIDM halos, using analytical approaches (gravothermal model; \cite{koda11, essig19, sq22}) or N-body simulations \citep{sameie20, zzc22, dnyang22, nadler23}, focus on the early stage of core-collapse.  There is a significant need to study of the end-state of core-collapsing SIDM halos, with many exciting questions to answer and new methods to develop: Will the core-collapse process halt with a dense, hot core that loses thermal contact with the outer part of the halo because the mean-free-path becomes so small? Will the core-collapse process quench itself for velocity-dependent SIDM models when the contracting core becomes too hot? Will the (sub)halo center contract to a supermassive black hole \citep{pollack15, wxfeng21, wxfeng22, Meshveliani23}? Will the SMBH keep accereting dark matter, and if so, at what rate? What is the `final' state of a core-collapsed (sub)halo, if there is a stable one?  While more studies are required to fully understand the end-state of core-collapse, our temporary treatment of freezing the self-interaction in core-collapsed subhalos serves as a conservative lower limit on their central density.

    \item \textit{Baryons} In this work, we focused on dark-matter-only results, so natural next steps are to include galaxies, either with an additional central galaxy potential to the host \citep{wxfeng21, ymzhong23}, or also stellar particles to subhalos. With an embedded central galaxy, we would be able to generate mocks of strongly lensed images using ray-tracing simulations, and thus more direct comparison with upcoming lensing observations. 
\end{itemize}

Core-collapse is the most exciting topic in SIDM studies today, because it implies that SIDM halos can span an enormous range in central density.  With the hierarchical simulation model that we introduced in this paper, we are able to efficiently explore core-collapse in a variety of environments, down to arbitrarily small halo mass, including the effects of evaporation.  The first results of our hierarchical simulation scheme show that subhalos span a wide range of central densities and slopes for fixed SIDM models, with the speed of core-collapse relative to isolated halos depending on a complicated interplay of physical processes.  Importantly, the velocity dependence of the SIDM cross section imprints itself in several metrics relevant for substructure lensing.  Intriguingly, the slope of density profiles of core-collapsed subhalos matches that found in a recent analysis of strong lenses \citep{gzhang23}.  Future work on the evolution of subhalos in SIDM cosmologies should focus on the detectability of these signatures with current and near-future observational facilities.

\begin{acknowledgments}
We thank Akaxia Cruz, Ethan Nadler, Yi-Ming Zhong, Shin'ichiro Ando, Gemma Zhang and Daneng Yang for useful discussions. 

This work was supported in part by the NASA Astrophysics
Theory Program, under grant 80NSSC18K1014.
F.-Y.C.-R. acknowledges the support of program HST-AR-17061.001-A whose support was provided by the National Aeronautical and Space Administration (NASA) through a grant from the Space Telescope Science Institute, which is operated by the Association of Universities for Research in Astronomy, Incorporated, under NASA contract NAS5-26555. Z.C. Zeng is partially supported by the Presidential Fellowship of the Ohio State University Graduate School.

The simulations in this work were conducted on Ohio Supercomputer Center \cite{osc}, mostly on the CCAPP condo. 
\end{acknowledgments}

\appendix

%\nocite{*}

\section{Resolution test}\label{appdx:res}

\begin{figure*}
    \centering
    \begin{subfigure}{0.42\textwidth}
        \centering
        \includegraphics[width=\textwidth, clip,trim=0.3cm 0cm 0.3cm 0cm]{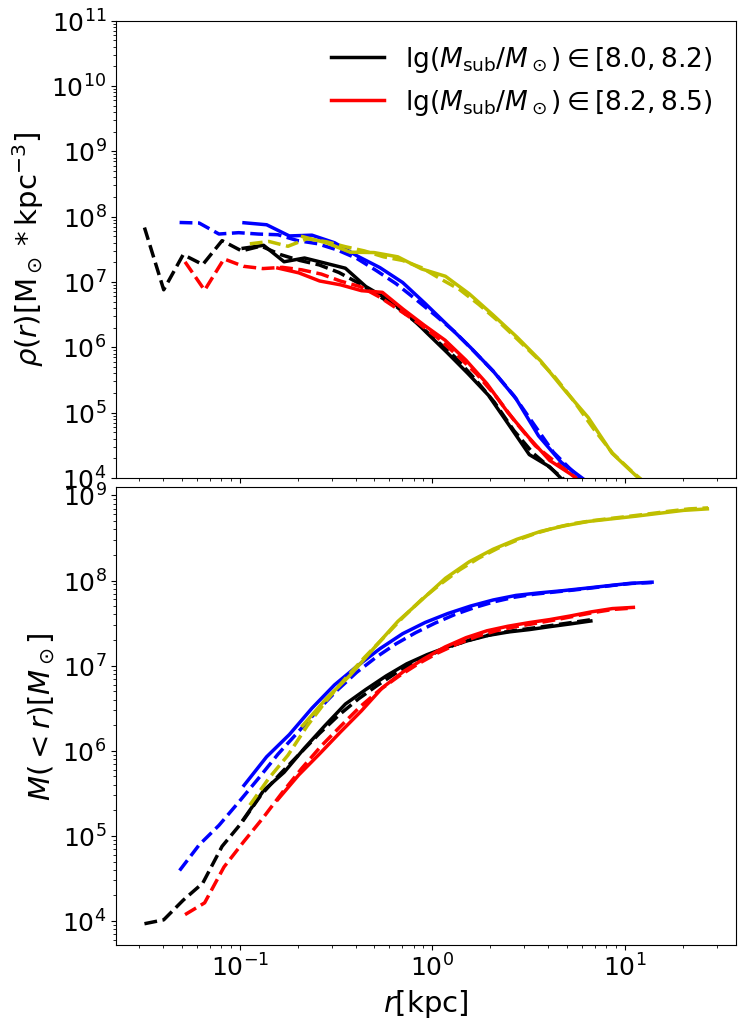} 
        \caption{Constant $\sigma/m$, ordinary orbit, cored}
    \end{subfigure}
    ~
    \begin{subfigure}{0.42\textwidth}
        \centering
        \includegraphics[width=\textwidth, clip,trim=0.3cm 0cm 0.3cm 0cm]{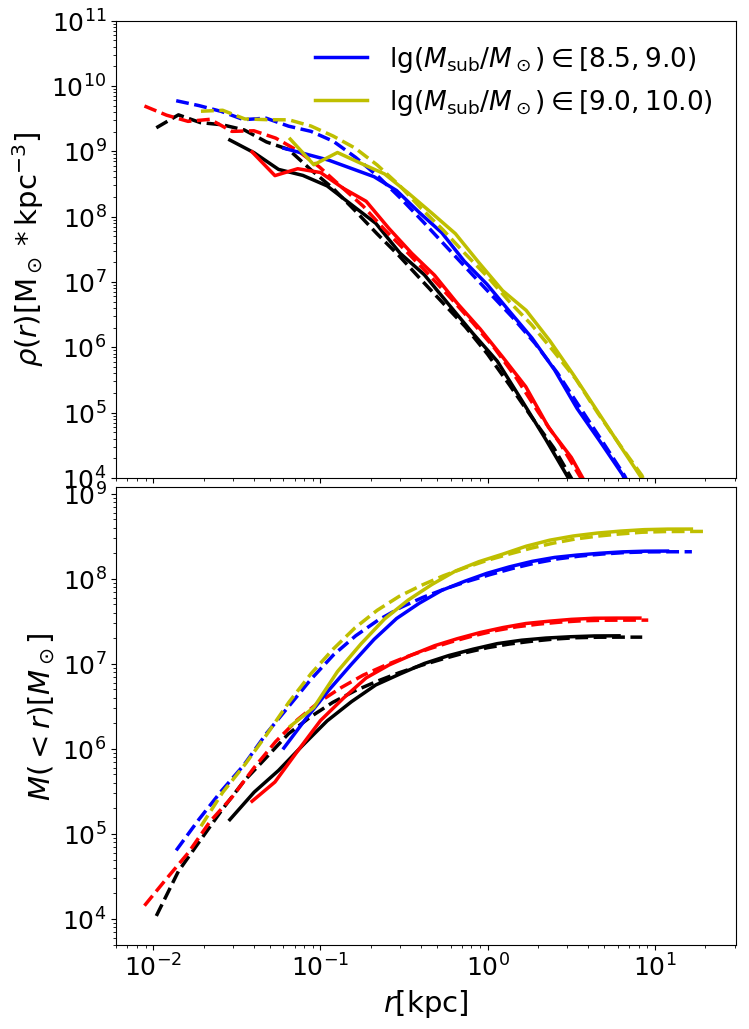} 
        \caption{Velocity-dependent $\sigma/m$, ordinary orbit, core-collapsed}
    \end{subfigure}
    ~
    \begin{subfigure}{0.42\textwidth}
        \centering
        \includegraphics[width=\textwidth, clip,trim=0.3cm 0cm 0.3cm 0cm]{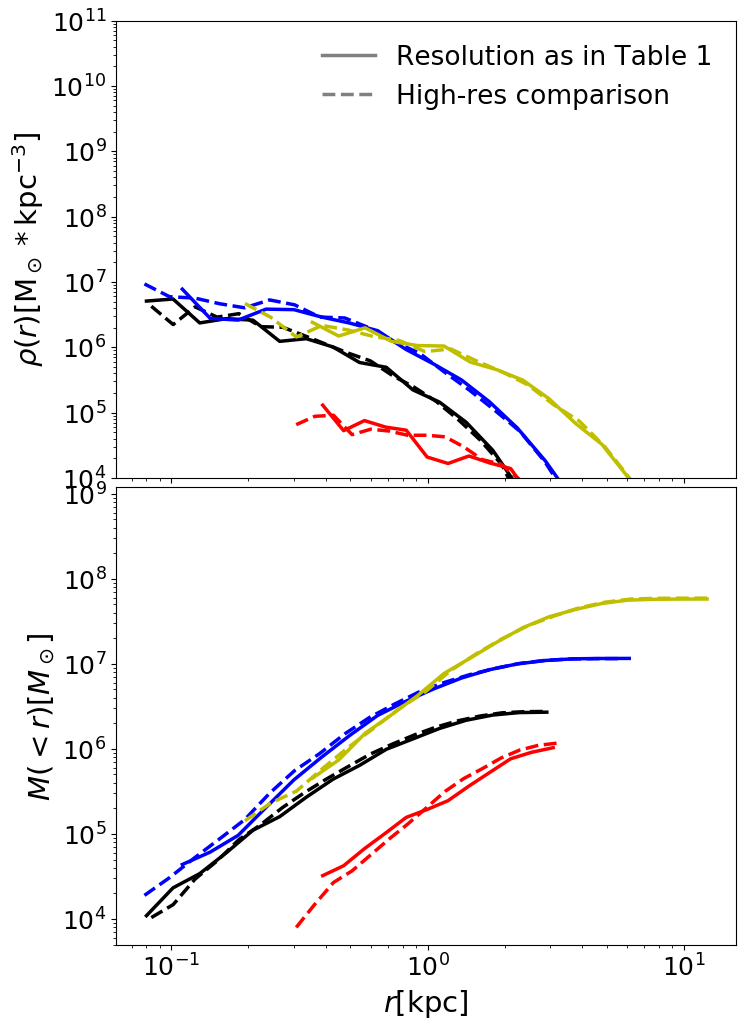} 
        \caption{Constant $\sigma/m$, heavily-stripping orbit, cored}
    \end{subfigure} 
    ~
    \begin{subfigure}{0.42\textwidth}
        \centering
        \includegraphics[width=\textwidth, clip,trim=0.3cm 0cm 0.3cm 0cm]{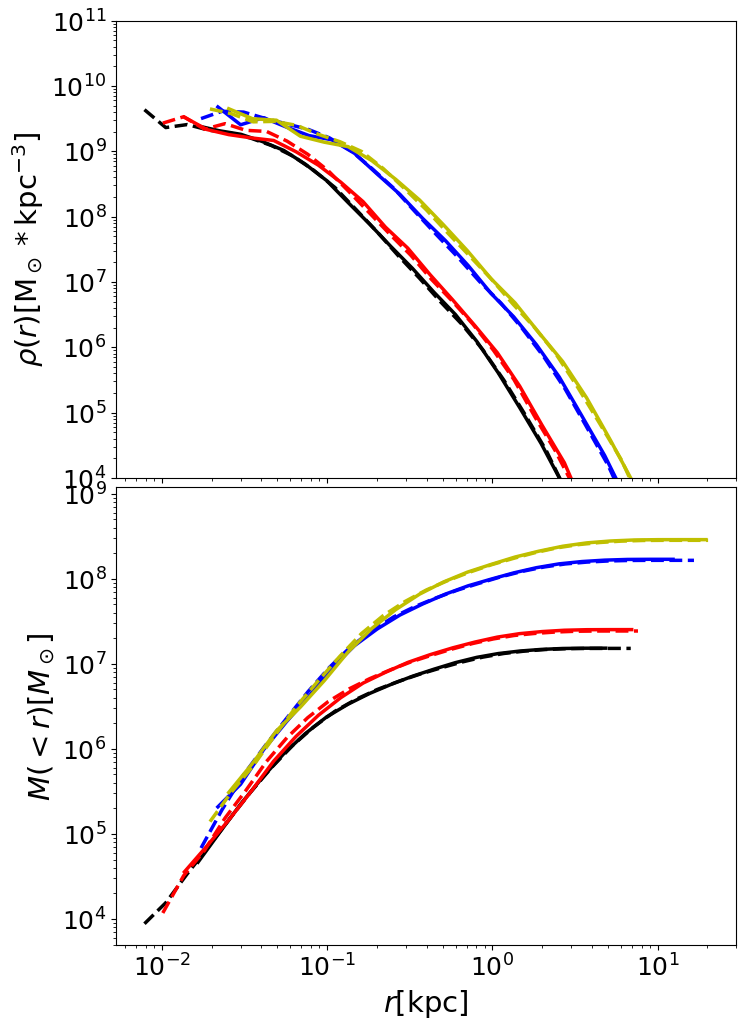} 
        \caption{Velocity-depedent $\sigma/m$, heavily-stripping orbit, core-collapsed}
    \end{subfigure} 
    
    \caption{Resolution convergence test with constant and velocity-dependent cross section SIDM, on ordinary orbits and heavily-stripping orbits (see Sec. \ref{sec:sim_method} for the definition of these two categories of orbits). All plots are at the end of the simulation $t=9$ Gyr. }
	\label{fig:res}
\end{figure*}

In Fig. \ref{fig:res}, we show the results of resolution tests that we use to determine the optimal resolution for our simulations, summarized in Table \ref{table:res}. For the tests, we sample a subhalo from each of the four infall-mass groups $[10^8, 10^{8.2}) M_\odot, [10^{8.2}, 10^{8.5}) M_\odot, [10^{8.5}, 10^{9}) M_\odot$, $[10^9, 10^{10}) M_\odot$, and simulate them with the presence of an analytic host potential. Four possible scenarios are considered, as correspondingly shown in Fig. \ref{fig:res}: a) SIDM with constant $\sigma/m$, subhalos on ordinary orbits; b) SIDM with velocity-dependent $\sigma/m$, with subhalos on ordinary orbits; c) SIDM with constant $\sigma/m$, subhalos on heavily-stripping orbits; d) SIDM with velocity-dependent $\sigma/m$, with subhalos on heavily-stripping orbits. In this resolution test, we choose $\sigma/m = 6\ \rm cm^2/g$ as the constant cross section, and a $\sigma_0-\omega$ model with $\sigma_0 = 200\ \rm cm^2/g, \omega=50\ \rm km/s$ as the velocity-dependent cross section. 
Here when selecting a specific 'ordinary orbit' for subplots a) and b) of Fig. \ref{fig:res} and a specific ``heavily-stripping orbit" for subplots c) and d), we deliberately choose the ones that tend to have more subhalo mass loss so that our resolution test is conservative and can be generalized to other orbits. For the constant cross section of $6\ \rm cm^2/g$, ordinary orbits leads to subhalo mass losses of $\lesssim 90\%$, and the heavily-stripping orbits lead to mass losses of $90\% \sim 99\%$. For example, the red line in the lower panel of Fig. \ref{fig:res} (c) shows a subhalo starting with a mass $\sim 10^{8.2} M_\odot$ and ending with $<10^6 M_\odot$. 

All the simulations in this resolution test are run from $t=0$ Gyr to $t=9$ Gyr, with a longer evolution time inside the host than in our main simulations. This is meant to ensure that the mass resolution we choose would produce robust results for subhalos on even more extreme orbits.

To conduct the resolution tests, we first prepare high-resolution runs for each subhalo in each scenario as a standard for comparison, then gradually reduce the number of particles at initialization, until the convergence in high/low resolution runs breaks down. Our criterion for good convergence requires that the subhalo's density/mass profile is in agreement with the high-resolution run at radii larger than 100 pc. We choose this convergence criterion because in our results sections (Sec.~\ref{sec:halo1} and \ref{sec:halo2}), we measure subhalo statistics at radii > 150 pc. The high-resolutions runs are simulated with each subhalo having $2\times 10^5$ particles for the ordinary orbit and $3\times 10^5$ particles for the heavily-stripping orbit, and the low-resolution runs that are well-converged are summarized in Table~\ref{table:res}.

We find that the primary challenge for particle resolution is different in the cases of cored (constant cross section) and core-collapsed (velocity-dependent cross sections) subhalos. For cored subhalos with constant SIDM cross section, the mass loss of subhalos can be extreme, especially on heavily-stripping orbits, which makes it difficult to properly resolve the subhalo with only a small fraction of remaining particles. Empirically, we estimate that $\sim 1000$ particles bound to a subhalo at the end of the simulation are needed to track the end-state of the subhalo with high fidelity according to the `100 pc' criterion. For core-collapsed subhalos, however, mass loss on heavily-stripping orbits is of less concern, as we can see from the comparison of subplots b) and d) in Fig. \ref{fig:res}. This arises from two physical effects.  First, the host-subhalo evaporation is weaker for this [$\sigma_0=200, \omega=50$] velocity-dependent SIDM model whose cross section drops fast at high relative velocities, than for constant SIDM cross sections.   Second, when the center of an SIDM (sub)halo collapses in the runaway process, it becomes so dense that tidal stripping from the host is less efficient. In this case, the challenge is to resolve the ultra-dense, collapsed center of the subhalo. Since we switch off the dark matter self-interaction after the core-collapse of a subhalo is triggered, the artificial gravitational two-body relaxation is a nuisance effect that becomes important, and is worse at lower resolution \citep{Diemand2004}. However, we find that this only causes a discrepancy in subhalo profiles in the very central region $r\lesssim 1\% r_{\rm vir}$, as shown in subplot b) of Fig. \ref{fig:res}, and thus does not affect our analysis results in Sec. \ref{sec:results}.

\section{Validation of the hierarchical scheme}\label{appdx:val}

\begin{figure*}
    \centering
    \begin{subfigure}{\columnwidth}
        \centering
        \includegraphics[width=\textwidth, clip,trim=0.3cm 0cm 0.3cm 0cm]{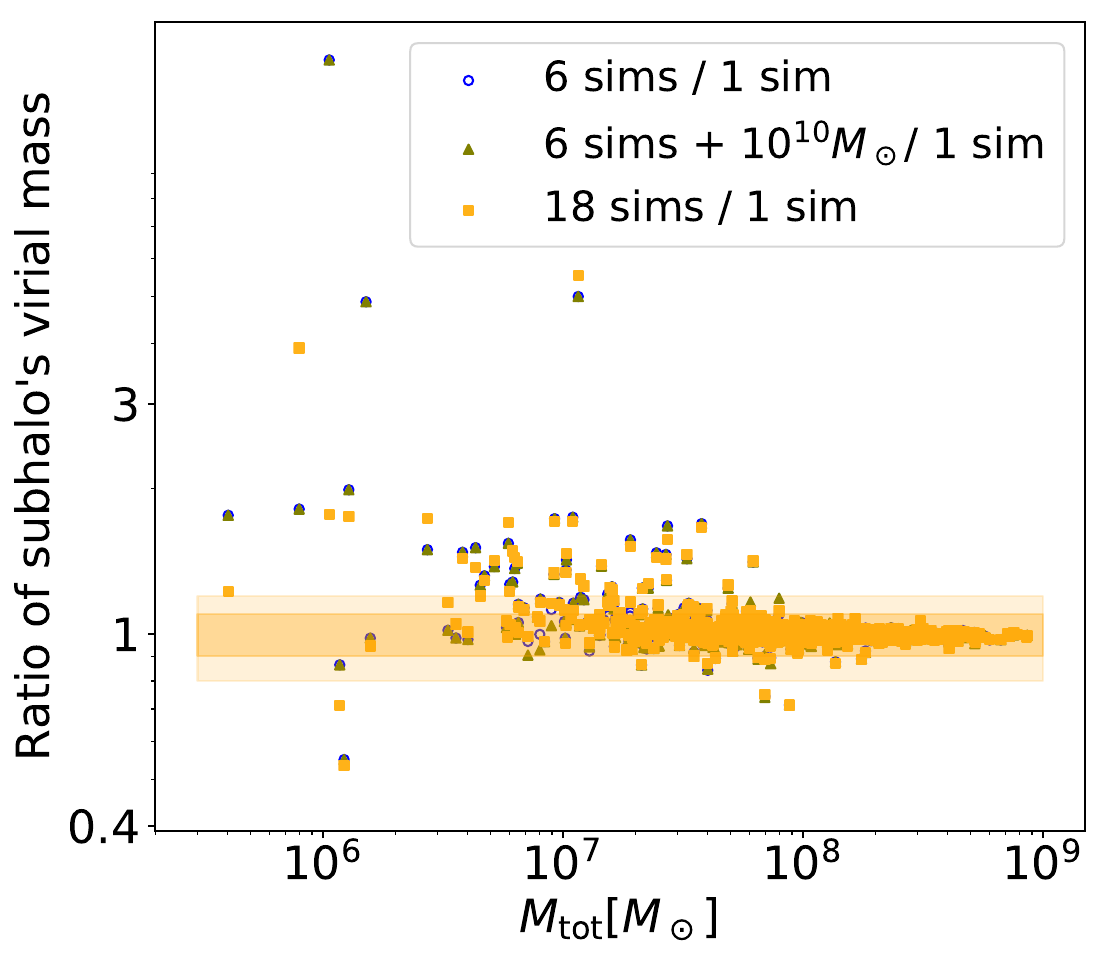} 
        \caption{}
    \end{subfigure}
    ~
    \begin{subfigure}{\columnwidth}
        \centering
        \includegraphics[width=\textwidth, clip,trim=0.3cm 0cm 0.3cm 0cm]{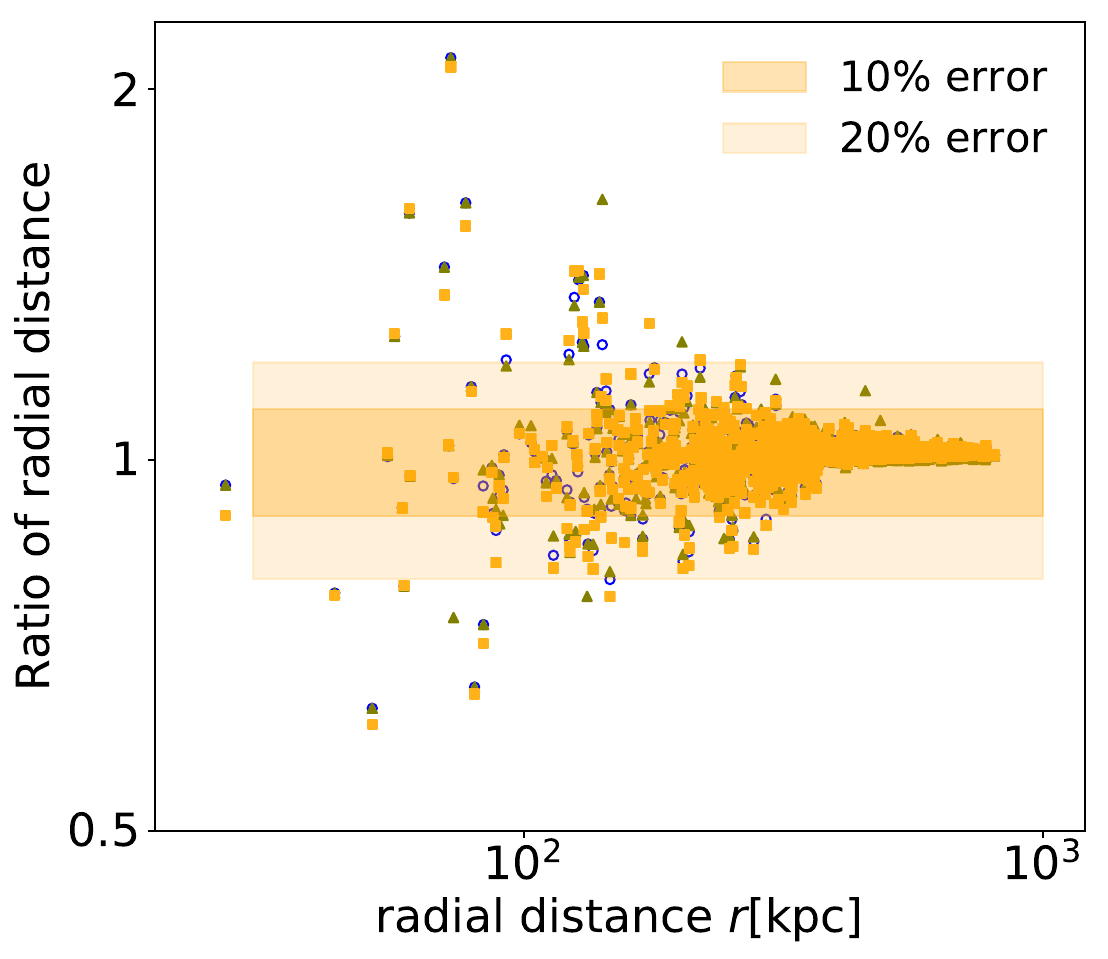} 
        \caption{}
    \end{subfigure}
    ~
    \begin{subfigure}{\columnwidth}
        \centering
        \includegraphics[width=\textwidth, clip,trim=0.3cm 0cm 0.3cm 0cm]{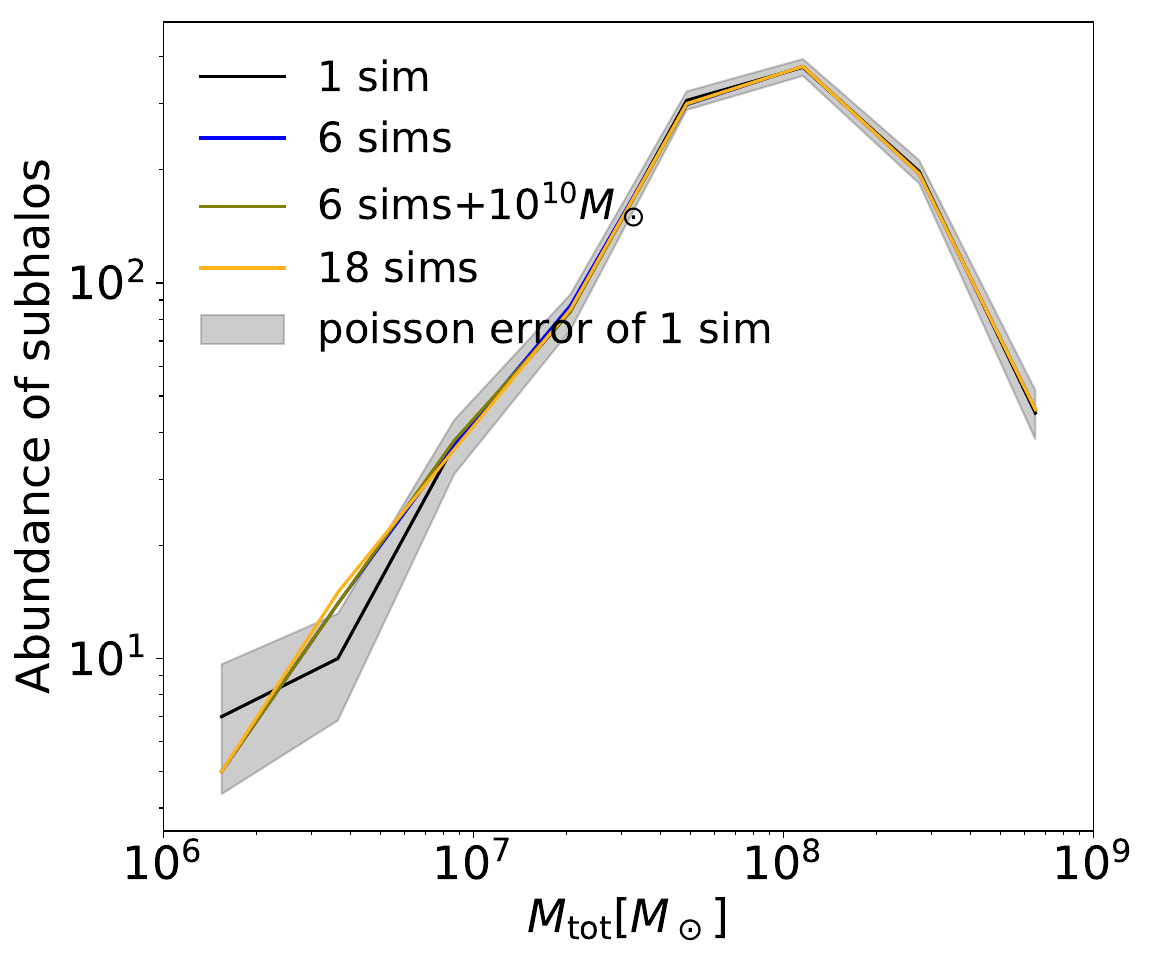} 
        \caption{}
    \end{subfigure}
    ~
    \begin{subfigure}{\columnwidth}
        \centering
        \includegraphics[width=\textwidth, clip,trim=0.3cm 0cm 0.3cm 0cm]{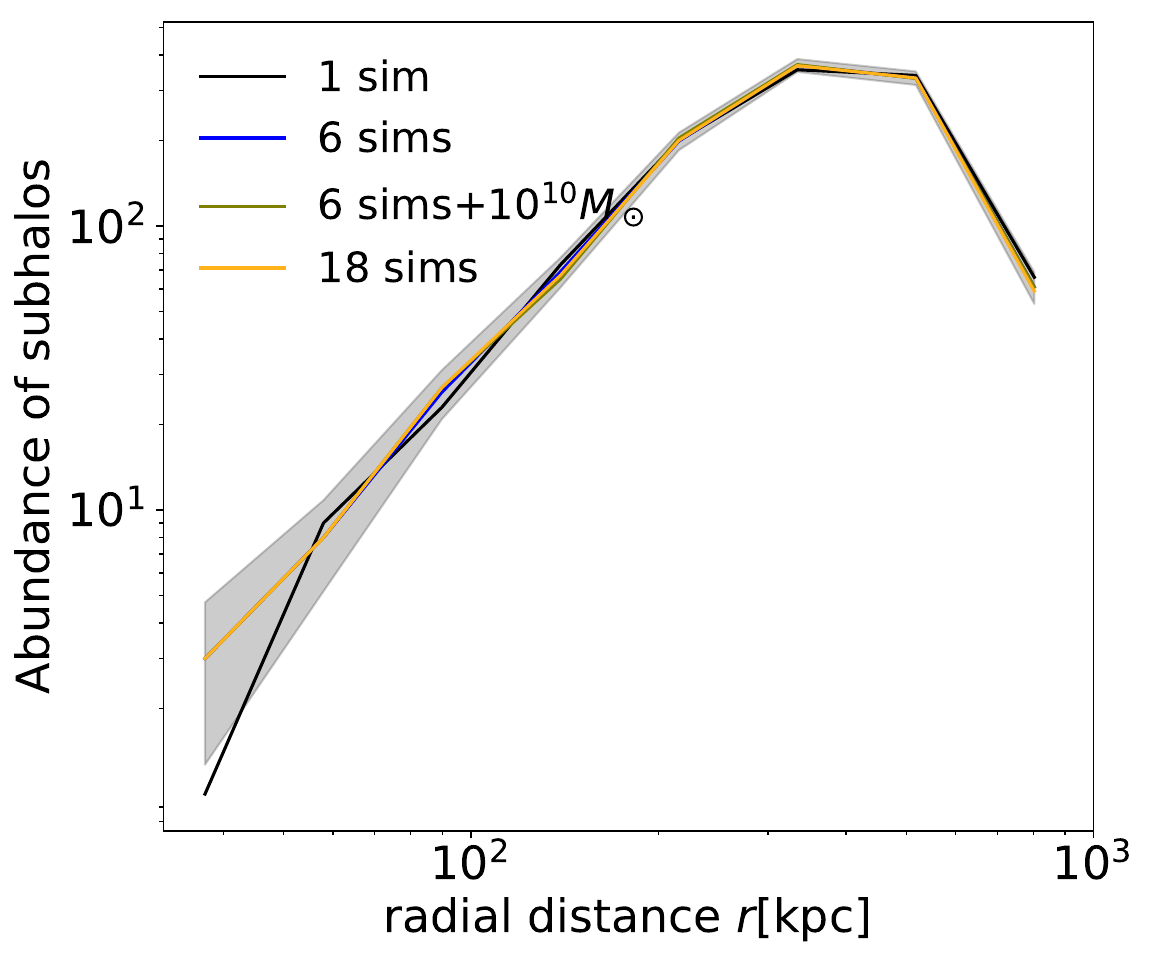} 
        \caption{}
    \end{subfigure}
    
    \caption{Validation for our hierarchical simulation scheme using subhalos with infall mass in the range $[10^8, 10^9]$. Three sets of simulations are compared here, either all the subhalos simulated in one big simulation, or divided into six simulations (with or without adding a $10^{10}M_\odot$ massive subhalo to each of the three o.o. simulations), or eighteen. All subplots show the end of the simulation. In a) and b) we show each subhalo's mass and radius from the host center in these three sets of simulations. In c) and d) we show the statistical of mass function and radial distribution. All panels are at the end of the simulation $t=9$ Gyr.}
	\label{fig:appdx-val}
\end{figure*}

A systematic of our hierarchical simulation scheme is that it neglects subhalo-subhalo interactions across different mass groups. Here we demonstrate that partially missing subhalo-subhalo interactions does not statistically impact the properties of subhalos.

We prepared four sets of simulations, including all subhalos from a merger tree that have infall masses in the range $[10^8, 10^9) M_\odot$.  The first has all subhalos in a single simulation, and its resolution is given by simulation particle mass $m_p = (10^8 / 30000) M_\odot$ (comparable to the $[10^8, 10^{8.2}) M_\odot$ group with o.o.).  The second follows our hierarchical scheme where the subhalos are split into six smaller simulations, $[10^8, 10^{8.2}) M_\odot, [10^{8.2}, 10^{8.5}) M_\odot, [10^{8.5}, 10^{9}) M_\odot$, on either ordinary or heavily-stripping orbits, with resolution assigned the same way as in Table \ref{table:res}.  The third set is the same as the second, but we add a relatively massive subhalo of $10^{10}M_\odot$ to each of the three o.o. simulations with infall mass $[10^8, 10^{8.2}) M_\odot, [10^{8.2}, 10^{8.5}) M_\odot$, and $[10^{8.5}, 10^{9}) M_\odot$, aiming to highlight the possible impact of missing more massive subhalos.  Finally, the fourth set of simulations is based on the second, but the subhalos in each simulation are further divided into three halo-formation-time bins and simulated separately. We denote these three sets of simulations as 1 sim, 6 sims, 6 sims + $10^{10}M_\odot$ and 18 sims hereafter. Note that for the 18 sims set, although we divide subhalos according to their formation time, the formation time is not used as the starting time of these validation simulations, unlike the pre-evolution part of our hierarchical framework described in Sec. \ref{sec:sim_method} and \ref{sec:step}. In the validation simulations for this appendix, these four sets of simulations are all started at $t=4$ Gyr, initialized with NFW profiles (parameters determined by \texttt{Galacticus} merger tree), and ended at $t=9$ Gyr, such that all subhalos to be matched and compared among these four setups have the same amount of evolution time. The SIDM model for these simulations is a constant cross section $\sigma_m = 6\ \rm cm^2/g$, for which the evaporation effect among subhalos is strong and can highlight any potential subhalo-subhalo interaction. We expect subhalo-subhalo SIDM interactions to be weaker for the other models we consider in this paper, since their cross sections at subhalo-subhalo pairwise velocities (comparable to subhalos' orbital velocities) are smaller. 

 At the beginning of the simulation, 1092 subhalos are initialized in each of the four sets of simulations. At the end of the simulation, there are around 1060 subhalos left that can be matched among these four. We conduct a subhalo-by-subhalo comparison, in terms of each subhalo's virial mass and radial distance from the host center, as shown in subplot a) and b) of Fig. \ref{fig:appdx-val}. We find that there are only a few outliers when comparing the split 6 sims, 6 sims + $10^{10}M_\odot$ and 18 sims against 1 sim. Overall we find that $\sim$94\% of subhalos have a mass discrepancy of $<10\%$ and $\sim$96\% of subhalos have a radial distance discrepancy $<10\%$, for all the three hierarchical sets against 1 sim.  The two fractions are $\sim96\%$ and $\sim98\%$ for the subhalo mass and radial distribution, respectively, for error tolerances of 20\%. We show the subhalo mass function and radial distribution in subplot c) and d) of Fig. \ref{fig:appdx-val}, where again we can see that the overall distribution of subhalos is not affected by the missing subhalo-subhalo interaction. Only a few subhalos at the very low mass end, which are very heavily stripped, show visible scatter among these simulations. 

Furthermore, as a comparison between the two split sets, 6 sims and 18 sims, we can see that their subhalo mass function and radial distribution are nearly identical, and the subhalo-by-subhalo trackers also show no sign of any systematic difference. This further hints that the missing subhalo-subhalo interaction is indeed a minimal effect, since splitting the whole population of subhalos into 6 or 18 simulations does not seem to matter at all. Any difference between the 1 sim and 6 sims or 18 sims are more likely to be caused by resolution and realization difference than the missing subhalo-subhalo interaction (recall that 6 sims and 18 sims have the same infall-mass binning so the subhalos are identical at the beginning of the simulation, but 1 sim have different mass resolution and hence also different realization of subhalos), which is also consistent with the Poisson error shown in Fig. \ref{fig:appdx-val}.

\section{Linking a halo's characteristic velocity for self-interaction to velocity dispersion} \label{appdx:window}

\begin{figure}
    \begin{subfigure}[t]{\columnwidth}
        \centering
        \includegraphics[width=\textwidth]{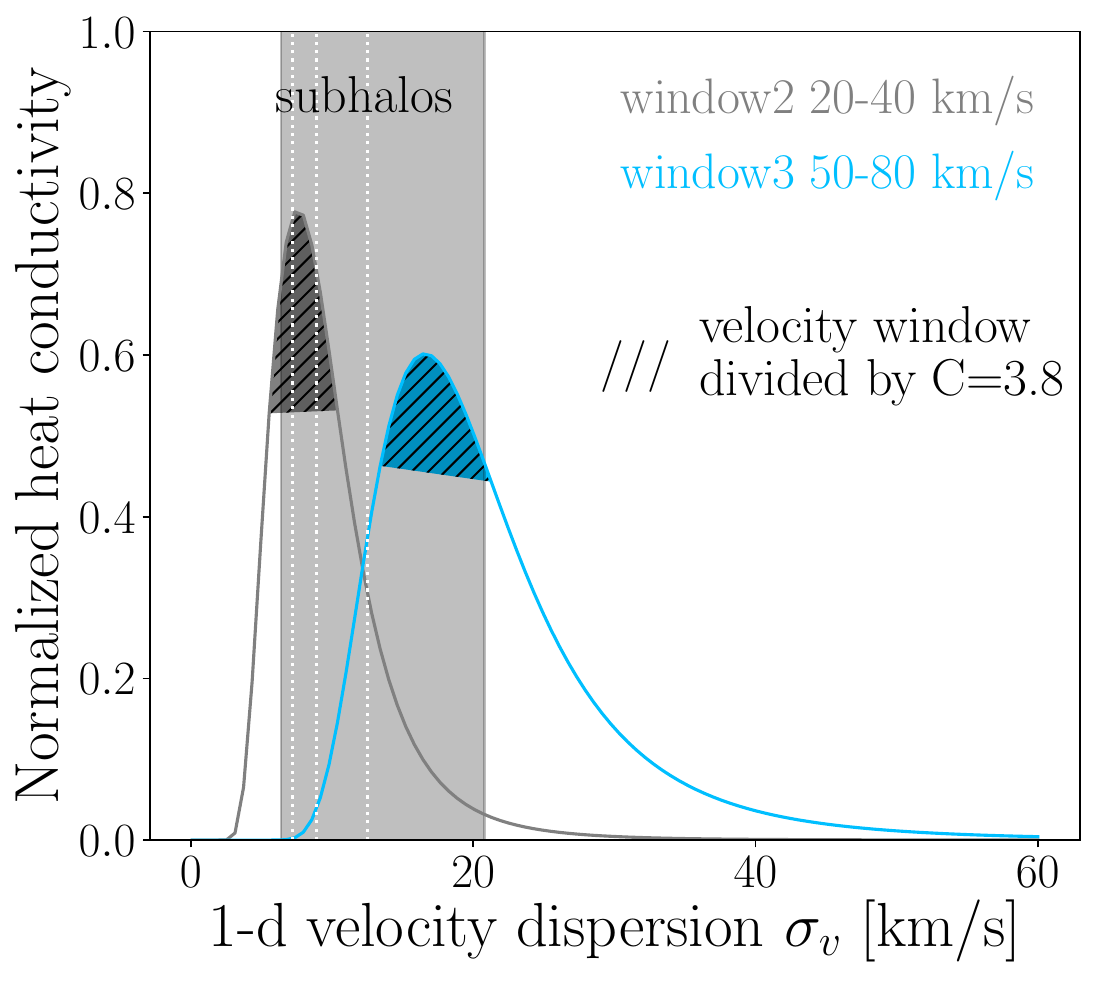}
    \end{subfigure}
    \caption{The analytically calculated SIDM heat conductivity of the `window function' models, as a function of halo's 1-d velocity dispersion $\sigma_v$. Results are normalized against the heat conductivity of a constant SIDM cross section $\sigma/m = $200 cm$^2$/g, which corresponds to the cross section inside the velocity window of the window function models. The $\sigma_v$ range for subhalos' infall mass from $10^8$ to $10^{10}M_\odot$ is shaded with gray color. The $\sigma_v$ range corresponding to velocity window range divided by $C=3.8$ is shown in regions with forward slash.}
    \label{fig:cfac}
\end{figure}

We have shown in Sec. \ref{sec:halo-artificial} that from simulation we estimate $3.0 < C < 4.2$, for the factor $C$ that characterizes the relation between a (sub)halo's typical velocity relevant for self-interaction and maximal 1-d velocity dispersion of a halo's CDM initial condition $\ev{v_{\rm rel}} = C \sigma_v$. In this appendix we approach the same question analytically with the gravothermal method.

If we consider the characteristic velocity for heat transfer inside a (sub)halo as the mean value of the particle-particle relative velocity, a simple averaging over the Maxwell-Boltzmann distribution gives
\begin{equation}
    \begin{split}
    \ev{v_{\rm rel}} &= \int \sqrt{v^2_1 + v^2_2 - v_1 v_2 \cos{\theta}} f_{\rm MB} (v_1) f_{\rm MB}(v_2) d^3 v_1 d^3 v_2 \\
   & = 2.26 \sigma_v,
    \end{split}
\end{equation}
where $v_1$ and $v_2$ are the velocities of any two scattering particles, and the Boltzmann distribution $f_{\rm MB} (v) = (\frac{m}{2\pi k_B T})^{3/2} \exp{(-\frac{mv^2}{2k_B T})}$ has the temperature $T$ determined by $\sigma_v = \sqrt{k_B T / m}$. This factor 2.26 from simple averaging does not agree with the result $3.0 < C < 4.2$ from simulation.
However, \cite{sq22} argued that the averaged heat conductivity in the long-mean-free-path regime 
\begin{equation} \label{eqn:conductivity}
    \ev{\kappa_{\rm lmfp}} \propto \frac{\ev{v_1 v_2 v^3_{12} \sigma(v_{12})}}{\ev{v_1 v_2 v^3_{12}}},
\end{equation}
where $v_{12}$ is the relative velocity between particle 1 and 2, would better characterize the heat transfer for a velocity-dependent SIDM model $\sigma(v_{12})$. In Fig. \ref{fig:cfac}, we show this numerically integrated $\ev{\kappa_{\rm lmfp}}$ of window2 and window3 models (window1 not used because it has a less sharp drop in cross section outside the velocity window, see Fig. \ref{fig:arti-x}), normalized by the $\ev{\kappa_{\rm lmfp}}$ of constant $\sigma/m = 200$ cm$^2$/g, as a function of the 1-d velocity dispersion $\sigma_v$. Note that the window functions, or in general any functions with step-like features in $\sigma(v_{12})$ cannot be easily integrated numerically, since the integral variables are $v_1$ and $v_2$ instead of $v_{12}$. Thus we use $[\tanh{(v_{12} - v_{\rm low})} \times \tanh{(v_{\rm high} - v_{12})} + 1]/2$) to represent window functions for this integral, where $v_{\rm low}$ and $v_{\rm high}$ are the bounds of the velocity window. From Fig. \ref{fig:cfac}, we find that this calculation of $\ev{\kappa_{\rm lmfp}}$ does successfully predict the corresponding subhalo mass/$\sigma_v$ bins that has strong core-collapse, where a window-model that has normalized heat conductivity $\gtrsim 0.4$ would have noticeable fraction of core-collapse in that infall-mass bin. 

Choosing an intermediate value $C=3.8$ from simulation constrained range $3.0<C<4.2$, we plot the range of $\sigma_v = [v_{\rm low}, v_{\rm high}] / C$ corresponding to each window model in Fig. \ref{fig:cfac}. We can see that the estimated $C=3.8$ from simulation correctly picks out the region of relatively high heat conductivity $\gtrsim 0.4$. Thus we conclude that both the analytical gravothermal approach and the estimated range $3.0<C<4.2$ or $C\approx3.8$ have correct predictions on the core-collapse mass-filtering effect by the window-type models.

We comment that our finding echoes the conclusion of \cite{o22} and \cite{sq22}, that it is the heat conductivity averaged over the whole (sub)halo that determines the speed of SIDM evolution, instead of the cross section at the average relative velocity.

\section{Animated time evolution of subhalos' inner mass and virial mass}\label{appdx:anime}

We compile the time-animated version from $t=4$ Gyr to $t=9$ Gyr of the $M_{\rm inner}-M_{\rm vir}$ plot in Fig. \ref{fig:scatter}, in \href{https://youtu.be/FA6j8hHZejM}{this link}, and \href{https://youtu.be/4nR-Rxd691c}{another version} where each subhalo's inner mass $M_{\rm inner}$ is not defined by a fixed aperture 200 pc but a fraction 2\% of its virial radius at any given time. The latter with the inner aperture scaling with subhalo mass may better describe low-mass subhalos that experience dramatic mass loss.

\section{Supplementary figures}\label{appdx:misc}
\begin{figure}
    \begin{subfigure}[t]{\columnwidth}
        \centering
        \includegraphics[width=\textwidth, clip,trim=0.1cm 0cm 0.1cm 0cm]{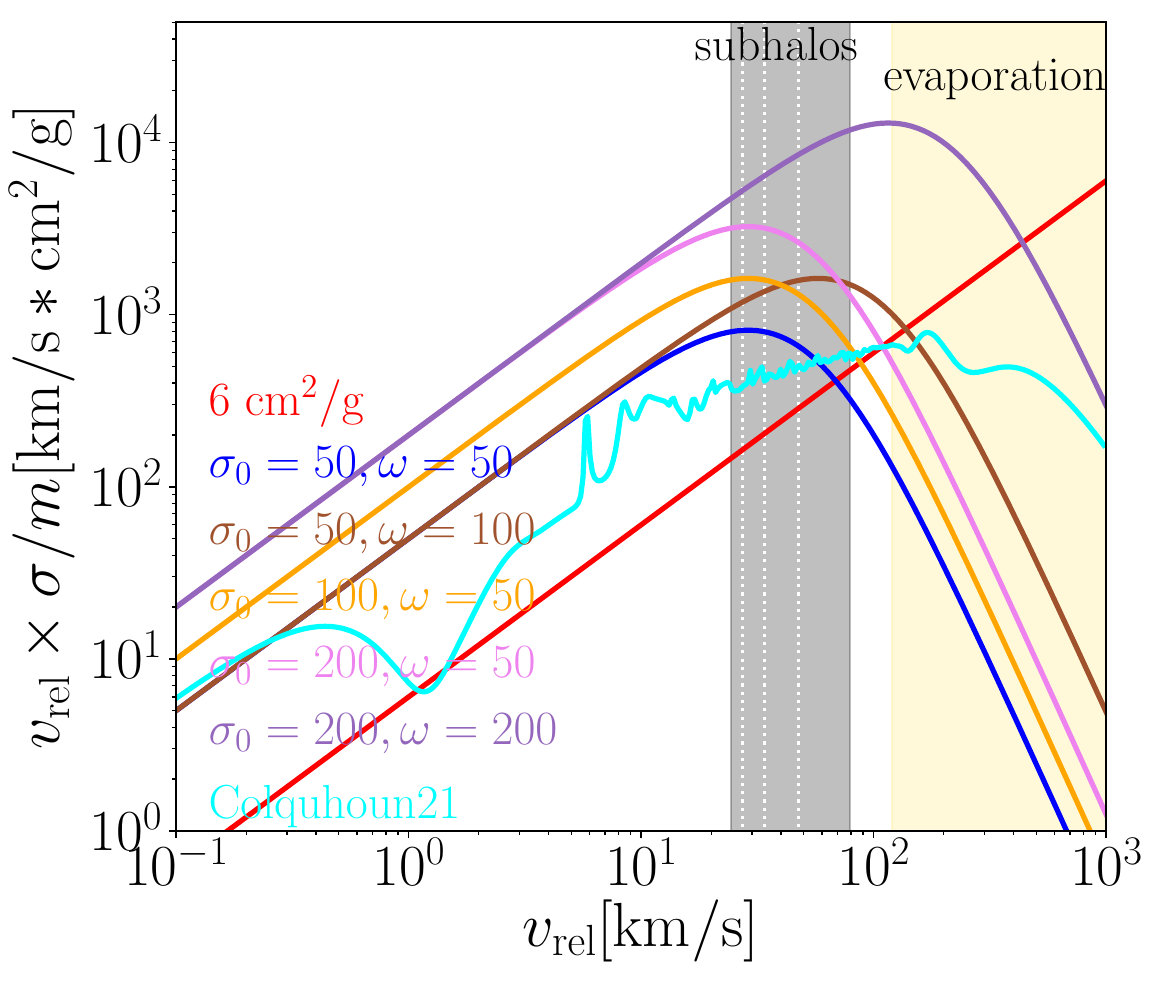}
        \caption{}
        \label{fig:sig-vx}
    \end{subfigure}
    ~
    \begin{subfigure}[t]{\columnwidth}
        \centering
        \includegraphics[width=\textwidth, clip,trim=0.1cm 0cm 0.1cm 0cm]{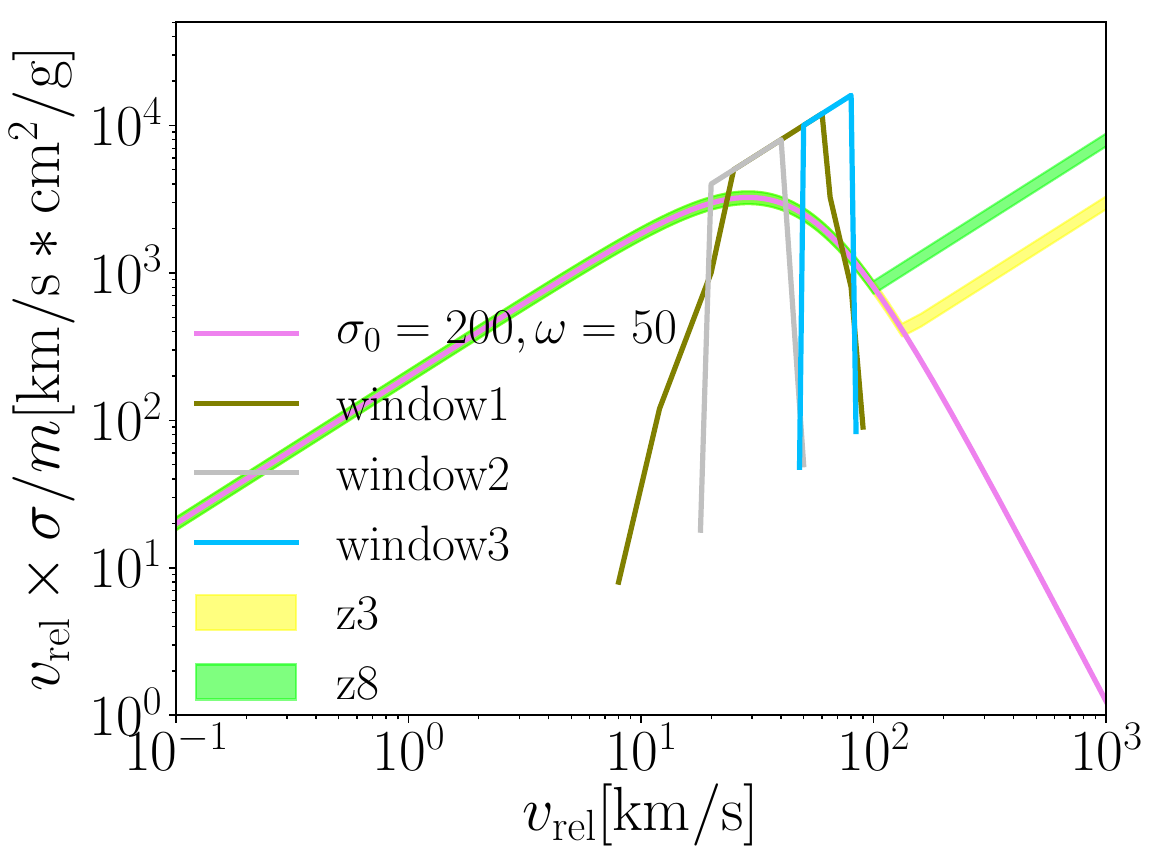}
        \caption{}
        \label{fig:arti-vx}
    \end{subfigure}
    \caption{The velocity dependent SIDM models we use in this work, but the y-axis shown in terms of $v_{\rm rel}\times \frac{\sigma}{m}$ instead of just the cross section that we show in the main text. Since the pairwise scattering probability is proportional to $v_{\rm rel}\times \frac{\sigma}{m}$, adding a weight of $v_{\rm rel}$ to cross section better reflects the actual strength of self-interaction. }
    \label{fig:vsigma-v}
\end{figure}

\begin{figure}
    \begin{subfigure}[t]{\columnwidth}
        \centering
        \includegraphics[width=\textwidth]{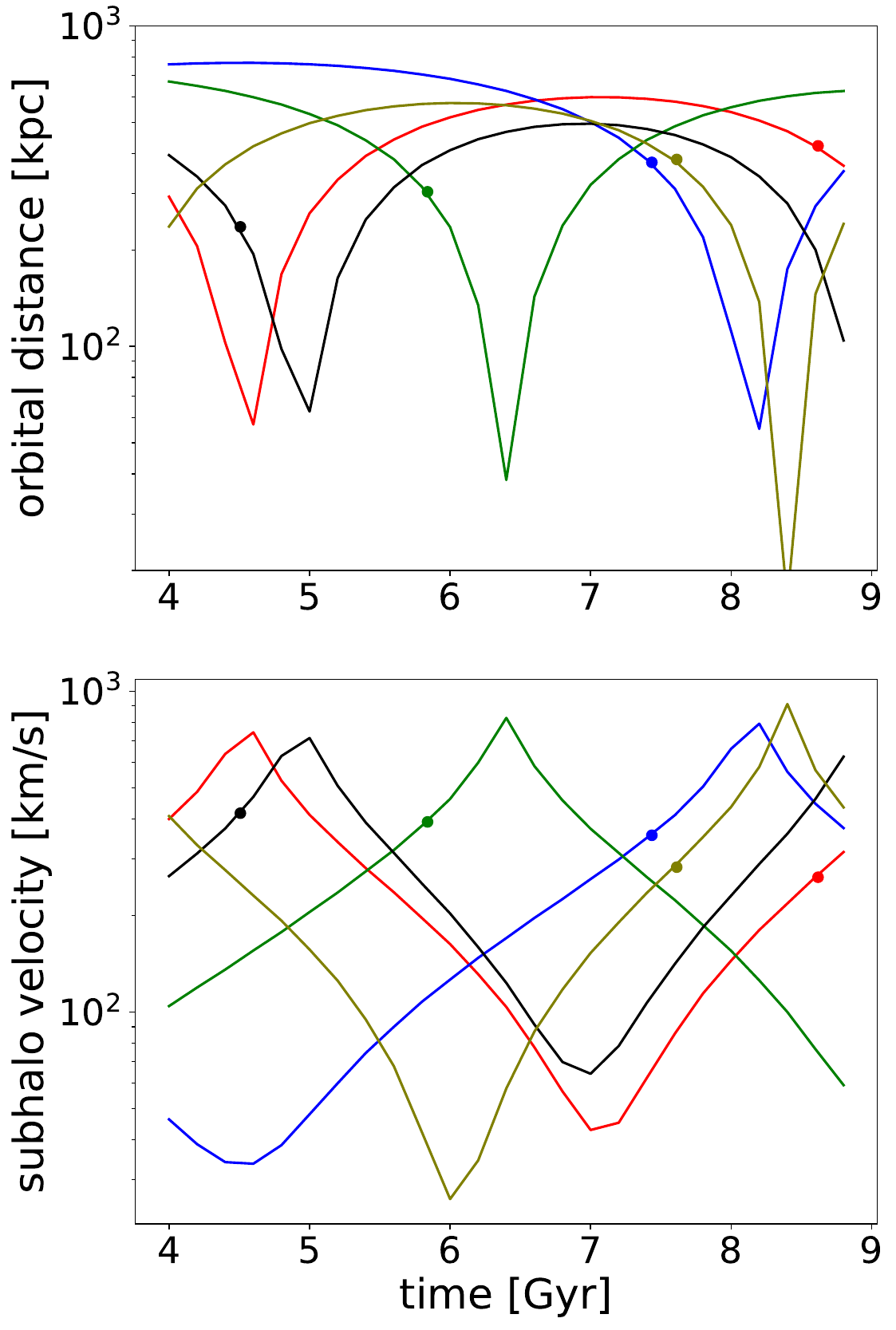}
    \end{subfigure}
    \caption{ The orbital distance (top panel) and orbital velocity (bottom panel) of five randomly selected subhalos. The lines are results from simulation, and the big dots mark their infall radii and velocities generated by \texttt{Galacticus} which serve as the input information to infer subhalos' motion. The dots lie almost exactly on the simulation lines, showing that our rewinding calculation of each subhalo's position and velocity at an earlier time is accurate --- when simulated forward in time, its position and velocity at the infall time both agree with the input from \texttt{Galacticus}. }
    \label{fig:orbit-check}
\end{figure}

\begin{figure}
    \begin{subfigure}[t]{\columnwidth}
        \centering
        \includegraphics[width=\textwidth]{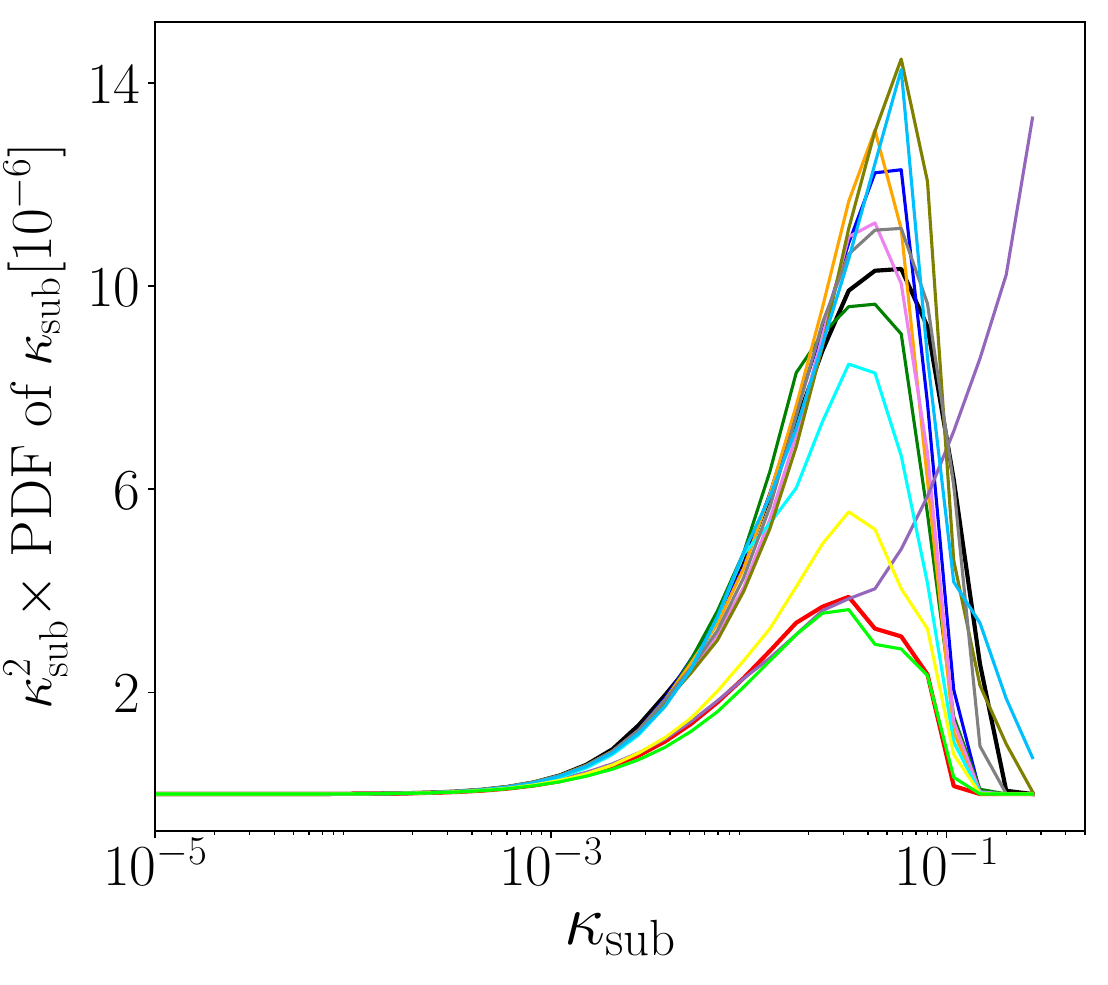}
    \end{subfigure}
    \caption{The PDF of the substructure lensing $\kappa_{\rm sub}$, which corresponds to Fig. \ref{fig:1pta} but with additional weight $\kappa_{\rm sub}^2$, so as to highlight the high density end. }
    \label{fig:kappa2-1pt}
\end{figure}

\begin{figure}
    \begin{subfigure}[t]{\columnwidth}
        \centering
        \includegraphics[width=\textwidth]{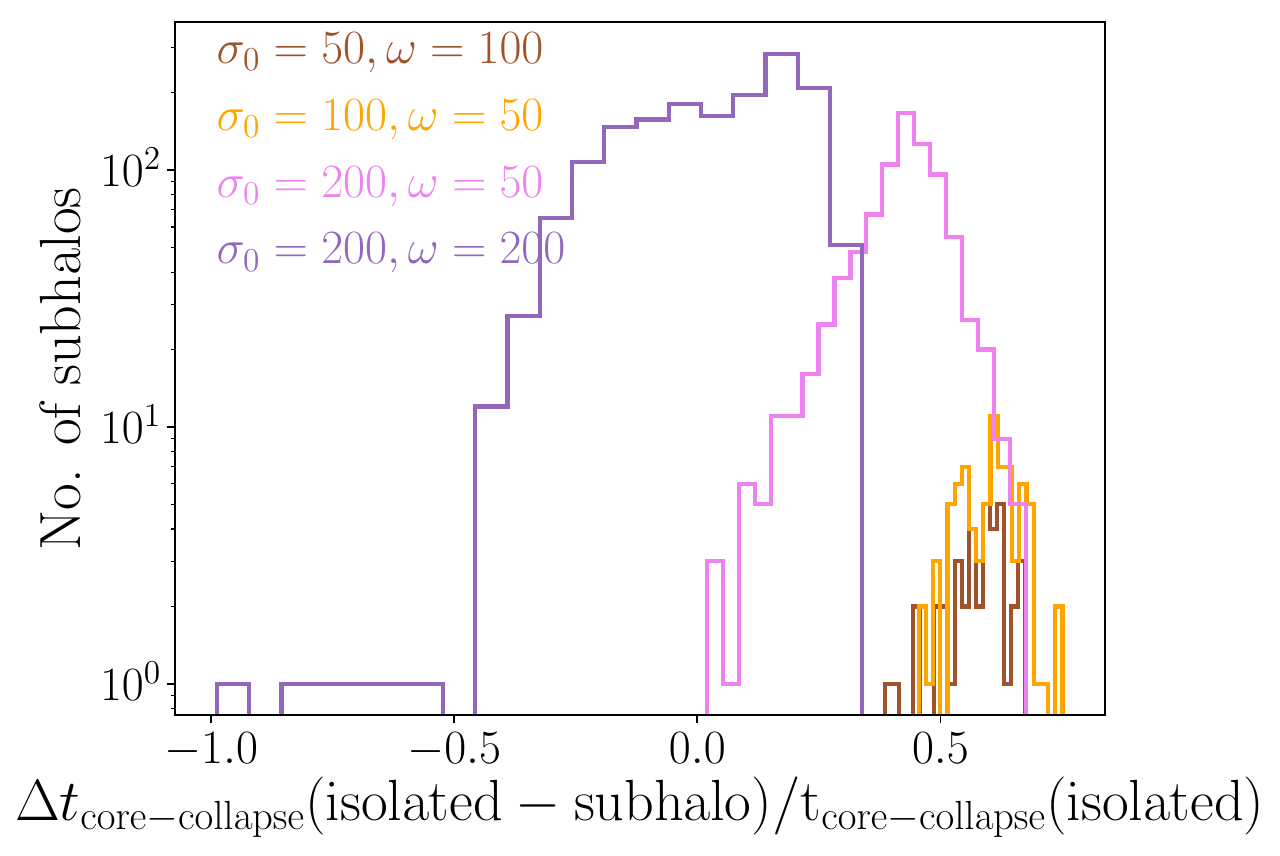}
    \end{subfigure}
    \caption{The time difference for core-collapse in field halos and subhalos, shown in the form as a deceleration (negative) or acceleration (positive) factor by the host field. }
    \label{fig:tdiff-frac}
\end{figure}

\begin{figure}
    \begin{subfigure}[t]{\columnwidth}
        \centering
        \includegraphics[width=\textwidth, clip,trim=0.1cm 0cm 0.1cm 0cm]{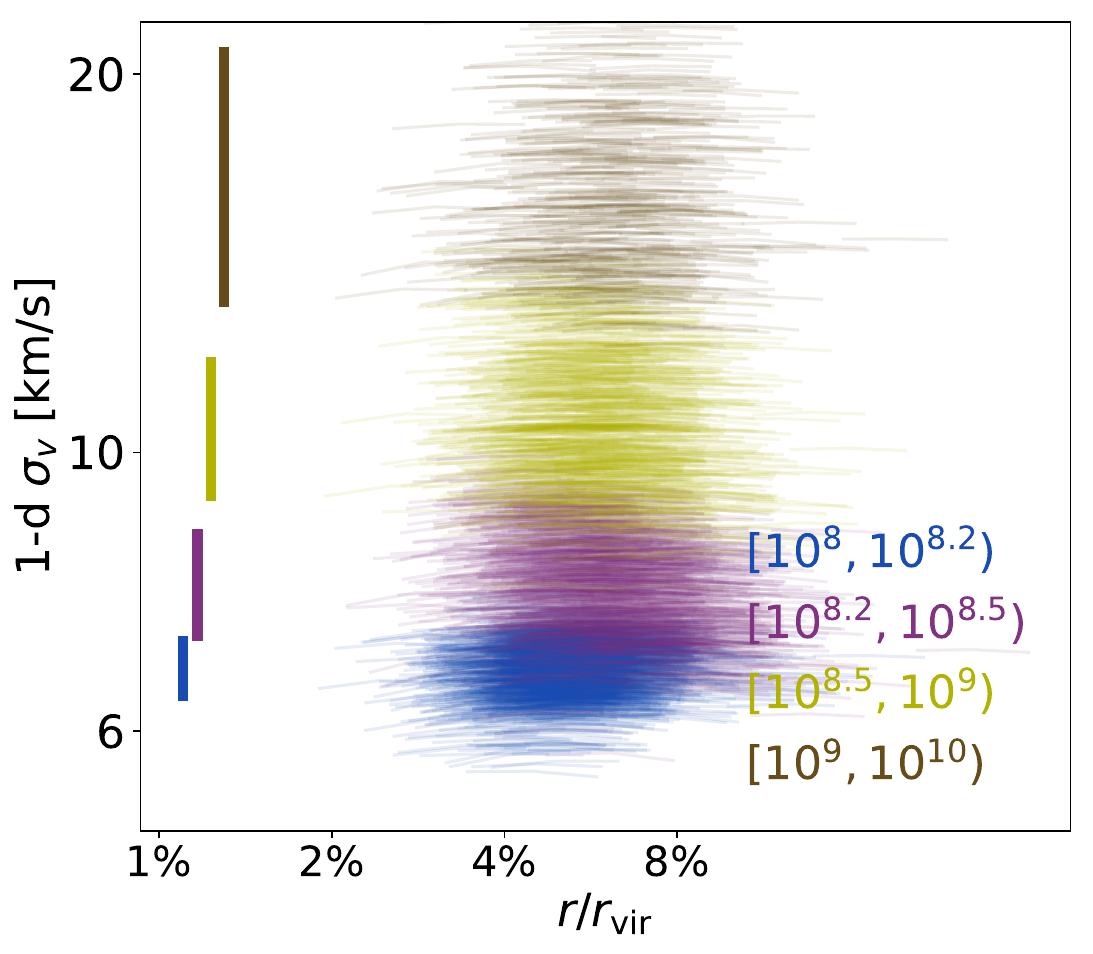}
        \caption{}
        \label{fig:sigmav1}
    \end{subfigure}
    ~
    \begin{subfigure}[t]{\columnwidth}
        \centering
        \includegraphics[width=\textwidth, clip,trim=0.1cm 0cm 0.1cm 0cm]{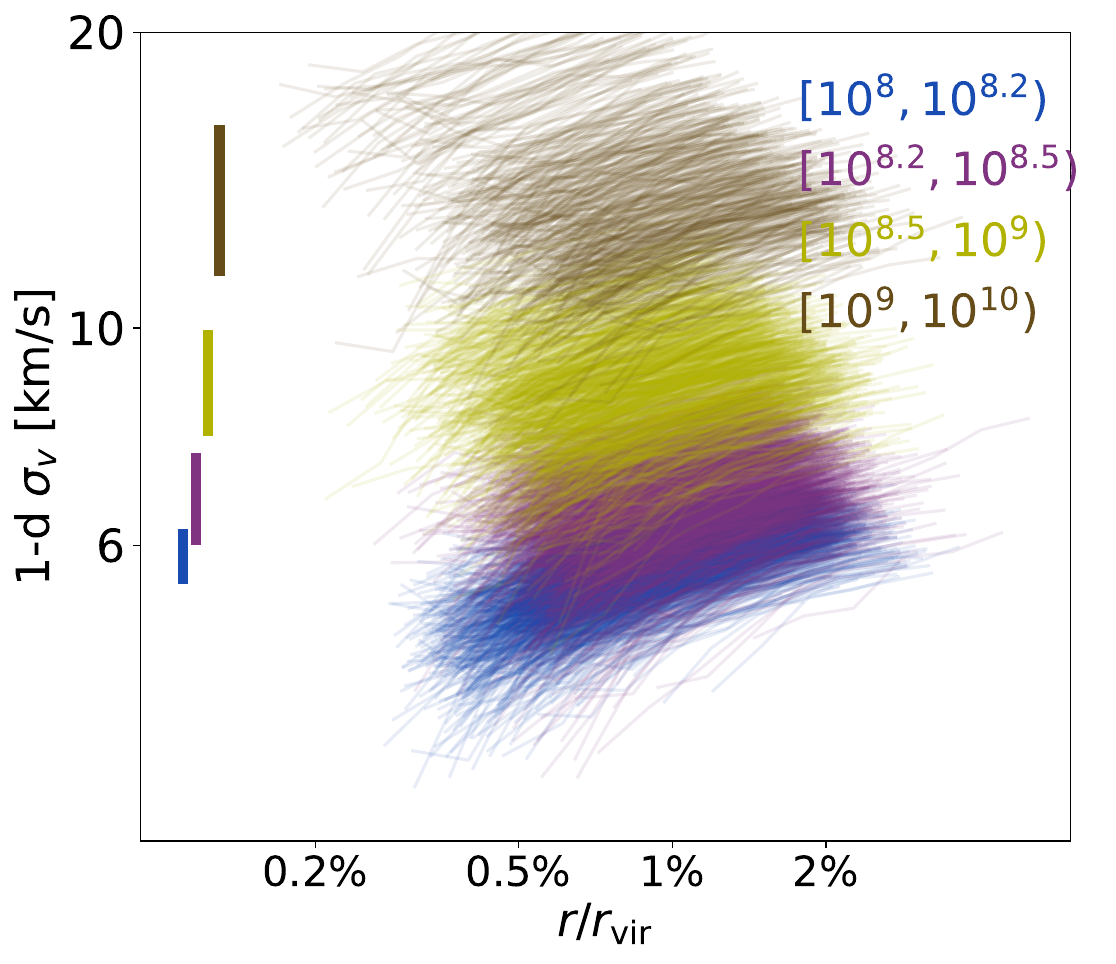}
        \caption{}
        \label{fig:sigmav2}
    \end{subfigure}
    \caption{The typical velocity dispersion of CDM subhalos in each infall-mass group, at $t=4$ Gyr, before host effects kick in. a) shows the $\sigma_v$ near its maximum value for each subhalo. b) shows the $\sigma_v$ near the center of each subhalo. The colored vertical bars mark the $\pm 1\sigma$ scatter of $\sigma_v$ in each panel, and those of panel a) are the range of $\sigma_v$ we use in Table \ref{table:arti}.}
    \label{fig:vdisp}
\end{figure}

We include variations of some figures in the main text in this section, which may be more relevant to other use cases: 
\begin{itemize}
    \item Fig. \ref{fig:vsigma-v} shows the cross section of various SIDM models used, but in the format of $v_{\rm rel} \times \sigma/m$ instead of the cross section $\sigma/m$ alone. Since the particle scattering rate/probability is proportional to $v_{\rm rel} \times \sigma/m$, this version gives more intuitive understanding of the strength of self-interaction at different velocity scales.
    \item Fig. \ref{fig:orbit-check} shows the validation of our rewinding algorithm, that the calculated initial position and velocity of randomly selected subhalos are correct, since their orbits in the simulation can match the input information of infall radii and velocities.
    \item Fig. \ref{fig:kappa2-1pt} is similar to the 1-point function of substructure lensing in Fig. \ref{fig:1pta}, but the y-axis has an additional weight of $\kappa_{\rm sub}^2$ that highlights the large $\kappa_{\rm sub}$ bins, to which substructure lensing is most sensitive. For example, the strong core-collapse in the $\{\sigma_0 = 200, \omega=200\}$ model (purple line) is featured by a sharp increase at high $\kappa_{\rm sub}$ bins, which is not visible in Fig. \ref{fig:1pta} but does show up in Fig. \ref{fig:kappa2-1pt} with the additional weight.
    \item Fig. \ref{fig:tdiff-frac} is similar to Fig. \ref{fig:tdiff}, but showing the time difference for core-collapse in field halos and subhalos as an acceleration (positive) or deceleration (negative) factor to the core-collapse time of field halos.
    \item Fig. \ref{fig:vdisp} shows the 1-d velocity dispersion profiles of each CDM subhalo at $t=4$ Gyr. Panel a) shows the velocity dispersion near its maximum value for each subhalo, and panel b) shows the velocity dispersion in center of the subhalo.
\end{itemize}

\bibliography{apssamp}% Produces the bibliography via BibTeX.

\end{document}